 \documentclass[12pt]{article}
 \usepackage{epsf,latexsym}
 \newcommand{\insertplot}[5]{\begin{figure}
 \hfill\hbox to 0.05in{\vbox to #5in{\vfill
 \inputplot{#1}{#4}{#5}}\hfill}
 \hfill\vspace{-.1in}
 \caption{#2}\label{#3}
 \end{figure}}
 \newcommand{\inputplot}[3]{
 \special{ps: plotfile #1}
}
\newcounter{fixy}
\begin{document}
\newenvironment{fixy}[1]{\setcounter{figure}{#1}}
{\addtocounter{fixy}{1}}
\renewcommand{\thefixy}{\arabic{fixy}}
\renewcommand{\thefigure}{\thefixy\alph{figure}}
\setcounter{fixy}{1}

\title{Sequences of Einstein-Yang-Mills-Dilaton Black Holes}
\vspace{1.5truecm}
\author{
{\bf Burkhard Kleihaus, Jutta Kunz and Abha Sood}\\
Fachbereich Physik, Universit\"at Oldenburg, Postfach 2503\\
D-26111 Oldenburg, Germany}

%\date{September 21, 1995}
%\date{Dezember 18, 1995}

\maketitle
\vspace{1.0truecm}

\begin{abstract}
Einstein-Yang-Mills-dilaton theory possesses sequences of
neutral static spherically symmetric black hole solutions.
The solutions depend on the dilaton coupling constant $\gamma$
and on the horizon.
The SU(2) solutions are labelled
by the number of nodes $n$ of the single gauge field function,
whereas the SO(3) solutions are labelled
by the nodes $(n_1,n_2)$ of both gauge field functions.
The SO(3) solutions form sequences characterized by
the node structure $(j,j+n)$, where $j$ is fixed.
The sequences of magnetically neutral solutions 
tend to magnetically charged limiting solutions.
For finite $j$ the SO(3) sequences tend
to magnetically charged Einstein-Yang-Mills-dilaton solutions
with $j$ nodes and charge $P=\sqrt{3}$.
For $j=0$ and $j \rightarrow \infty$
the SO(3) sequences tend to Einstein-Maxwell-dilaton solutions
with magnetic charges $P=\sqrt{3}$ and $P=2$,
respectively.
The latter also represent the scaled limiting solutions
of the SU(2) sequence.
The convergence of the global properties of the black hole solutions,
such as mass, dilaton charge
and Hawking temperature, is exponential.
The degree of convergence of the matter and metric functions
of the black hole solutions is related to the relative 
location of the horizon to the
nodes of the corresponding regular solutions.
\end{abstract}
\vfill
\noindent {Preprint hep-th/9605109} \hfill\break
\vfill\eject

\section{Introduction}

Motivated by higher-dimensional unified theories,
such as Kaluza-Klein theory or superstring theory,
where a scalar dilaton field arises naturally,
static black hole solutions have been found
in Einstein-Maxwell-dilaton (EMD) theory \cite{gib1,str}.
In contrast to the static charged black holes of Einstein-Maxwell 
theory,
referred to as Reissner-Nordstr\o m (RN) black holes,
the charged EMD black holes exist for arbitrarily small
event horizon, $x_{\rm H}> 0$. 
The ``extremal'' solutions of EMD theory,
obtained in the limit $x_{\rm H}=0$,
possess a naked singularity at the origin.

Recently static black hole solutions have also been
studied in Einstein-Yang-Mills-dilaton (EYMD) theory
\cite{don,lav2,maeda,neill,kks3}.
SU(2) EYMD theory
possesses a sequence of magnetically neutral
static spherically symmetric black hole solutions,
labelled by the number of nodes $n$
of the single gauge field function.
The solutions exist for arbitrary event horizon $x_{\rm H}> 0$ 
\cite{don,lav2,maeda,neill,kks3}.
Here, in the limit $x_{\rm H} \rightarrow 0$
regular particle-like solutions are obtained
\cite{don,lav2,maeda,bizon3,neill,kks3}.

For finite dilaton coupling constant $\gamma$, the sequence
of regular neutral SU(2) EYMD solutions tends to the
``extremal'' EMD solution with the same dilaton coupling
constant $\gamma$ and with charge $P=1$
for large $n$ \cite{lav2,bizon3}.
For finite event horizon $x_{\rm H}$ 
and dilaton coupling constant $\gamma$, the corresponding sequence
of neutral black hole solutions of SU(2) EYMD theory
also tends to a limiting solution for large $n$.
This limiting solution is
the EMD black hole solution with the same event horizon $x_{\rm H}$,
the same dilaton coupling constant $\gamma$,
and with magnetic charge $P=1$ \cite{kks3}.

The magnetically neutral black hole solutions of EYMD theory
have many features in common with those
of Einstein-Yang-Mills (EYM) theory
\cite{volkov,bizon1,kuenzle1,kks2}.
In fact, in the limit of vanishing dilaton coupling constant 
$\gamma$,
the EYMD black hole solutions approach those of EYM theory.
In SU(2) EYM theory, the limiting solutions are the RN
black holes with charge $P=1$ and event horizon $x_{\rm H}$,
for $x_{\rm H} > 1$ \cite{bfm,sw,bfm2,sw2}.
For sequences of black holes with $x_{\rm H} < 1$
and for the regular sequence \cite{bm},
the limiting solutions are different \cite{bfm,sw,bfm2,sw2}.

The non-abelian SU(2) black hole solutions
do not possess a global YM charge.
Since they are characterized not only by their mass 
but in addition by an interger $n$, they
represent (unstable \cite{strau2,volkov5,lav,volkov4,strau3})
counterexamples to the ``no-hair conjecture''.
In contrast, SU(2) black holes with non-zero YM charge are 
uniquely characterized by their mass and charge.
Indeed, for SU(2) all static spherically symmetric
EYM black hole solutions with non-zero YM charge are embedded
Reissner-Nordstr\o m  solutions \cite{ersh,popp}.
This ``non-abelian baldness theorem'' no longer holds
for SU(3), where black hole solutions with both Coulomb-like
and essentially non-abelian gauge field configurations
exist \cite{volkov3,gal}.
While these black hole solutions \cite{volkov3,gal}
correspond to SU(2)$\times$U(1) solutions,
the properties of genuine SU(3) black hole 
solutions have remained largely unknown.

The genuine magnetically neutral static spherically symmetric 
black hole solutions of SU(3) EYM and EYMD theory
are based on the SO(3) embedding.
These solutions can be labelled by the nodes $(n_1,n_2)$
of both gauge field functions \cite{kuenzle,kks2,kks3}.
We have constructed SU(3) EYM and EYMD
black hole solutions with a small number of nodes previously
\cite{kks2,kks3},
and we have identified two sequences of solutions
together with their limiting solutions \cite{kks3}.
These sequences are the sequence $(0,n)$, containing the two lowest 
genuine SU(3) solutions,
and the sequence $(n,n)$ of scaled SU(2) solutions.
The limiting solutions of these two sequences
are again EMD solutions with the same event horizon
and dilaton coupling constant, but with magnetic
charges $P=\sqrt{3}$ and $P=2$, respectively.
The limiting solutions for $\gamma=0$ 
and  $x_{\rm H} > P$
are the corresponding RN black holes.

For SU(3) EYMD theory it is not a priori clear,
in which way the black hole solutions 
with general node structure $(n_1,n_2)$
assemble to form sequences, converging to limiting solutions.
One may expect though, that the limiting solutions 
of the general sequences of genuine SU(3) solutions
are again magnetically charged solutions.

Here we construct genuine SU(3) black hole solutions
with general node structure $(n_1,n_2)$.
We show, that these solutions fall into sequences
with node structure $(j,j+n)$,
and we identify the limiting solutions 
of the new sequences with fixed finite $j$ with the
known 
magnetically charged SU(3) black hole solutions
\cite{gal,volkov3}.
We study the convergence of the matter and metric functions 
and of the global properties for these sequences of solutions.

In section 2 we derive the SU(3) EYMD equations of motion
based on spherically symmetric ans\"atze for the
metric and the matter fields.
We discuss the boundary conditions
and obtain relations between the metric and the dilaton field.
In section 3 we consider SU(2) EYMD black hole
solutions. We investigate their dependence on the
parameters $x_{\rm H}$ and $\gamma$.
We relate the degree of convergence of the matter and
metric functions of the black hole solutions 
to the relative location of the horizon to the nodes
of the corresponding regular solutions.
We show, that the global properties of the solutions,
such as mass, dilaton charge and Hawking temperature
\cite{moss,don,maeda,kks3},
converge exponentially to those of the limiting solutions.
In section 4 we discuss SU(3) EYMD black hole solutions.
We identify the general set of sequences of 
genuine SU(3) solutions.
We demonstrate that the convergence properties
of these sequences are analogous to those
of the SU(2) sequence.
In section 5 we present our conclusions.
In appendices A and B we review the limiting magnetically
charged black hole solutions of EMD and EYMD theory.
We demonstrate that the ``extremal'' EMD and EYMD solutions
agree in many respects.

\section{SU(3) Einstein-Yang-Mills-Dilaton Equations of Motion}

\subsection{SU(3) Einstein-Yang-Mills-Dilaton action}

We consider the SU(3) Einstein-Yang-Mills-dilaton action
\begin{equation}
S=S_G+S_M=\int L_G \sqrt{-g} d^4x + \int L_M \sqrt{-g} d^4x
\ \label{action}  \end{equation}
with
\begin{equation}
L_G=\frac{1}{16\pi G}R
\ , \end{equation}
and matter Lagrangian
\begin{equation}
L_M=-\frac{1}{2}\partial_\mu \Phi \partial^\mu \Phi
 -e^{2 \kappa \Phi }\frac{1}{2} {\rm Tr} (F_{\mu\nu} F^{\mu\nu})
\ , \label{lagm} \end{equation}
where
\begin{equation}
F_{\mu\nu}= \partial_\mu A_\nu - \partial_\nu A_\mu
            - i g [A_\mu,A_\nu]
\ , \end{equation}
\begin{equation}
A_\mu = \frac{1}{2} \lambda^a A_\mu^a
\ , \end{equation}
and $g$ and $\kappa$ are the gauge and dilaton coupling constants,
respectively.

Variation of the action eq.~(\ref{action}) with respect to the 
metric
$g^{\mu\nu}$ leads to the Einstein equations
\begin{equation}
G_{\mu\nu}= R_{\mu\nu}-\frac{1}{2}g_{\mu\nu}R = 8\pi G T_{\mu\nu}
\   \end{equation}
with stress-energy tensor
\begin{eqnarray}
T_{\mu\nu}&=&g_{\mu\nu}L_M -2\frac{\partial L_M}{\partial g^{\mu\nu}} 
 \nonumber \\
  &=& \partial_\mu \Phi \partial_\nu \Phi 
   -\frac{1}{2} g_{\mu\nu} \partial_\alpha \Phi \partial^\alpha \Phi
    +2e^{2 \kappa \Phi}  {\rm Tr} 
    ( F_{\mu\alpha} F_{\nu\beta} g^{\alpha\beta}
   -\frac{1}{4} g_{\mu\nu} F_{\alpha\beta} F^{\alpha\beta})
\ , \end{eqnarray}
and variation with respect to the gauge field $A_\mu$ 
and the dilaton field $\Phi$ 
leads to the matter field equations.

\subsection{Static spherically symmetric ans\"atze}

To construct static spherically symmetric black hole solutions
we employ Schwarz\-schild-like coordinates and adopt
the spherically symmetric metric
\begin{equation}
ds^2=g_{\mu\nu}dx^\mu dx^\nu=
  -A^2N dt^2 + N^{-1} dr^2 + r^2 (d\theta^2 + \sin^2\theta d\phi^2)
\ , \label{metric} \end{equation}
with
\begin{equation}
N=1-\frac{2m}{r}
\ . \end{equation}

For SU(2) EYMD theory, the static spherically symmetric ansatz
for the gauge field with vanishing time component is,
%\cite{bm},
\begin{eqnarray}
A_0 &=& 0
\ , \nonumber\\
A_i &=& \frac{1-w(r)}{2rg} (\vec e_r \times \vec \tau)_i
\ , \label{su2} \end{eqnarray}
with the SU(2) Pauli matrices
$\vec \tau = (\tau_1,\tau_2,\tau_3)$.
For SU(3) EYMD theory,
generalized spherical symmetry for the gauge field is realized
by embedding the SU(2) or the SO(3) generators in SU(3).
In the SU(2)-embedding, the ansatz is given by (\ref{su2})
with $\vec \tau$ replaced by
$\vec \lambda = (\lambda_1,\lambda_2,\lambda_3)$.
In the SO(3)-embedding,
the corresponding ansatz for the gauge field
with vanishing time component is
\begin{eqnarray}
A_0 &=& 0
\ , \nonumber\\
A_i &=& \frac{2-K(r)}{2rg} (\vec e_r \times \vec \Lambda)_i
       +\frac{H(r)}{2rg} \Bigl[ (\vec e_r \times \vec \Lambda)_i,
       \vec e_r \cdot \vec \Lambda) \Bigr]_+
\ , \label{so3} \end{eqnarray}
where $[\ , \ ]_+$ denotes the anticommutator, and
$\vec \Lambda = (\lambda_7,-\lambda_5,\lambda_2)$.
For the dilaton field we take $\Phi = \Phi(r)$.

\subsection{Field equations}

For the above ans\"atze we now evaluate the $tt$ and $rr$
components of the Einstein equations.
The metric (\ref{metric}) yields the components of
the Einstein tensor
\begin{equation}
G_{tt}=\frac{2m'A^2N}{r^2}=8\pi G T_{tt}=-8\pi GA^2NL_M
\ , \label{tt} \end{equation}
and
\begin{equation}
G_{rr} =  -\frac{G_{tt}}{A^2N^2}+\frac{2}{r} \frac{A'}{A} 
   =8\pi G T_{rr}
\ , \label{rr} \end{equation}
where the prime indicates the derivative with respect to $r$.

The spherically symmetric ans\"atze for the fields,
eqs. (\ref{su2}) and (\ref{so3}),
yield the $tt$ and $rr$ components of the
stress-energy tensor,
$T_{tt}=-A^2 N L_M$, with
\begin{eqnarray}
L_M &=& -\frac{1}{2} N \Phi'^2 
    - \frac{e^{2 \kappa \Phi}}{g^2 r^2} \left[ N w'^2
    + \frac{1}{2 r^2} \left(w^2-1 \right)^2  \right] 
\   \end{eqnarray}
and $T_{rr}$
\begin{eqnarray}
T_{rr} 
 &=& \frac{1}{N}L_M + \Phi'^2
    + \frac{2e^{2 \kappa \Phi}}{g^2 r^2} w'^2
\   \end{eqnarray}
for the SU(2) embedding, and with 
\begin{eqnarray}
L_M &=& - \frac{1}{2} N \Phi'^2 
    - \frac{e^{2 \kappa \Phi}}{g^2 r^2} \left[ N (K'^2 +H'^2)
    + \frac{1}{8 r^2} \left( 
     \left(K^2+H^2-4 \right)^2 +12 K^2 H^2 \right) \right] 
\   \end{eqnarray}
and
\begin{eqnarray}
T_{rr}  
 &=& \frac{1}{N}L_M +  \Phi'^2
    + \frac{2e^{2 \kappa \Phi}}{g^2 r^2} (K'^2 +H'^2)
\   \end{eqnarray}
for the SO(3) embedding.

Let us now introduce the dimensionless coordinate $x$,
\begin{equation}
x=\frac{g}{\sqrt{4\pi G}} r,
\end{equation}
(the prime indicating now the derivative
with respect to $x$),
the dimensionless mass function $\mu(x)$,
\begin{equation}
\mu=\frac{g}{\sqrt{4\pi G}} m = \frac{g m_{\rm Pl}}{\sqrt{4\pi}} m
\  , \end{equation}
the dimensionless dilaton field $\phi$,
\begin{equation}
\phi = \sqrt{4\pi G} \Phi
\  , \end{equation}
and the dimensionless coupling constant $\gamma$,
\begin{equation}
\gamma =\kappa/\sqrt{4\pi G}
\  . \end{equation}
The choice $\gamma=1$ corresponds to string theory,
whereas $4+n$ dimensional Kaluza-Klein theory
has $\gamma^2=(2+n)/n$ \cite{gib1}.

With these definitions we then obtain the 
dimensionless EYMD equations.
For SU(2) \cite{don,lav2,maeda,bizon3,neill}
the equations for the metric functions are
\begin{equation}
\mu'= \frac{1}{2} N x^2 \phi'^2 + e^{2 \gamma \phi } 
\left[ N w'^2 
  + \frac{1}{2 x^2} \left(w^2-1 \right)^2 \right]
\ , \label{eqmu2} \end{equation}
\begin{equation}
 A' =  \frac{2}{x} \left[\frac{1}{2} x^2 \phi'^2 
                 + e^{2 \gamma \phi } w'^2 \right] A
\ , \label{eqa2} \end{equation}
and the equations for the matter field functions are
\begin{equation}
(e^{2 \gamma \phi } ANw')' 
= \frac{e^{2 \gamma \phi }}{x^2} A w \left( w^2-1 \right)
\ , \label{eqw} \end{equation}
\begin{equation}
(AN x^2 \phi')' = 2 \gamma A e^{2 \gamma \phi } 
\left[ N w'^2
  + \frac{1}{2 x^2} \left(w^2-1 \right)^2\right]
\ . \label{eqdil2} \end{equation}

For SO(3) \cite{kks3},
the equations for the metric functions are
\begin{equation}
\mu'= \frac{1}{2} N x^2 \phi'^2 + e^{2 \gamma \phi } 
\left[ N (K'^2 +H'^2)
  + \frac{1}{8 x^2} \left(
     \left(K^2+H^2-4 \right)^2 +12 K^2 H^2 \right)\right]
\ , \label{eqmu3} \end{equation}
\begin{equation}
 A' =  \frac{2}{x} \left[\frac{1}{2} x^2 \phi'^2 
                 + e^{2 \gamma \phi } (K'^2 +H'^2) \right] A
\ , \label{eqa3} \end{equation}
and the equations for the matter field functions are
\begin{equation}
(e^{2 \gamma \phi } ANK')' 
= \frac{e^{2 \gamma \phi }}{4 x^2} A K \left( K^2+7H^2-4 \right)
\ , \label{eqk} \end{equation}
\begin{equation}
(e^{2 \gamma \phi } ANH')' 
= \frac{e^{2 \gamma \phi }}{4 x^2} A H \left( H^2+7K^2-4 \right)
\ , \label{eqh} \end{equation}
\begin{equation}
(AN x^2 \phi')' = 2 \gamma A e^{2 \gamma \phi } 
\left[ N (K'^2 +H'^2)
  + \frac{1}{8 x^2} \left(
     \left(K^2+H^2-4 \right)^2 +12 K^2 H^2 \right)\right]
\ . \label{eqdil3} \end{equation}

With help of eq.~(\ref{eqa2}) for SU(2) 
or eq.~(\ref{eqa3}) for SO(3)
the metric function $A$
can be eliminated from the matter field equations.
Note that the SU(2) equations are symmetric with respect to
the transformation $w(x) \rightarrow -w(x)$,
and the SO(3) equations are symmetric with respect to both
an interchange of the functions $K(x)$ and $H(x)$,
and to the transformations $K(x) \rightarrow -K(x)$,
and $H(x) \rightarrow -H(x)$,
thereby yielding degenerate solutions.

\subsection{Boundary conditions}

Let us now consider the boundary conditions
for the black hole solutions of EYMD theory.
Requiring asymptotically flat solutions implies
that the metric functions $A$ and $\mu$ both
must approach a constant at infinity.
We here adopt
\begin{equation}
A(\infty)=1
\ , \end{equation}
thus fixing the time coordinate.
For magnetically neutral solutions
the gauge field functions approach a vacuum configuration,
i.~e.~for SU(2)
\begin{equation}
w(\infty)=\pm 1 
\ , \label{bc0} \end{equation}
and for SO(3)
\begin{equation}
K(\infty)=\pm 2 \ , \quad H(\infty)=0
\ , \label{bc1} \end{equation}
\begin{equation}
K(\infty)=0 \ , \quad H(\infty)=\pm 2
\ . \label{bc2} \end{equation}
For the dilaton field we choose \cite{don,lav2,bizon3,neill,kks3,fn0}
\begin{equation}
\phi(\infty) = 0
\ . \label{bc3} \end{equation}
Magnetically charged solutions require different
boundary conditions for the gauge field functions.

The existence of a regular event horizon at $x_{\rm H}$
requires for the metric functions
\begin{equation}
\mu(x_{\rm H})= \frac{x_{\rm H}}{2}
\ , \end{equation}
and $A(x_{\rm H}) < \infty $.
The matter functions must satisfy at the horizon $x_{\rm H}$
\begin{eqnarray}
 N'w'&=& \frac{1}{x^2} w \left( w^2-1 \right)
\ , \\
 N'\phi'&=& \frac{\gamma e^{2 \gamma \phi } }{x^4}
     \left(w^2-1 \right)^2
\   \end{eqnarray}
for SU(2) and
\begin{eqnarray}
 N'K'&=& \frac{1}{4 x^2} K \left( K^2+7H^2-4 \right)
\ , \\
 N'H'&=& \frac{1}{4 x^2} H \left( H^2+7K^2-4 \right)
\ , \\
 N'\phi'&=& \frac{\gamma e^{2 \gamma \phi } }{4 x^4}
    \left[ \left(K^2+H^2-4 \right)^2 +12 K^2 H^2 \right]
\   \end{eqnarray}
for SO(3).

On considering the regular solutions,
we observe that they satisfy the same boundary conditions at infinity
as the black hole solutions.
However, at the origin, regularity of the solutions requires
\begin{equation}
\mu(0)=0
\ , \end{equation}
and the gauge field functions must satisfy
\begin{equation}
w(0)=\pm 1 
\ , \label{bc4} \end{equation}
for SU(2) and 
\begin{equation}
K(0)=\pm 2 \ , \quad H(0)=0
\ , \label{bc5} \end{equation}
\begin{equation}
K(0)=0 \ , \quad H(0)=\pm 2
\ , \label{bc6} \end{equation}
for SO(3),
and the dilaton field satisfies
\begin{equation}
\phi'(0) = 0
\ . \label{bc7} \end{equation}
As in EYM theory \cite{bm,kks2},
it is sufficient to consider solutions
with $w(0)=1$ for SU(2) and
with $K(0)=2$ and $H(0)=0$ for SO(3).

\subsection{Relations between metric and dilaton field}

Let us now derive relations between the dilaton field
and the metric functions for purely magnetic (or electric)
gauge fields of a general gauge group.
For that purpose, we return to eq.~(\ref{lagm}) and
introduce the gauge field Lagrangian $L_A$
\begin{equation}
L_M=-\frac{1}{2}\partial_\mu \Phi \partial^\mu \Phi
 +e^{2 \kappa \Phi } L_A
\ . \label{laga} \end{equation}
 From the matter field action we then obtain
the equation of motion for a static spherically symmetric dilaton 
field
\begin{equation}
(AN r^2 \Phi')' = - 2 \kappa A e^{2 \kappa \Phi } r^2 L_A
\ , \label{eqdil4} \end{equation}
where for the moment we retain the dimensioned quantities.
Next we return to the $tt$ component of the
Einstein equations, eq.~(\ref{tt}),
\begin{equation}
m'=4\pi G \left( \frac{1}{2} N r^2 \Phi'^2 
               - \eta e^{2 \kappa \Phi} r^2 L_A \right)
\ , \label{ttt} \end{equation}
where $\eta=1$ ($\eta=-1$) for purely magnetic (electric) gauge 
fields.
With the help of this equation (\ref{ttt}),
we eliminate the gauge field from the dilaton equation (\ref{eqdil4})
\begin{equation}
(AN r^2 \Phi')' = 2 \eta \kappa A \left( \frac{m'}{4 \pi G} 
   - \frac{1}{2} N r^2 \Phi'^2 \right)
\ . \label{eqdil6} \end{equation}
We then consider the contracted Einstein equations.
For purely magnetic (or electric) gauge fields
the contracted gauge field stress energy tensor vanishes,
yielding for the curvature scalar
\begin{equation}
 R = 8 \pi G N \Phi'^2 
\ . \label{R} \end{equation}
Thus the r.h.s of the dilaton equation (\ref{eqdil6})
can be expressed purely in terms of metric functions
\begin{equation}
(AN r^2 \Phi')' = \eta \frac{2 \kappa}{4 \pi G} 
 \left( A m' - \frac{1}{4} A r^2 R \right)
\ . \label{eqdil7} \end{equation}
With
\begin{equation}
 A r^2 R = - \left[ r^2 (2 A' N + A N') \right]' + 4 A m'
\   \label{R1} \end{equation}
we then obtain for the dilaton field the equation 
\begin{equation}
(AN x^2 \phi')' = \frac{\eta}{2} \gamma 
  \left( x^2 (2 A' N + A N') \right)'
\ , \label{rel1} \end{equation}
where we have returned to dimensionless quantities.
This equation can be integrated, yielding
\begin{equation}
 \phi' = \frac{\eta}{2} \gamma 
  \left( \ln ( A^2 N) \right)' + \frac{C}{A N x^2}
\ , \label{rel2} \end{equation}
where $C$ is an integration constant.
In the following we only consider purely magnetic gauge fields
($\eta=1$).

We are now interested in global relations
between the metric and the dilaton field,
involving the mass $\mu(\infty)$
and the dilaton charge $D$, which we define via
\begin{equation}
\phi(x) \stackrel {x\rightarrow \infty} {\longrightarrow}
-\frac{D}{x}
\ . \label{bc8} \end{equation}
We further introduce the dimensionless Hawking temperature $T$
for the metric (\ref{metric}) \cite{moss,don,maeda,kks3}
\begin{equation}
T = T_{\rm S} A (1-2 \mu')|_{x_{\rm H}}
  = T_{\rm S} x_{\rm H} A N' |_{x_{\rm H}}
\ , \label{thaw} \end{equation}
where $T_{\rm S} = (4 \pi x_{\rm H})^{-1}$
is the Hawking temperature of the Schwarzschild black hole.

We now integrate eq.~(\ref{rel1}) from the horizon to infinity.
The asymptotic behaviour of the dilaton field
is given by eq.~(\ref{bc8}).
The metric functions approach constants asymptotically,
$A(\infty)=1$ and $N(\infty)=1$ (see sect.~2.4),
while their derivatives tend to zero \cite{don,lav2,bizon3}
\begin{equation}
A'(x) \stackrel {x\rightarrow \infty} {\longrightarrow} 
 O\left( \frac{1}{x^3} \right) \ , \ \ \
N'(x) \stackrel {x\rightarrow \infty} {\longrightarrow} 
 \frac{2 \mu(\infty)}{x^2} +
 O\left( \frac{1}{x^3} \right)
\ . \label{bc10} \end{equation}
At the horizon, the metric function $N(x_{\rm H})$ vanishes,
while $N'(x_{\rm H})$ is related 
to the Hawking temperature (\ref{thaw}).
These boundary conditions then yield for the black holes 
the relation \cite{kks3,fn1}
\begin{equation}
D = \gamma\left(\mu(\infty)-\frac{1}{2}x_{\rm H}^2 A N' |_{x_{\rm H}}
  \right) 
  = \gamma \mu(\infty) \left( 1 - \frac{\mu_{\rm S}}{\mu(\infty)}
   \frac{T}{T_{\rm S}}  \right)
\ , \label{res3} \end{equation}
where $\mu_{\rm S}= x_{\rm H}/2$
is the mass of the Schwarzschild black hole.
Relation (\ref{res3}) holds for static,
spherically symmetric, magnetic gauge fields
of general gauge groups (provided the metric functions
have the proper asymptotic behaviour).
Thus besides the magnetically neutral SU(2) and SO(3) black holes
considered here, relation (\ref{res3})
also holds for the magnetically charged EMD
black holes \cite{gib1,str} and the magnetically charged
SU(3) EYMD black holes \cite{gal}.

The integration constant $C$ in relation (\ref{rel2})
is given by
\begin{equation}
C=D-\gamma \mu(\infty)
\ . \label{C} \end{equation}
According to (\ref{res3}) it does not vanish for black holes 
in general \cite{fn2}.
Integration of eq.~(\ref{rel2}) therefore
does not yield a simple expression for black holes.

Let us now consider the above relations for the regular solutions.
By integrating eq.~(\ref{rel1}) from zero to infinity
we obtain the simple relation \cite{fn3}
\begin{equation}
D = \gamma \mu(\infty)
\ . \label{res1} \end{equation}
Consequently the integration constant $C$ in eq.~(\ref{rel2})
vanishes for regular solutions.
Integrating eq.~(\ref{rel2}) 
from $x$ to infinity therefore gives the simple relation 
\cite{fn3,fn2}
\begin{equation}
\phi(x) = \frac{1}{2} \gamma \ln( A^2 N)
   =\frac{1}{2} \gamma \ln(-g_{tt})
\ . \label{res2} \end{equation}
Again, these relations are valid for 
static, spherically symmetric, magnetic gauge fields
of general gauge groups.

\section{SU(2) Einstein-Yang-Mills-Dilaton Black Holes}

SU(2) EYMD theory possesses a sequence of magnetically neutral
static spherically symmetric black hole solutions,
which can be labelled by the number of nodes $n$
of the gauge field function, $w_n(x)$
\cite{don,lav2,maeda,neill,kks3}.
These black hole solutions exist 
for arbitrary dilaton coupling $\gamma$ and
for arbitrary event horizon $x_{\rm H}> 0$.
In the limit $x_{\rm H} \rightarrow 0$,
regular particle-like solutions are obtained
\cite{don,lav2,maeda,bizon3,neill,kks3}.

The sequences of SU(2) EYMD solutions converge to limiting solutions.
In the following, we demonstrate numerically
that in the limit $n \rightarrow \infty$,
given a dilaton coupling constant $\gamma$
and an event horizon $x_{\rm H}$,
the corresponding sequence of neutral SU(2) EYMD black hole solutions 
tends to the charged EMD black hole solution
with the same dilaton coupling constant $\gamma$,
the same event horizon $x_{\rm H}$
and with magnetic charge $P=1$ \cite{kks3}.
This generalizes the earlier observation \cite{lav2,bizon3},
that for a given coupling constant $\gamma$,
the sequence of regular neutral SU(2) EYMD 
solutions converges to the charged ``extremal'' EMD
solution with the same coupling constant $\gamma$
and with magnetic charge $P=1$.
(For completeness
we review the EMD black hole solutions in appendix A.)

\boldmath
\subsection{$\gamma=1$}
\unboldmath

Since the coupling constant $\gamma=1$ corresponds to
the case of the low energy effective action of string theory,
it deserves special attention. SU(2) black hole solutions
for $\gamma=1$ have been studied in \cite{don,neill},
where in \cite{neill} the effect of a small dilaton
mass has additionally been considered.

We first discuss the properties of the black hole solutions
of the stringy SU(2) sequence for a given event horizon.
We relate the degree of convergence of the $n$-th EYMD solution
with respect to the limiting EMD solution 
to the location of the event horizon.
We observe, that the smaller the event horizon $x_{\rm H}$, 
the larger is the value of the lowest $n$
of the solution for which the functions have converged well.

Let us begin with a small event horizon $x_{\rm H}=0.01$.
In Figs.~1a-e we present the lowest black hole solutions 
with an odd number of nodes, $n=1$, 3, 5 and 7
for this event horizon.
(Table~1 shows their dimensionless mass.)
In particular, we present the EYMD matter functions $w_n$ and 
$\phi_n$
and metric functions $N_n$, $A_n$ and $P_n^2$,
where the charge function $P_n^2$ is obtained from
the metric coefficient $g_{tt}$ according to \cite{don}
\begin{equation}
g_{tt} = - \left( 1 - \frac{2\mu(\infty)}{x} \sqrt{1+\frac{D^2}{x^2}} 
 +\frac{P_n^2}{x^2} \right)
\ , \label{p2eq} \end{equation}
such that $P=1$ in the abelian limiting case.
For comparison we also show the corresponding functions
of the limiting EMD black hole solution,
having the same horizon but magnetic charge $P=1$.

Fig.~1a shows the EYMD gauge field functions $w_n$
and the EMD function $w_\infty \equiv 0$.
The value of $w_n$ at the horizon decreases with increasing $n$
towards the limiting value of zero.
The location of the innermost node of $w_n$ decreases with $n$,
while the location of the outermost node of $w_n$ increases with $n$,
and thus $w_n$ approaches the limiting value of zero in an
increasingly larger region inbetween.
The gauge field function $w_7$ is already close to zero
over a region of many orders of magnitude.
Obviously, with increasing $n$
the EYMD gauge field functions $w_n$ tend (nonuniformly) to
the EMD function $w_\infty \equiv 0$.

As seen in Fig.~1b, with increasing $n$
the EYMD dilaton functions $\phi_n$
also tend to the EMD function $\phi_\infty $.
The EYMD dilaton functions $\phi_n$
differ significantly from the EMD function $\phi_\infty $ only
in an interior region, which decreases with $n$.
Already for $\phi_7$ there is no notable deviation
from the limiting function to be observed.

Turning to the metric functions,
we observe a similar pattern of convergence.
As seen in Fig.~1c,
the EYMD functions $N_n$ tend to the EMD function $N_\infty $.
The limiting EMD function $N_\infty $ rises from zero at the event 
horizon
to a plateau $N_\infty  \approx 1/4$, and then rises further.
Again, significant deviations between the EYMD and the EMD functions
occur only in an interior region,
which decreases with $n$.
There the EYMD functions $N_n$ develop a peak,
decreasing in size with $n$.
Again, for $N_7$ there is no notable deviation
from the limiting function to be observed.

As seen in Fig.~1d,
the EYMD metric functions $A_n$ also tend to the EMD function 
$A_\infty $.
Again, significant deviations between the EYMD and the EMD functions
occur only in a decreasing interior region,
and for $A_7$ there is no notable deviation
from the limiting function to be observed.

In Fig.~1e the EYMD magnetic charge functions $P_n^2$
are seen and compared to the constant
EMD magnetic charge, $P_\infty  \equiv 1$.
The magnetic charge $P_n$ tends to the limiting
value $P_\infty $ in a large inner region, roughly
limited by the outermost node of the EYMD gauge field function
$w_n$. Beyond this region, the magnetic charge function
decays rapidly to zero.
(Only for $n=1$ there is a significant deviation
from the limiting magnetic charge even in the inner region.)

To understand this relation between the gauge field functions and the
charge functions,
we insert the asymptotic expansion
(for large $x$) of the EYMD functions in eq.~(\ref{p2eq}).
This yields for the magnetic charge function 
\begin{equation}
P^2_n  = \frac{4}{3} \mu (\infty) D^2 \frac{1}{x}
 + (w_{n (1)}^2 +\frac{2}{3} \mu^2 (\infty) D^2)\frac{1}{x^2} 
 + O(\frac{1}{x^3})
\end{equation}
where the constant $w_{n (1)}$ is the 
($1/x$) expansion coefficient of the gauge field function $w_n$,
$w_n = (-1)^n + w_{n (1)}/x + O(1/x^2)$. 
The value of $w_{n (1)}$ is larger than $\mu (\infty)$ and $D$
by six orders of magnitude,
therefore the magnetic charge function has (up to a sign)
approximately the same expansion coefficient as the gauge field 
function. 
The region where $w_n$ reaches its
asymptotic value is thus the same as the region
where $P_n$ decays to zero, 
i.~e.~the region beyond the outermost node of the gauge field 
function. 
Indeed, comparing the functions $P_n^2$ and $(w_n+1)^2$,
shown in Fig.~1e for $n=7$,
we observe, that they are very close beyond the
outermost node of the gauge field function.

To gain more insight into the above observed pattern
of convergence, let us consider for a moment the regular solutions.
In Figs.~2a-b we show the regular EYMD gauge field functions $w_n$
and metric functions $N_n$ for $n=1,3,5$ and $7$.
The sequence of regular solutions converges to the
``extremal'' EMD solution \cite{lav2,bizon3}, 
which is also shown. 

Let us now discuss some features of these solutions.
With increasing $n$, the innermost node $z_n^{(1)}$
of the EYMD gauge field functions $w_n$ decreases exponentially
to zero, as seen in Fig.~2a (see also Table~2), 
whereas the outermost node increases exponentially.
The boundary conditions for the gauge field functions $w_n$
do not allow for a uniform convergence
to the EMD function $w_\infty  \equiv 0$.
But the region where both EYMD and EMD functions are close
increases exponentially with $n$.
The EYMD metric functions $N_n$
approach the value one at the origin, whereas the EMD function 
$N_\infty $
monotonically approaches one quarter.
Therefore there is again no uniform convergence of the EYMD functions
to the limiting EMD function possible.
But, as seen in Fig.~2b,
the region, where the EYMD and EMD functions closely agree,
again increases exponentially with $n$.
We observe again that strong deviations of the functions $N_n$ 
from the limiting function $N_\infty $ 
occur only in an interior region, 
decreasing exponentially with $n$.
In the vicinity of the innermost node $z_n^{(1)}$ of $w_n$,
the function $N_n$ has already come quite close to $N_\infty $, 
and beyond the second node $z_n^{(2)}$ of $w_n$ both functions
closely agree (with the exception of $N_1$).

The EYMD dilaton functions $\phi_n$ approach constant values
close to the origin which decrease linearly with $n$.
Further from the origin they approach the limiting
EMD function $\phi_\infty $.
The interior regions, where the functions $\phi_n$
deviate significantly from the limiting solution $\phi_\infty $,
are limited by the respective  
location of the innermost node $z_n^{(1)}$ of $w_n$
(with the exception of $\phi_1$).
The EYMD metric functions $A_n$ exhibit an analogous
pattern with respect to their convergence to the EMD function 
$A_\infty $.
They deviate significantly from the limiting function
in an interior region, extending slightly beyond the innermost
node $z_n^{(1)}$ of $w_n$ (with the exception of $A_1$).

Turning now back to the EYMD black hole solutions
we observe, that the regular solutions have considerable
influence on the black hole solutions for small horizon size
and that the innermost nodes 
of the regular EYMD gauge field functions
largely determine the pattern of convergence.	
The location of these innermost nodes 
$z_n^{(1)}$ is presented in Table~2.
In particular we note, that for the black hole solutions
of Figs.~1, the location of the innermost node of
the regular gauge field function with seven nodes 
is smaller than the event horizon $x_{\rm H}=0.01$.
This fact has important consequences.

Considering the black hole gauge field functions $w_n$,
we see that they are large at the horizon as long
as the horizon is small compared to the location of the innermost
 node
$z_n^{(1)}$ of the corresponding $n$-th regular gauge field function.
Thus $w_1-w_5$ are large at the horizon,
whereas $w_7$ is small. Consequently, 
$w_7$ deviates strongly from the limiting function 
only in the far region where, beyond the last node, 
it must approach its boundary value.
The black hole dilaton functions $\phi_n$
deviate significantly from the limiting function
only in the interior region between the event horizon and the 
location
of the innermost node $z_n^{(1)}$ 
of the corresponding $n$-th regular gauge
field function.
Thus there is no notable deviation left for $\phi_7$.
The same is true for the metric functions $N_7$ and $A_7$.
Further, we now recognize that the peak in the functions $N_1-N_5$,
located shortly behind the event horizon, represents the tendency
of these functions 
towards the corresponding regular metric functions.
Also the functions $N_n$ and $A_n$ for $n<7$ deviate significantly 
from
the limiting functions only in the interior region
between the event horizon and (slightly beyond) the location
of the innermost node 
$z_n^{(1)}$ of the corresponding $n$-th regular gauge
field function.

Choosing different values for the event horizon $x_{\rm H}$
we realize that the above observations are rather general.
Considering solutions with event horizon $x_{\rm H}=0.1$,
we see that already the location of the innermost
node of the fourth regular gauge field function is smaller
than $x_{\rm H}$.
Consequently, there are already for $\phi_4$, $N_4$ and $A_4$
no notable deviations from the limiting functions left.
The convergence of the gauge field functions is somewhat
slower, though $w_5$ is already small at the event horizon.
Increasing the event horizon further to $x_{\rm H}=1$,
shows that only the location of 
the innermost (and only) node of the first regular
solution is still larger than the event horizon.
Consequently, there are now 
already for $\phi_2$, $N_2$ and $A_2$
no notable deviations from the limiting functions left,
while $w_3$ is small at the event horizon.
For still larger horizons, $x_{\rm H}\ge 10$,
no notable deviations are left even for the functions
$\phi_1$, $N_1$ and $A_1$.

We demonstrate this dependence of the degree of convergence
on the event horizon in Fig.~3, where we show
the gauge field function $w_7$
for the event horizons $x_{\rm H} = 0$, 0.01, 0.1 and 1.
A more quantitative analysis 
of the dependence of the degree of convergence
of the functions is shown in Figs.~4a-b.
There the functions $\Delta w_n$ and $\Delta \phi_n$,
defined as the relative
deviations of the EYMD functions $(1-w_n)$ and $\phi_n$
from the EMD functions $(1-w_\infty )$ and $\phi_\infty $ at the 
horizon
\begin{equation}
\Delta w_n(x_{\rm H}) = 
\frac{|(1-w_n(x_{\rm H})) - (1-w_\infty (x_{\rm H}))|}
  {(1-w_\infty (x_{\rm H}))}
= |w_n(x_{\rm H}) - w_\infty (x_{\rm H})| 
\ , \label{delw} \end{equation}
\begin{equation}
\Delta \phi_n(x_{\rm H}) = \frac{|\phi_n(x_{\rm H}) - \phi_\infty 
(x_{\rm H})|}
  {|\phi_\infty (x_{\rm H})|}
\ , \label{delp} \end{equation}
are shown as functions of the horizon for $n=1-5$. 
Also shown are the locations of the two innermost nodes $z_n^{(1)}$
and $z_n^{(2)}$ of the corresponding regular solutions.

In Fig.~4a we observe, that the function $\Delta w_n(x_{\rm H})$
crosses the line indicating the location of the innermost node 
of the $n$-th regular solution, $z_n^{(1)}$, at a value of 0.27,
for large $n$.
$\Delta w_n(x_{\rm H})$ crosses the line indicating the location of 
the 
innermost node
of the $(n-1)$-th regular solution, $z_{n-1}^{(1)}$
(which is located approximately
halfway between the innermost and the second node), at a value of
 $0.05$.
(It crosses the line indicating the location of the innermost node
of the $(n+1)$-th regular solution, $z_{n+1}^{(1)}$ at $0.75$ .)
Considering the function $\Delta \phi_n$,
shown in Fig.~4b, we see, that $\Delta \phi_n$
is much smaller than $\Delta w_n$. 
In particular, the function $\Delta \phi_n$ crosses the line
indicating $z_{n}^{(1)}$ at a value of approximately 0.01
for the larger $n$.
We observe, that for the larger $n$,
the crossings lie again on straight lines,
which however have a small positive slope.
Concluding we observe, that the solutions are converged well to 
the limiting solution for $x_{\rm H} \gg z_n^{(1)}$.

Let us now consider the global properties
of the solutions, the mass, the dilaton charge and
the Hawking temperature.
We recall, that these global properties are not independent,
but related by relation (\ref{res3}).
The dimensionless mass $\mu$,
the dilaton charge $D$ and the Hawking temperature $T/T_{\rm S}$
for the black hole solutions
with $n=1-7$ and event horizons $x_{\rm H} = 0$, 0.01, 0.1 and 1
are given in Table~1.

To discuss the inverse Hawking temperature $\beta=T^{-1}$
of the black hole solutions as a function of their mass 
$\mu(\infty)$,
we first recall the case of the limiting EMD solution,
which is special for $\gamma=1$.
Here the inverse Hawking temperature
is a straight line as a function of mass,
$T^{-1}=4 \pi X_+=8 \pi \mu(\infty)$ (see appendix A).
Note, that this is the same relation as the one obtained 
for Schwarzschild black holes.
However, the EMD curve does not extend to the origin, since
the ``extremal'' EMD solution has a finite mass and a finite
temperature. For $P=1$ this lower limiting mass is $1/\sqrt{2}$.
In Fig.~5 we show the inverse Hawking temperature
as a function of the mass for the EYMD black holes
with $n=1-4$.
As expected, the curves converge rapidly towards the limiting
EMD curve, also shown.

Let us now consider the convergence of the global properties
of the black hole solutions in more detail.
The mass, the dilaton charge and the Hawking temperature
converge exponentially to the corresponding properties
of the limiting EMD solution.
To demonstrate this, we define $\Delta \mu_n$ as the deviation 
of the mass
of the $n$-th EYMD solution from the mass of the limiting EMD
solution and make an exponential ansatz for the $n$-dependence
\begin{equation}
\Delta \mu_n = \mu_\infty(\infty) - \mu_n(\infty)
= a_\mu e^{-\alpha_\mu n}
\ . \label{muexp} \end{equation}
We define $\Delta D_n$ and $\Delta T_n$ for
the dilaton charge and the Hawking temperature analogously,
\begin{equation}
\Delta D_n = D_\infty - D_n 
= a_D e^{-\alpha_D n}
\ , \label{Dexp} \end{equation}
\begin{equation}
\Delta T_n=
\frac{T_n}{T_{\rm S}} - \frac{T_\infty}{T_{\rm S}}
= a_T e^{-\alpha_T n}
\ . \label{Texp} \end{equation}
The coefficients $a$ and $\alpha$ still depend on the parameters, 
the horizon $x_{\rm H}$ and the dilaton coupling constant $\gamma$.

A logarithmic convergence
then requires the function $\ln {(\Delta \mu_n)}$ to be a straight 
line
as a function of $n$.
This is indeed the case, as seen in Fig.~6, for the regular
solutions and for the black hole solutions with $x_{\rm H} \ge 1$.
However for small $x_{\rm H}$ and small values of $n$,
we observe deviations from straight lines.
These deviations are again related to the relative location
of the horizon to the innermost nodes
$z_n^{(1)}$ of the regular solutions.
When the location of the innermost node
of the $n$-th regular solution becomes smaller than the horizon,
the values of $\ln {(\Delta \mu_n)}$
for the solutions with larger $n$ fall on a straight line,
whereas those for smaller $n$ are close to the straight line
of the regular solutions.

We observe, that the straight lines for the various finite values
of the horizon are parallel, indicating the same coefficient
$\alpha$ for all $x_{\rm H}$. 
(The coefficient $\alpha$ is different for the regular solutions.)
For small horizons the lines
are roughly equally spaced; the same is true for large horizons.
This suggests one to parametrize the coefficients $\alpha$ and $a$
with the following $x_{\rm H}$ dependence
\begin{equation}
\alpha(x_{\rm H}) = {\rm const.} \ , \ \ \
a(x_{\rm H}) = \frac{b}{(x_{\rm H})^\delta}
\ . \label{delta} \end{equation}
Approximately, we find $\alpha_\mu=3.6$,
$b_\mu=1.2$ and $\delta_\mu=2$ for small values of the horizon,
$x_{\rm H}\le 1$, and the same $\alpha_\mu$ but
$b_\mu=0.8$ and $\delta_\mu=1$ for large values of the horizon,
$x_{\rm H} \gg 1$.

The functions $\ln {(\Delta D_n)}$ and $\ln {(\Delta T_n)}$ 
follow a similar pattern as the function $\ln {(\Delta \mu_n)}$.
The coefficient $\alpha$ is approximately the same
for all three functions, as expected, because of relation 
(\ref{res3}).
Also the coefficients $b$ and $\delta$
for the three functions are related, satisfying
relation (\ref{res3}). 
In particular, $\delta_T = \delta_\mu +1$.
The constants $\alpha$, $b$ and $\delta$
for the mass, the dilaton charge and the Hawking temperature 
obtained by a mean square fit to our numerical data 
are given in Table~3.

In fact for large $x_{\rm H}$ one obtains an
analytic handle on some of the coefficients.
For instance, it follows analytically, that 
\begin{equation}
\delta_\mu = \delta_D = 1 \ , \ \ \ \delta_T=2
\ , \end{equation}
and, allowing for general $\gamma$, one finds
\begin{equation}
b_D= 2 \gamma b_\mu \ , \ \ \ b_T= 2 b_\mu
\ , \end{equation}
which is indeed confirmed by the calculations.

\boldmath
\subsection{$\gamma \ne 1$}
\unboldmath

In the following we discuss the convergence
of the sequences of EYMD black hole solutions
to the corresponding limiting EMD solutions
for $\gamma \ne 1$.  Note, that the values of $\gamma$
in $4+n$-dimensional Kaluza-Klein theory,
$\gamma^2=(2+n)/n$ \cite{gib1}
(with $n=1$, 2 and 4 for U(1), SU(2) and SU(3), respectively),
are in the range $\gamma > 1$.
The limit $\gamma \rightarrow 0$ needs special consideration.

The EYMD black hole solutions and their limiting
EMD solutions depend smoothly 
on the dilaton coupling constant for finite $\gamma$.
Not surprisingly therefore, they follow
the same pattern of convergence for any finite $\gamma$
as for $\gamma = 1$.  
However, the degree of convergence of an EYMD black hole
solution now also depends on $\gamma$. 
To analyze this dependence, let us inspect Table~2,
where the location of the innermost node $z_n^{(1)}$
of the seven lowest regular EYMD gauge field functions is presented
for the dilaton coupling constants $\gamma=0.1$, 0.5, 1 and 2.
For finite $\gamma$, the sequences of nodes $z_n^{(1)}$
approach zero exponentially with increasing $n$, 
\begin{equation}
z_n^{(1)}(\gamma) = d_{\gamma} e^{-c_{\gamma} n}
\ . \label{nodes} \end{equation}
The coefficients $d_\gamma$ and $c_\gamma$
increase with $\gamma$.
Thus the nodes $z_n^{(1)}(\gamma)$
converge faster to zero for $\gamma >1$
and slower for $\gamma < 1$, as compared to $\gamma=1$.

The implications for the convergence of the 
sequences of EYMD black hole solutions are obvious.
Let $n$ be the number of nodes of the first EYMD black hole solution 
of a sequence with fixed horizon $x_{\rm H}$,
for which the functions have largely assumed their limiting values.
Then $n$ becomes smaller, the larger the value of the
dilaton coupling constant $\gamma$.
This is demonstrated in Figs.~7a-b,
where we show the EYMD gauge field function $w_7$
and the metric function $N_7$ of the black hole solutions
with dilaton coupling constants $\gamma=0$, 0.5, 1 and 2
and event horizon $x_{\rm H} = 0.01$,
as well as the corresponding EMD solutions \cite{fn7}.
According to Table~2, the location of the
innermost node $z_7^{(1)}(\gamma)$ of the regular EYMD 
gauge field functions is larger than this horizon
for $\gamma=0$ and 0.5,
so the corresponding EYMD black hole functions 
show large deviations from the limiting EMD functions
in the inner regions. In contrast, for $\gamma=1$ and 2,
where $z_7^{(1)}(\gamma) < x_{\rm H}$,
the EYMD functions have largely assumed their limiting EMD values.

The value $\gamma=0$ is special, since EYMD theory then
reduces to EYM theory.
The limiting behaviour of the sequences of EYM black hole 
solutions with horizon $x_{\rm H}$
has been discussed before \cite{bfm,sw,bfm2,sw2}.
For $x_{\rm H}>1$ the limiting solutions are
the RN black hole solutions with charge $P=1$.
For black hole solutions with smaller event horizons
and for the regular solutions 
the limiting solutions are more complicated,
having a mass $\mu(\infty)=1$, independent of the horizon 
\cite{bfm,sw,bfm2,sw2}.
In this case the solutions must be considered separately in
two regions, the inner region
$x<1$ and the outer region $x>1$. In the outer region
the limiting solution is given by the extremal RN black hole
solution, while in the inner region the solution
is of an oscillating type with $w_\infty \ne 0$
\cite{bfm,sw,bfm2,sw2}.

Because of this different limiting behaviour of the black hole
solutions for $\gamma=0$ (and $x_{\rm H} < 1$)
the limit $\gamma \rightarrow 0$ is non-trivial.
We will consider this limit separately elsewhere \cite{we}.

Let us now consider the inverse Hawking temperature $\beta$
for general $\gamma$ 
beginning with the limiting EMD black hole solutions.
For $\gamma > 1$, the temperature in the ``extremal'' limit
diverges, i.~e.~$\beta$ is zero. 
The curve $\beta$ versus mass then rises monotonically
and approaches the Schwarzschild curve from below
for large values of the horizon.
For $0<\gamma<1$ the temperature in the ``extremal'' limit
is zero, i.~e.~$\beta$ diverges in the ``extremal'' limit.
The curve $\beta$ versus mass then falls to a minimum
and approaches the Schwarzschild curve from above
for large values of the horizon.
For $\gamma=0$ the Reissner-Nordstr\o m case is obtained.
In contrast to the ``extremal'' EMD solutions,
the extremal RN solution has a finite horizon,
$x_{\rm H}=\mu=P=1$.

In Fig.~8, we show the convergence
of the inverse temperature versus mass curves
for the sequences of EYMD black holes with $n=1-4$
to the corresponding curves for the limiting EMD black hole solutions
for $\gamma=0.5$ and 2.
(The EYM case $\gamma=0$ is shown in Fig.~5).
Depending on the dilaton coupling constant $\gamma$
and the number of nodes $n$,
phase transitions can occur \cite{maeda,fn5,kks3}.
We observe, that for $\gamma \ge 1$ no phase transitions occur,
since the inverse temperature curves 
rise monotonically for all $n$ like the
corresponding limiting EMD curves.
In contrast, for $\gamma < 1$, phase transitions do
occur for all members of a sequence beyond
some critical $n_{\rm cr}(\gamma)$.
The reason for this behaviour is, 
that for $\gamma < 1$ the inverse temperature curve
of the black hole solutions of a sequence
approaches the inverse temperature curve of the corresponding
limiting EMD black hole solutions from below.
But while there is one phase transition along the limiting EMD curve,
which starts from $\beta=\infty$,
there are two phase transitions along the EYMD curves (beyond
some critical $n_{\rm cr}$),
which start from $\beta=0$ and therefore develop two extrema;
a maximum and a minimum.
The value $\gamma=1$ is special in EYMD and EMD theory.
Only for $\gamma < 1$ phase transitions can occur.

\section{SU(3) Einstein-Yang-Mills-Dilaton Black Holes}

Let us now turn to the magnetically neutral
static spherically symmetric black hole solutions of
SU(3) EYMD theory.
Besides the solutions discussed in section 3,
corresponding to the SU(2) embedding,
SU(3) EYMD theory possesses magnetically neutral
static spherically symmetric black hole solutions,
corresponding to the SO(3) embedding.
As in EYM theory \cite{kuenzle,kks2}, these SO(3) solutions
can be labelled by the nodes
of both gauge field functions \cite{kks3}.
Here we adopt the classification of
the solutions with respect to 
the node structure of the functions $(u_1,u_2)$
\cite{kuenzle,kks2,kks3}, 
\begin{equation}
u_1(x)= \frac{K(x)+H(x)}{2} \ , \ \ \
u_2(x)= \frac{K(x)-H(x)}{2}
\ . \end{equation}
We denote the number of nodes of the functions $u_1$ and $u_2$
by $n_1$ and $n_2$, respectively,
and the total number of nodes by $n_T$.
(For SO(3) solutions $n_T=n_1+n_2$.)
Due to the
symmetry $H(x) \rightarrow -H(x)$ (see section 2.3),
solutions with node structures $(n_1,n_2)$ and
$(n_2,n_1)$ are equivalent, and it is sufficient
to consider solutions with $n_1 \leq n_2$.

Previously, we have only constructed magnetically neutral 
SU(3) EYMD and EYM black hole solutions 
with a small number of nodes \cite{kks3,kks2}.
The lowest SO(3) black hole solution
has $n_T=1$ and node structure $(0,1)$, 
the second solution has $n_T=2$ and node structure $(0,2)$.
The third and fourth solution have also $n_T=2$,
but both have the same node structure $(1,1)$.
Therfore the node structure $(n_1,n_2)$
does not seem to classify the SO(3) solutions uniquely.
However, the third solution is a scaled SU(2) solution,
where $u_1=u_2=w$, and only the fourth solution
is a genuine SO(3) solution, where the two gauge field
functions are not proportional.
The two next solutions have $n_T=3$ and node structure $(0,3)$
and $(1,2)$, respectively.
We have listed the lowest solutions with $n_T \le 5$ 
in Table 4, 
where their mass is shown 
for $\gamma=1$ and several values of the horizon,
$x_{\rm H}=0$, 0.5, 1, 2.

In our previous analysis \cite{kks2,kks3},
we have identified two sequences of solutions:
the genuine SO(3) sequence $(0,n)$, 
containing the lowest SO(3) solution,
and the sequence $(n,n)$ of scaled SU(2) solutions.
As in the SU(2) case,
the limiting solutions of these two neutral sequences
are magnetically charged EMD solutions,
with the same event horizon
and the same dilaton coupling constant. 
However, their magnetic charges are
$P=\sqrt{3}$ and $P=2$, respectively \cite{kks3}.
We discuss the convergence of these sequences in detail below.

From our previous analysis \cite{kks3,kks2} 
it has not been clear yet,
in which way SO(3) black hole solutions 
with general node structure $(n_1,n_2)$
assemble to form sequences,
which converge to limiting solutions.
Below we construct sequences from
the neutral SU(3) black hole solutions
with general node structure $(n_1,n_2)$,
and we identify the limiting solutions of these
neutral sequences with the
known magnetically charged SU(3) black hole solutions 
\cite{gal,volkov3}.

\boldmath
\subsection{$(n,n)$ sequences}
\unboldmath

The solutions with node
structure $(n,n)$ often come in pairs (see Table~4).
One solution of the pair $(n,n)$ exists always, this
is the scaled SU(2) solution. The second solution of the pair
with node structure $(n,n)$ is a genuine SO(3) solution,
whose existence depends on the parameters $\gamma$ and
$x_{\rm H}$. 

\boldmath
\subsubsection{Scaled SU(2) solutions}
\unboldmath

Comparison of the equations of motion of the SO(3) embedding
and those of the SU(2) embedding 
shows, that to each SU(2) solution
there corresponds a ``scaled SU(2)'' solution 
of the SO(3) system
with precisely double the mass of its SU(2) counterpart.
Introducing
\begin{equation}
x = 2 \tilde x
\ , \end{equation}
\begin{equation}
\mu(x) = 2 \tilde \mu(\tilde x)
\ , \end{equation}
\begin{equation}
K(x) = 2 w(\tilde x) \ , \ \ \ H(x)=0
\ , \end{equation}
and 
\begin{equation}
\phi(x) = \tilde \phi(\tilde x)
\ , \end{equation}
the functions $\tilde \mu$, $w$ and $\tilde \phi$
satisfy the SU(2) equations with coordinate $\tilde x$.
Obviously, these scaled SU(2) solutions,
having node structure $(n,n)$, then converge
to the corresponding EMD solutions
with scaled magnetic charge $P=2$.
Thus these solutions and their properties
are obtained by simply scaling the solutions
of section 3.

\boldmath
\subsubsection{Genuine SO(3) solutions}
\unboldmath

Although the two gauge field functions $u_1$ and $u_2$ 
of the genuine SO(3) solutions have the same node structure 
$(n,n)$ as the scaled SU(2) solutions,
they are not proportional to each other.
We now discuss the dependence of the genuine SO(3)
solutions on the dilaton coupling constant and on the horizon.
The regular solutions exist for all values of the dilaton coupling
constant and (supposedly) for all values of $n$.
Considering the SO(3) solution with node structure $(n,n)$
as a function of the horizon for a fixed dilaton coupling constant, 
we observe, that it merges into the corresponding ``scaled SU(2)''
solution at a critical value of the horizon 
$x_{\rm H}^{\rm cr}(\gamma)$. This critical behaviour was first
observed for vanishing dilaton coupling \cite{kks2}.
In the parameter ranges where the genuine SO(3) solutions
exist, they have a slightly higher energy than their
``scaled SU(2)''counterparts.

In Fig.~9, we show as an example the matter functions $H$ and $K$
of the genuine SO(3) solution $(5,5)$ for $\gamma=1$
and several values of the horizon. At the critical value
$x_{\rm H}^{\rm cr}=0.03$ the solution becomes identical
to the ``scaled SU(2)'' solution, i.~e.~the gauge field
function $H$ vanishes identically.
In Fig.~10, we show the critical value of the horizon
$x_{\rm H}^{\rm cr}$
as a function of the dilaton coupling constant $\gamma$
for the lowest odd solutions, $n=1$, 3 and 5.
The critical horizon first decreases as a function of the
dilaton coupling constant, reaches a minimum and then increases
linearly with $\gamma$, $x_{\rm H}^{\rm cr}= b(n) \gamma$ 
(with $b(1)=0.677$, $b(3)= 0.0151$, $b(5)=0.0004$)
for $\gamma \gg 1$.
Only for $\gamma=0$ the critical value of the horizon
$x_{\rm H}^{\rm cr}(n)$ converges with increasing $n$ to a finite 
limit,
$x_{\rm H}^{\rm cr}(\infty)=1.3122$.

We further show in Fig.~10, the innermost nodes of the corresponding
regular genuine SO(3) solutions as a function of the dilaton
coupling constant.
We observe, that for each value of $n$, the two curves closely
follow each other.
In general,
the critical value of the horizon is slightly smaller than
the innermost node of the corresponding regular SO(3) solution.
Only for $n=1$ the two curves cross at $\gamma=0.9576$.
(The curves with the innermost nodes of the corresponding
regular ``scaled SU(2)'' solutions also have the same shape
but lie somewhat higher.)
 From our previous considerations on SU(2) black holes, 
this behaviour is not unexpected.
There, when the horizon of the black hole 
becomes larger than the innermost node of the regular solution,
the solution loses most of its structure.
Here, since the genuine SO(3) black hole solutions have 
more structure than the ``scaled SU(2)'' black hole solutions,
this additional structure disappears, and thus the genuine
SO(3) solution disappears, when the horizon comes close to the
vicinity of the innermost node of the regular SO(3) solution.

This observation demonstrates even more 
the importance of the innermost nodes
of the regular solutions for the black hole solutions. 
For the SU(2) solutions, the innermost nodes of
the regular solutions rule
the pattern of convergence of the sequences of black hole solutions.
For the genuine $(n,n)$ SO(3) solutions,
they determine the range of existence
of the corresponding black hole solutions.

\boldmath
\subsection{$(0,n)$ sequences}
\unboldmath

The two lowest neutral black hole solutions of the SO(3) embedding
have node structure $(0,1)$ and $(0,2)$.
They are the first members of the sequence $(0,n)$.
For a given event horizon and dilaton coupling constant,
this sequence tends to the EMD black hole solution
with the same event horizon, the same dilaton coupling constant
and with magnetic charge $P=\sqrt{3}$ \cite{kks3}.
Table~5 shows the dimensionless mass,
the dilaton charge and the Hawking temperature
for the lowest solutions of this sequence
for $\gamma=1$ and horizon $x_{\rm H}=1$.

Previously, we have only demonstrated the 
convergence of the global properties of this sequence.
Here we consider in addition, the convergence
of the spatial functions of the solutions of this sequence 
with increasing node number $n$.
As in the SU(2) case, the location of the innermost node
of the regular solution with $n$ nodes 
determines the degree of convergence
of the corresponding black hole solution with $n$ nodes 
and horizon $x_{\rm H}$. 
(Remember, that the nodes of the function $u_2(x)$ correspond
to the intersections of the functions $K(x)$ and $H(x)$.)
We observe that 
the black hole solution with $n$ nodes has already largely
converged to the corresponding limiting EMD solution,
if the innermost node of the regular solution with $n$ nodes
is smaller than the horizon. It has not yet converged
well, if the innermost node of the regular solution
is larger than the horizon.
Thus the pattern of convergence of the $(0,n)$ sequence
is fully analogous to the SU(2) case.
We demonstrate this in the following with a few examples.

Let us begin again with the case $\gamma=1$,
the case of the low energy effective action of string theory.
In Figs.~11a-c, we show 
the matter functions $K_n$ and $H_n$,
and the metric functions $N_n$ and $P_n^2$
for the small horizon $x_{\rm H}=0.2$ for $n=1$, $3$ and $5$.
(The matter functions $\phi_n$ and the metric functions $A_n$ 
are analogous to the SU(2) case and contain no relevant new 
information.)
 From Fig.~11a, we observe, that with increasing $n$,
both EYMD gauge field functions $K_n$ and $H_n$ approach
the same limiting function
\begin{equation}
K_n (x) \stackrel {n\rightarrow \infty} {\longrightarrow}
\frac{f(x)}{\sqrt{2}} \ , \ \ \
H_n (x) \stackrel {n\rightarrow \infty} {\longrightarrow}
\frac{f(x)}{\sqrt{2}}
\ , \label{lim1} \end{equation}
where $f$ is constant 
\begin{equation}
f(x) \equiv 1
\ . \label{lim2} \end{equation}
To relate this limiting function $f$ to the 
corresponding EMD solution with charge $P=\sqrt{3}$
we inspect the EYMD equations of motion
for the case $K=H=f/\sqrt{2}$.
(These equations are presented in appendix B.)
Indeed, for $f \equiv 1$ these equations
reduce to the EMD equations of motion
for a black hole with charge $P=\sqrt{3}$.

Let us now consider the degree of convergence of the
black hole solutions of Figs.~11a-c with $x_{\rm H}=0.2$.
For that purpose, we inspect the location of the innermost nodes 
of the gauge field functions
of the corresponding regular solutions, which are given in Table~6.
As for the regular SU(2) solutions, 
with increasing $n$ the innermost nodes
decrease exponentially to zero.
For the example of Figs.~11a-c, the innermost nodes of the
regular solutions are larger than
the horizon for $n=1$ and 3 and smaller than the horizon
for $n=5$.
We conclude, that the solution with $n=5$ 
should be close to the limiting EMD solution, also shown in 
Figs.~11a-c,
whereas the solutions with $n=3$ and $n=1$ 
should still show large deviations from the limiting solution.
This is indeed the case. The functions $\phi_5$, $N_5$ and $A_5$
are almost indiscernible from the limiting EMD functions,
whereas the functions $\phi_3$, $N_3$ and $A_3$
deviate significantly from the limiting functions
in the region interior to the innermost node.
The gauge field functions $K_n$ and $H_n$ 
converge more slowly (and non-uniformly because of the boundary
conditions at infinity), but $K_5 \approx H_5 \approx 1/\sqrt{2}$
already in a large region of space.
Also the charge function $P_5^2$ has assumed its limiting value
$P^2=3$ in a large region. 
In contrast, the charge function $P_3^2$
still deviates slightly from the limiting value
in the inner region and $P_1^2$ deviates strongly.

We demonstrate the dependence of the black hole solutions
on the horizon $x_{\rm H}$ in Fig.~12.
There the gauge field functions $K_5$ and $H_5$ are shown
for the event horizons $x_{\rm H} = 0$, $0.5$, $1$ and $2$.
For the larger horizons, even the gauge field
functions have converged  well in the inner region.
(For the regular solutions
the boundary conditions inhibit 
a uniform convergence at the origin.)
As for the SU(2) solutions, 
a more quantitative analysis 
of the degree of convergence
of the solutions is obtained by studying
the functions $\Delta K_n$, $\Delta H_n$
and $\Delta \phi_n$,
where the functions $\Delta K_n$ and $\Delta H_n$,
are defined as relative differences
analogously to $\Delta \phi_n$ (eq.~(\ref{delp})).
We observe, 
that the function $\Delta K_n(x_{\rm H})$ 
crosses the line indicating the location of the innermost node 
of the $n$-th regular solution, $z_n^{(1)}$, at the value $0.75$
for the larger $n$, while
the function $\Delta H_n(x_{\rm H})$ crosses at the value $0.60$.
Both functions have fallen to the value $0.05$,
when they cross the line indicating the location of the second node
of the $n$-th regular solution, $z_n^{(2)}$.
The function $\Delta \phi_n$
is again much smaller than $\Delta K_n$  or $\Delta H_n$
and crosses the line
indicating $z_{n}^{(1)}$ at a value on the order of 0.05
for the larger $n$.

The SO(3) solutions depend smoothly on the dilaton coupling constant
$\gamma$.
The $\gamma$ dependence of the convergence properties
of the sequences of black hole solutions is analogous to the SU(2) 
case.
With increasing $\gamma$, the convergence of the sequences
is faster, with decreasing $\gamma$ the convergence is slower.
This is seen from Table~6, where the innermost nodes
of the lowest regular solutions of the $(0,n)$ sequence
are given for $\gamma=0.1$, 0.5, 1 and 2.

Again, the limit $\gamma \rightarrow 0$ 
needs special consideration and will be discussed elsewhere 
\cite{we}.
However, we already note, that 
for $\gamma=0$ the limiting solution 
is more complicated and analogous to the limiting SU(2) solution.
For $x_{\rm H}>P$ ($P=\sqrt{3}$), the limiting solution
is the RN solution, while for $x_{\rm H}<P$
the limiting solution is the extremal RN solution
for $x>P$ and apparently an oscillating solution for $x<P$.

Let us now consider the convergence of 
the global properties of the solutions.
Turning to the thermal properties of the black hole solutions first,
we show in Fig.~13 the inverse Hawking temperature $\beta$
as a function of the mass for the EYMD black holes for $n=1-4$
and $\gamma=1$ as well as $\gamma=0$.
As in the SU(2) case,
the EYMD inverse temperature curves converge rapidly towards 
the limiting EMD and RN inverse temperature curves.
Consequently, we also observe phase transitions,
depending on the dilaton coupling constant
and on the number of nodes.
For instance, for $\gamma=0$ and $n=1$, there are two critical values
of the mass, $\mu_1=1.561$ and $\mu_2=1.588$.
The critical behaviour disappears beyond $\gamma=0.0293$
for the lowest solution, while it occurs up to
larger values of $\gamma$ for the solutions with larger $n$.
For $\gamma \ge 1$ no phase transitions occur.

The convergence of mass, dilaton charge and Hawking temperature
is exponential, analogous to the SU(2) case. For the dilaton
coupling constant $\gamma=1$, these global properties
are given in Table~5 for $n=1-5$ together with their limiting
values
and the constants of the exponential approximation formulae 
eqs. (\ref{muexp})-(\ref{Texp}) \cite{fn4}.
For large values of the horizon, the $x_{\rm H}$-dependence in the
approximation formulae can be extracted according to 
eq.~(\ref{delta}).
The corresponding parameters are shown in Table 7.

\boldmath
\subsection{$(1,1+n)$ sequences}
\unboldmath

Let us now address the question, in which way
the solutions with general node structure $(n_1,n_2)$ assemble
into sequences and to what limiting solutions
these sequences converge.
The existence of the sequence $(0,n)$,
where one gauge field function has no node, whereas the other
function has an increasing number of nodes,
suggests one to examine the set of solutions, where
one of the gauge field functions has one node, whereas the other
function has an increasing number of nodes.
We thus consider the set of solutions $(1,1+n)$.
The first solution of this set is the solution $(1,2)$; it has $n=1$.
(The solution $(1,0)$ is equivalent
to the solution $(0,1)$ and belongs to the
$(0,n)$ sequence. The solutions $(1,1)$
are members of the $(n,n)$ sequences.)
In the following, we demonstrate that 
for any value of the dilaton coupling constant $\gamma$
and of the horizon $x_{\rm H}$,
the set of solutions $(1,1+n)$ indeed forms a sequence,
tending to a limiting solution.

To identify the limiting solution, let us consider
an example.
In Figs.~14a-c we show
the matter functions $K_n$ and $H_n$,
and the metric functions $N_n$ and $P_n^2$
for the horizon $x_{\rm H}=0.2$ and $n=1$, $3$ and $5$
for $\gamma=1$.
As for the $(0,n)$ sequence,
in the limit $n \rightarrow \infty$
both EYMD gauge field functions $K_n$ and $H_n$ approach
a single function $f(x)$.
However, in contrast to the $(0,n)$ sequence,
this function is not a constant; it depends on $x$
and has one node.
Obviously the limiting solution cannot be a magnetically charged
EMD black hole.
Instead the limiting solution
turns out to be the lowest magnetically charged SU(3) 
black hole solution found in \cite{gal}.
This limiting solution is also shown in Figs.~14a-c.

In fact, the simpler system of equations
obtained with only one gauge field function,
(see appendix B)
\begin{equation}
K(x) \equiv H(x)  \equiv f(x) /\sqrt{2}
\ , \label{simple} \end{equation}
possesses SU(3) black hole solutions
with magnetic charge $P=\sqrt{3}$.
Such charged black hole solutions have been constructed
before for the dilaton coupling constants
$\gamma=1$ \cite{gal} and $\gamma=0$ \cite{volkov3}.
The equations for these charged black hole solutions
closely resemble the SU(2) equations
(and correspond to the ${\rm SU(2)} \times {\rm U(1)}$ equations).
Not surprisingly therefore,
the charged black hole solutions form a sequence, which can be
labelled by the number of nodes $j$ of the single gauge field 
function $f$
\cite{gal,volkov3}.
Like the EMD black holes, 
for finite dilaton coupling constant $\gamma$,
these charged black hole solutions with $j$ nodes
exist down to arbitrary small values of the horizon.
In the ``extremal'' limit $x_{\rm H}=0$,
the solutions present naked singularties
like their EMD counterparts.
(The ``extremal'' limit is discussed in appendix B.)
In contrast, for vanishing coupling constant $\gamma$,
the extremal black hole solutions have a finite
horizon, $x_{\rm H}=P$ ($P=\sqrt{3}$),
analogous to the RN case \cite{volkov3}.

Turning back to the $(1,1+n)$ sequence of EYMD solutions,
we observe, that
for any finite value of the dilaton coupling constant
and any finite event horizon,
the sequence of neutral black hole solutions tends to the
magnetically charged SU(3) solution with one node, $j=1$.
This limiting solution has the same dilaton coupling constant,
the same horizon, and it has magnetic charge $P=\sqrt{3}$.
For finite dilaton coupling constant $\gamma$,
the ``extremal'' solution with $j=1$
represents the limiting solution 
of the regular neutral SO(3) sequence $(1,1+n)$.
However, for zero coupling constant $\gamma$,
the extremal black hole solution has a finite horizon, $x_{\rm H}=P$,
therefore the limiting solution is again more
complicated for values of the horizon
smaller than $P$ \cite{we}.

The convergence properties of the solutions
of the $(1,1+n)$ sequence are again analogous to
those of the SO(3) $(0,n)$ sequence and the SU(2) sequence.
The location of the horizon with respect to the
innermost nodes of the function $u_2$ 
of the regular solutions
determines the degree of convergence of the black hole solutions.
The innermost nodes of the regular solutions are
shown in Table~6 for several values of the dilaton coupling constant
$\gamma$. For the example of Figs.~14a-c,
the horizon is slightly smaller than the
innermost node of the regular solution with $n=3$.
Consequently, the black hole solution with $n=5$
has already converged well, the solution with $n=3$
has almost converged, and the solution with $n=1$
is far from convergence.
Note the interesting shape of the limiting 
charge function $P_\infty^2(x)$, which is already approached well
in the inner region by $P_5^2$ and in a smaller region
by $P_3^2$.
The limiting charge function has two plateaus,
an inner plateau at 3.790 and an outer
plateau at the magnetic charge of the solution, $P^2=3$.

The global properties of the lowest solutions of
the $(1,1+n)$ sequence 
for $\gamma = 1$ and $x_{\rm H} = 1$
are given in Table~5.
The inverse temperature as a function of mass is shown for
the lowest solutions
of the $(1,1+n)$ sequence and for the charged limiting
$j=1$ solution in Fig.~15 for $\gamma=1$
and $\gamma=0$.
The exponential convergence of the global properties is 
demonstrated in Table~5 and Table~7 for $\gamma = 1$.

\boldmath
\subsection{$(j,j+n)$ sequences}
\unboldmath
 
 From the above results, it is clear, in which way the remaining
SO(3) solutions fall into sequences.
The next sequence is formed by the solutions $(2,2+n)$,
beginning with the solution $(2,3)$.
In Figs.~16a-b, we show 
the matter functions $K_n$ and $H_n$,
and the charge functions $P_n^2$
for $n=1$, $3$ and $5$, the horizon $x_{\rm H}=0.2$,
and the dilaton coupling constant $\gamma=1$.
In the limit $n \rightarrow \infty$,
both EYMD gauge field functions $K_n$ and $H_n$ approach
a single function $f(x)$ which has two nodes.
This limiting function is the gauge field function of the 
magnetically 
charged SU(3) EYMD black hole solution with two nodes, $j=2$,
and charge $P = \sqrt{3}$.

In general, we observe, that 
for any finite dilaton coupling constant $\gamma$,
the $(2,2+n)$ sequence of EYMD solutions
tends to the 
corresponding magnetically charged SU(3) solution with two nodes, 
$j=2$,
and magnetic charge $P=\sqrt{3}$.
Considering the convergence properties of the solutions
of the $(2,2+n)$ sequence, we observe that they are again analogous 
to
those of the above sequences $(0,n)$ and $(1,1+n)$.
The location of the horizon with respect to the
innermost nodes of the regular solutions
determines the degree of convergence of the black hole solutions.

In the example of Figs.~16a-b,
the horizon is located well behind the
innermost node of the regular solution with $n=3$,
as can be seen in Table~6 \cite{fn8}.
Consequently, the black hole solutions with $n=5$ and $n=3$
have already converged well,
whereas the solution with $n=1$ is still far from convergence.
In particular, also the charge function $P_\infty^2$ of the limiting 
solution
is well approached in the inner region by $P_5^2$ and $P_3^2$.
We note, that the inner plateau of the limiting charge function
has increased in height to 3.964 and has also increased in size.
We further note, that the deviations $\Delta K_n$ 
fall on top of each other
for $j=0$, $1$ and $2$, 
when regarded as functions of the scaled horizon 
$x_{\rm H}/z_5^{(1)}(j)$, and the same is true for the deviations
 $\Delta H_n$. 

The global properties of the lowest solutions of
the $(2,2+n)$ sequence are given in Table~5.
The exponential convergence with $n$ of the global properties is
demonstrated in Tables~5 and 7,
where we anticipate also an exponential convergence with $j$.

By generalizing the above observations we see, that
(apart from the $(n,n)$ solutions considered separately above)
the neutral SO(3) EYMD black hole solutions assemble
into sequences $(j,j+n)$, with fixed index $j\ge 0$
and running index $n \ge 1$.
For any finite value of the dilaton coupling constant
and any finite event horizon,
the sequence $(j,j+n)$ of neutral black hole solutions tends to the
magnetically charged SU(3) solution with $j$ nodes,
the same dilaton coupling constant,
the same horizon, and with magnetic charge $P=\sqrt{3}$.
For finite dilaton coupling constant $\gamma$,
the ``extremal'' solution with $j$ nodes
represents the limiting solution 
of the regular neutral SO(3) sequence $(j,j+n)$.
(For coupling constant $\gamma = 0$,
the limiting solution is the RN solution if $x_{\rm H} > P$; it is
again more complicated,
if $x_{\rm H}<P$ \cite{we}.)

Let us finally consider the sequence of the sequences $(j,j+n)$
of SO(3) solutions, with running index $j$.
The lowest SO(3) sequence has $j=0$.
It approaches an EMD solution with magnetic
charge $P=\sqrt{3}$ since $f \equiv 1$.
The $(j,j+n)$ sequences approach the EYMD solutions
with magnetic charge $P=\sqrt{3}$ and $j$ nodes.
In the limit $j \rightarrow \infty$,
the limiting solutions of the sequences
tend to a limiting solution themselves.
This limiting solution has
$f \equiv 0$, and therefore it is again an EMD solution. 
But this limiting EMD solution has magnetic charge $P=2$.
(It is also identical to the limiting solution of the
sequence of scaled SU(2) solutions.)
The inverse temperature curves of the limiting
charged solutions for $j=1-4$ are shown in Fig.~17,
together with the curves for $j=0$ and $j=\infty$
for $\gamma=1$ and $\gamma=0$.

\section{Conclusion}

We have considered sequences of magnetically neutral,
static, spherically symmetric black hole solutions
of SU(3) Einstein-Yang-Mills-dilaton theory,
based on the SU(2) embedding
and on the SO(3) embedding.
Such sequences of black hole solutions exist for arbitrary
dilaton coupling constant $\gamma$
and arbitrary event horizon $x_{\rm H}$.
In the limit of vanishing event horizon,
sequences of regular EYMD solutions are found.
In the limit of vanishing dilaton coupling constant,
sequences of EYM solutions are obtained.

The equations of motion for the EYMD system,
together with the boundary conditions,
allow for relations between the metric
and the dilaton field.
Black hole solutions satisfy a relation between
mass and dilaton charge,
$D=\gamma (\mu(\infty)-2\pi x_{\rm H}^2 T)$,
where $T$ is the Hawking temperature.
Regular solutions satisfy two simple relations,
$D=\gamma \mu(\infty)$
and $\phi(x)=\gamma \ln(\sqrt{-g_{tt}})$.
These relations hold for static,
spherically symmetric solutions
with magnetic gauge fields of general gauge groups.

The members of the SU(2) EYMD sequences
are labelled by the number of nodes $n$ of the single
gauge field function.
In the limit of large $n$,
for a given dilaton coupling constant $\gamma$ 
and a given horizon $x_{\rm H}$,
the corresponding sequence of black hole solutions
tends to the EMD black hole solution \cite{gib1,str} 
with magnetic charge $P=1$,
the same coupling constant and the same horizon.
The corresponding sequence of regular EYMD solutions
tends to the corresponding ``extremal'' EMD solution.

The solutions of the SO(3) embedding can be labelled
by their node structure $(n_1,n_2)$,
representing the number of nodes
of both gauge field functions.
The SO(3) solutions fall into sequences of two types.
Most solutions belong to the first type,
having node structure $(j,j+n)$,
where $j$ labels the sequence itself and
$n$ labels the members of the sequence,
i.~e.~the lowest sequence has $j=0$, the next $j=1$, etc.
The solutions of the second type
have node structure $(n,n)$.
Beside the ``scaled SU(2)'' solutions, there exist
also genuine SO(3) solutions of this type.
With regard to the genuine SO(3) solutions,
the above labelling of the solutions by their
node structure appears to be unique.

In the limit of large $n$,
for a given dilaton coupling constant $\gamma$ 
and a given horizon $x_{\rm H}$,
the corresponding sequence of black hole solutions
with node structure $(j,j+n)$ and finite $j$
tends to the magnetically charged SU(3) EYMD black hole solution
with $j$ nodes \cite{gal},
which has the same coupling constant, the same horizon
and magnetic charge $P=\sqrt{3}$.
The correspoding sequences of regular solutions
tend to the corresponding ``extremal'' 
magnetically charged EYMD solutions.

For $j=0$ and $j \rightarrow \infty$,
the magnetically charged EYMD SU(3) black hole solutions have
constant gauge field functions. 
Therefore the limiting solutions in these cases are 
again charged EMD black hole solutions,
having magnetic charges $P=\sqrt{3}$
and $P=2$, respectively.

The genuine SO(3) solutions of type $(n,n)$
differ from those of type $(j,j+n)$ in an important aspect.
They do not exist for all values of the dilaton coupling
constant $\gamma$ and the horizon $x_{\rm H}$.
Instead, for each value of the coupling constant,
there exists a critical value of the horizon,
where the genuine SO(3) solution $(n,n)$ merges into
the corresponding ``scaled SU(2)'' solution.

In the case of vanishing dilaton coupling constant,
the sequences of neutral EYM black hole solutions tend
with increasing $n$ 
to magnetically charged EYM black hole solutions \cite{volkov3}
and to RN black hole solutions (for $(0,n)$ and scaled SU(2) 
$(n,n)$),
as long as the horizon is larger than the magnetic charge.
If the horizon is smaller than the magnetic charge,
the limiting solutions are more complicated
\cite{bfm,sw,bfm2,sw2,we}.

All sequences of black hole solutions
exhibit the same pattern of convergence.
The degree of convergence of a given EYMD black hole
solution, labelled by $n$
within the corresponding sequence,
depends on the relative location of the horizon
to the innermost node of the corresponding
regular EYMD solution.
If the horizon is larger than the location
of this node, then the functions of the EYMD black hole solution
have largely converged to the functions of the limiting solution.
If the horizon is smaller than the location
of this innermost node, then the functions of the EYMD black hole
solution still differ appreciably from those of the limiting 
solution, 
with a tendency to approach the functions of the regular
EYMD solutions for small $x$.

The location of the innermost nodes of the regular EYMD solutions
converges to zero exponentially, for all sequences.
The larger (smaller) the dilaton coupling constant,
the faster (slower) the convergence.
For $\gamma = 0$
the location of the innermost nodes converges to a finite value
for the various sequences.

For all sequences, the convergence
of the global properties of the EYMD solutions, 
such as mass, dilaton charge and Hawking temperature,
to those of the limiting charged solutions
is exponential.
The coefficients depend on the dilaton coupling constant
$\gamma$, yielding a fast convergence for large
$\gamma$ and a slow convergence for small $\gamma$.

Obviously, the thermodynamic properties 
of the EYMD black hole solutions
also tend to those of the corresponding limiting solutions
for large $n$.
The occurrence of phase transitions,
where the specific heat changes sign,
depends on the dilaton coupling constant $\gamma$.
Only for $\gamma < 1$ phase transitions can occur.

Addressing the stability of the solutions, we note that
the SU(2) and SU(3) EYM black hole solutions are unstable
\cite{strau2,volkov5,volkov4,lav2,strau3,kks2},
and so are the SU(2) EYMD black hole solutions \cite{lav2}.
We therefore conjecture, that the genuine
SO(3) EYMD black hole solutions are also unstable.
It appears to be interesting to study
the number of unstable modes of the genuine SO(3) 
black hole solutions
and look for a relation to the number of nodes
of the solutions, in analogy to the SU(2) case, where
the black hole solution with $n$ nodes 
has $2n$ unstable modes \cite{lav,volkov4}.

{\sl Acknowledgement}

We dedicate this work to Larry Wilets
on the occasion of his retirement.
We gratefully acknowledge discussions with M. Volkov.

\section{Appendix A: Einstein-Maxwell-Dilaton Black Holes}

We here briefly recall the well-known
static, spherically symmetric EMD black hole solutions
with magnetic charge $P$ \cite{gib1,str}.
Following the notation of \cite{str},
we introduce the coordinate $X$ via
\begin{equation}
 x = X \left( 1- \frac{X_-}{X} \right)^{\frac{
 \gamma^2}{1+\gamma^2}}
\ , \end{equation}
$X=X_-$ then corresponds to the origin $x=0$.
(The coordinates $x$ and $X$ correspond
to $R$ and $r$ of \cite{str}, respectively.)
The metric then takes the form
\begin{equation}
ds^2= -\lambda^2 dt^2 + \lambda^{-2} dX^2 
      + x^2 (d\theta^2 + \sin^2\theta d\phi^2)
\ , \label{metricemd} \end{equation}
with
\begin{equation}
\lambda^2=\left(1-\frac{X_+}{X}\right)
          \left(1-\frac{X_-}{X}\right)^{(1-\gamma^2)/(1+\gamma^2)}
\ . \end{equation}
The non-vanishing component of 
the Maxwell field strength tensor is given by
\begin{equation}
F_{\theta \phi}=P \sin \theta
\   \label{ftp} \end{equation}
and the dilaton field by
\begin{equation}
e^{2\phi}= \left(1-\frac{X_-}{X}\right)^{2 \gamma/(1+\gamma^2)}
\ . \end{equation}

The parameters $X_+$ and $X_-$ 
are determined by the regular event horizon $x_{\rm H}$ 
\begin{equation}
 x_{\rm H} = X_+ \left( 1- \frac{X_-}{X_+} \right)^{\frac{
 \gamma^2}{1+\gamma^2}}
\   \end{equation}
and by the magnetic charge $P$
\begin{equation}
P= \left( \frac{X_+ X_-}{1+\gamma^2} \right)^{\frac{1}{2}}
\ . \end{equation}

The black hole solutions have mass $\mu(\infty)$
\begin{equation}
\mu(\infty)=\frac{X_+}{2} + \left( \frac{1-\gamma^2}{1+\gamma^2}
 \right) \frac{X_-}{2}
\   \end{equation}
and dilaton charge $D$
\begin{equation}
D=\frac{\gamma X_-}{1+\gamma^2}
\ . \end{equation}
Thus they satisfy a quadratic relation between mass,
dilaton charge and magnetic charge 
\begin{equation}
\mu^2(\infty)+D^2-P^2 = \left( \frac{X_+-X_-}{2} \right)^2
\ . \label{mdp} \end{equation}

The EMD black hole solutions
exist for arbitrary event horizon $x_{\rm H}$.
The limit $x_{\rm H} \rightarrow 0$ is known as ``extremal'' limit.
The ``extremal'' solutions have naked singularities
at the origin.
For finite horizon the solutions satisfy relation (\ref{res3}),
whereas in the ``extremal'' limit they satisfy relations (\ref{res1})
and (\ref{res2}).
For ``extremal'' EMD solutions
the quadratic relation (\ref{mdp}) between mass,
dilaton charge and magnetic charge simplifies to
\begin{equation}
\mu=P/\sqrt{1+\gamma^2}
\ . \label{mdpex} \end{equation}

For comparison let us also note the metric functions
$A$ and $N$ of the metric (\ref{metric})
\begin{equation}
N = \frac{\lambda^2}{A^2} \ , \ \ \ 
A = \frac{\partial X}{\partial x}
\ . \end{equation}
For the ``extremal'' solutions
\begin{equation}
N(0)=\frac{\gamma^4}{(1+\gamma^2)^2}
\ . \end{equation}

\boldmath
\subsection{$\gamma=1$}
\unboldmath

For $\gamma=1$ the above formulae simplify considerably.
The relation between the coordinates $x$ and $X$
is easily inverted
\begin{equation}
 X = \frac{X_- + \sqrt{X_-^2 + 4 x^2}}{2}
\ . \end{equation}
Eliminating the parameter $X_-$ via
\begin{equation}
X_- = \frac{2 P^2}{X_+}
\ , \end{equation}
the parameter $X_+$ is determined by 
\begin{equation}
X_+=\sqrt{x_{\rm H}^2+2P^2}
\   \end{equation}
for a given horizon $x_{\rm H}$ and magnetic charge $P$.
The mass, dilaton charge and Hawking temperature then are
\begin{equation}
\mu(\infty)=X_+/2
\ , \end{equation}
\begin{equation}
D=P^2/X_+ =\frac{P^2}{2 \mu(\infty)}
\ , \end{equation}
and 
\begin{equation}
T/T_{\rm S}=x_{\rm H}/X_+=\frac{x_{\rm H}}{2 \mu(\infty)}
\ . \end{equation}

\boldmath
\subsection{$\gamma=0$}
\unboldmath

For $\gamma=0$ the dilaton field decouples. The black hole
solutions with magnetic charge $P$ are simply the
Reissner-Nordstr\o m solutions
with metric (\ref{metric}), where
\begin{equation}
N=\left(1-\frac{2 \mu}{x} +\frac{P^2}{x^2}\right) \ , \ \ \
A\equiv 1
\ . \label{mRN} \end{equation}
The non-vanishing component of 
the Maxwell field strength tensor is given by eq.~(\ref{ftp})
as above.
For magnetically neutral black holes 
the metric (\ref{mRN}) reduces to the Schwarzschild metric,
where $P=0$.

\section{Appendix B: Charged SU(3) Einstein-Yang-Mills Black Holes}

A restricted subset of SU(3) 
EYMD black hole solutions with $K \equiv H$
has been studied previously \cite{gal}. 
These solutions
correspond to magnetically charged black holes \cite{gal}. 
To derive their equations of motion, let us put
$ K=H=\frac{f}{\sqrt{2}}$
in the SO(3) equations, eqs.~(\ref{eqmu3})-(\ref{eqdil3}).
This yields for the metric functions \cite{fn6}
\begin{equation}
\mu'= \frac{1}{2} N x^2 \phi'^2 + e^{2 \gamma \phi } 
\left[ N f'^2
  + \frac{1}{2 x^2}  \left(f^2-1 \right)^2 
  + \frac{3}{2 x^2} \right]
\ , \label{eqmu1} \end{equation}
\begin{equation}
 A' =  \frac{2}{x} \left[\frac{1}{2} x^2 \phi'^2 
                 + e^{2 \gamma \phi } f'^2  \right] A
\ . \label{eqa1} \end{equation}
For the matter field functions we obtain the equations
\begin{equation}
(e^{2 \gamma \phi } ANf')' 
= e^{2 \gamma \phi } \frac{1}{x^2} A f \left( f^2-1 \right)
\ , \label{eqf1} \end{equation}
\begin{equation}
(AN x^2 \phi')' = 2 \gamma A e^{2 \gamma \phi } 
\left[ N f'^2
  + \frac{1}{2 x^2} \left(f^2-1 \right)^2 
  + \frac{3}{2 x^2} \right]
\ . \label{eqdil1} \end{equation}
These equations differ from the SU(2) equations only
by the presence of the term $\frac{3}{2 x^2}$
in the equations of $\mu$ and $\phi$.
In fact they correspond to magnetic ${\rm SU(2)} \times {\rm U(1)}$ 
equations.
In general a term of the form
$\frac{P^2}{2 x^2}$ is present for magnetically
charged solutions with charge $P$.
For the asymptotic behaviour of the gauge field function
\begin{equation}
f(x) \stackrel {x\rightarrow \infty} {\longrightarrow} 1
\ , \label{bc11} \end{equation}
the above solutions have magnetic charge $P=\sqrt{3}$,
while for 
\begin{equation}
f(x) \stackrel {x\rightarrow \infty} {\longrightarrow} 0
\ , \label{bc12} \end{equation}
their magnetic charge is $P=2$.

Obviously, the condition $K\equiv H$ is in conflict with
the boundary conditions (\ref{bc1}) and (\ref{bc2}),
imposed on the gauge field functions
of neutral black hole solutions at infinity.
Neither can the boundary conditions (\ref{bc5}) and (\ref{bc6}),
imposed for the gauge field function of regular solutions 
at the origin, be satisfied for solutions with $K\equiv H$.
Therefore these solutions have no regular limit \cite{volkov3}.
For $\gamma=0$ there exists for any $n$
an extremal black hole solution
with finite horizon $x_{\rm H}=P$, in close analogy to the
extremal RN solution \cite{volkov3}.
For $\gamma \ne 0$ the black hole solutions
exist for any $n$ down to arbitrary small horizon.
In the limit $x_{\rm H} \rightarrow 0$
an ``extremal'' solution is obtained \cite{gal}.

In the following we show, that this ``extremal'' solution is
closely analogous to the ``extremal'' EMD solution.
To study the limit $x \rightarrow 0$,
let us introduce new functions $G$ and $\Gamma$ \cite{gal} by 
\begin{equation}
G = e^{\gamma \phi}\ , \ \ \ \Gamma =\gamma x\phi'=\frac{x G'}{G} \ ,
\end{equation}
with 
\begin{equation}
G \stackrel{x \rightarrow 0}{\longrightarrow}  G_0 x \ ,
 \ \ \ \Gamma \stackrel{x \rightarrow 0}{\longrightarrow} 1 \ .
\end{equation} 
Inserting the small $x$ expansion for the functions $G$ and $f$ 
in eqs.~(\ref{eqmu1}), (\ref{eqa1}) and (\ref{eqdil1}), 
we obtain to lowest order
\begin{equation}
\mu' = \frac{1}{2} \frac{N}{\gamma^2} + \frac{1}{2} G_0^2 P^2 \ ,
\label{aseqmu1}
\end{equation}
\begin{equation}
\frac{x A'}{A} = \frac{1}{\gamma^2} \ 
\label{aseqa1}
\end{equation}
and
\begin{equation}
(A N x)' = \gamma^2 A G_0^2 P^2 \ .
\label{aseqdil1}
\end{equation}
There are no contributions from the gauge field function to this 
order.
Solving eqs.~(\ref{aseqmu1}) and (\ref{aseqa1}) gives
to lowest order in $x$  
\begin{equation}
\mu = \frac{1}{2} \ \frac{\gamma^2}{1+\gamma^2}
 \ (\frac{1}{\gamma^2} + G_0^2 P^2) \ x \ ,
\end{equation}
\begin{equation}
A = A_0 x^{\frac{1}{\gamma^2}} \ ,  \label{eqA0}
\end{equation}
where $A_0$ is an integration constant.
Eq. (\ref{aseqdil1}) then leads to 
\begin{equation}
G_0^2 P^2 = \frac{1}{1+\gamma^2} \ .
\end{equation}
Thus we find to lowest order in $x$ 
\begin{equation}
\mu = \frac{1}{2} \frac{1+2\gamma^2}{(1+\gamma^2)^2} x \ ,
\end{equation}
\begin{equation}
G = \frac{1}{\sqrt{P^2(1+\gamma^2)}} x \ ,
\label{gzero}
\end{equation}
and at the origin 
\begin{equation}
N(0) = (\frac{\gamma^2}{1+\gamma^2})^2 \ .
\label{nzero}
\end{equation}
To determine $A_0$ we note, that eq.~(\ref{res2}) holds also
for the ``extremal" EYMD solutions. Rewriting eq.~(\ref{res2})
in terms of $G(x)$ yields 
\begin{equation}
G = (A^2 N)^{\gamma^2/2} \ .
\label{exmetdil}
\end{equation}
Inserting the small $x$ expansions,
eqs. (\ref{eqA0}), (\ref{gzero}) and (\ref{nzero}), we find
the constant $A_0$, 
\begin{equation}
A_0 =  \frac{1+\gamma^2}{\gamma^2}
                 \ (\sqrt{1+\gamma^2} P)^{-\frac{1}{\gamma^2}}
                  \ .
\label{azero}
\end{equation}
$G_0$, $A_0$ and $N(0)$ 
coincide with the corresponding expressions for the 
``extremal'' EMD solutions in the limit $X_+ = X_-$.

Using the functions $G$ and $\Gamma$ instead of $\phi$ and $\phi'$,
we have constructed the ``extremal'' solutions numerically with the
boundary conditions 
\begin{equation}
G  \stackrel {x\rightarrow \infty} {\longrightarrow} 1 \ , \ \ 
\Gamma(0) = 1 \ ,
\end{equation} 
and the boundary conditions of the regular solutions for
the other functions.

Let us finally calculate the inverse Hawking temperature 
of the ``extremal" solutions. 
At the horizon the equations for $\mu$ and $\Gamma$ are given by
\begin{equation}
\mu_{\rm H}' = G_{\rm H}^2 [\ \ \ ]_{\rm H} \ , \ \ 
\Gamma_{\rm H} = \frac{2 \gamma^2 G^2_{\rm H}}{x_{\rm H} N_{\rm H}'} 
                 [\ \ \ ]_{\rm H} \ . \label{eqGamxh}
\end{equation}
(The index ${\rm H}$ indicates the value of the functions at the
 horizon,
 $[\ \ \ ]_{\rm H}$ abbreviates an irrelevant expression.)
Eliminating $G_{\rm H}^2 [\ \ \ ]_{\rm H}$ in eq.~(\ref{eqGamxh})
we find 
\begin{equation}
\mu_{\rm H}' =  \frac{\Gamma_{\rm H}}{\Gamma_{\rm H}+\gamma^2}  
\frac{\mu_{\rm H}}{x_{\rm H}}  \ . 
\end{equation}
In the limit $x_{\rm H} \rightarrow 0$
with $\Gamma_{\rm H} \rightarrow 1$ and 
${\mu_{\rm H}} = \frac{x_{\rm H}}{2}$
 we obtain
\begin{equation}
\mu_{\rm H}' \  
\stackrel {x_{\rm H}\rightarrow 0} {\longrightarrow} \
\frac{1}{2} \frac{1}{1+\gamma^2} \ , \ \  {\rm i. e. } \ \ 
(xN')_{\rm H} \ 
\stackrel {x_{\rm H}\rightarrow 0} {\longrightarrow} \
\frac{\gamma^2}{1+\gamma^2} \ . 
\end{equation}
(Note, that the limit $x \rightarrow 0$ for the ``extremal" solutions 
coincides with the limit $x_{\rm H} \rightarrow 0$ of the 
black hole solutions for the functions $G$, $\Gamma$, $A$, $f$, 
but not for 
$N$.)

Using the small $x$ behaviour of $A$, eqs.~(\ref{eqA0}) and 
(\ref{azero}),
the inverse Hawking temperature becomes in the
limit of vanishing horizon 
\begin{equation}
\beta(x_{\rm H}) \
\stackrel {x_{\rm H}\rightarrow 0} {\longrightarrow} \
4 \pi \ 
      \ \left(\sqrt{1+\gamma^2} P \right)^{\frac{1}{\gamma^2}}
      \ x_{\rm H}^{\frac{\gamma^2-1}{\gamma^2}} \ .
\end{equation}
Thus for the ``extremal" EYMD solutions 
$\beta$ is finite for $\gamma = 1$,
it diverges for $\gamma < 1$ and vanishes for $\gamma > 1$,
completely analogous to
the inverse Hawking temperature of the ``extremal" EMD solutions.

The limit of small horizon for the EYMD black hole solutions
on the one hand and the limit of small $x$ for the ``extremal" 
 EYMD solutions
on the other hand coincide with the limiting behaviour of the 
corresponding 
EMD solutions. 
The reason is clearly, that there is no explicit contribution from
the gauge field in this limit.
Therefore, for $\gamma =1$, $\beta$ has the same value for the
``extremal" EMD solution and for the 
``extremal" EYMD solutions with $j$ nodes, as seen in Fig.~17.

\newpage

\begin{table}[p!]
\begin{center}
$\mu(\infty)$ \vspace{.5cm}\\ 
\begin{tabular}{|c|ccccc|}
 \hline
  $n$   & $x_{\rm H}$ & 
$ .0$     &      $ .01$    &      $ .1$     &      $1.0$     \\
\hline  
  $1$ & & $ .57699$ & $ .57896$ &      $ .59726$ &      $ .83671$ \\
  $2$ & & $ .68483$ & $ .68567$ &      $ .69381$ &      $ .86507$ \\
  $3$ & & $ .70345$ & $ .70379$ &      $ .70755$ &      $ .86600$ \\
  $4$ & & $ .70651$ & $ .70665$ &      $ .70882$ &      $ .86602$ \\
  $5$ & & $ .70701$ & $ .70707$ &      $ .70887$ &      $ .86603$ \\
  $6$ & & $ .70709$ & $ .70712$ &      $ .70887$ &      $ .86603$ \\
  $7$ & & $ .70710$ & $ .70712$ &      $ .70887$ &      $ .86603$ \\
\hline  
\end{tabular} \vspace{.5cm}\\
$D$ \vspace{.5cm}\\ 
\begin{tabular}{|c|ccccc|}
 \hline
  $n$   & $x_{\rm H}$ & 
$ .0$     &      $ .01$    &      $ .1$     &      $1.0$     \\
\hline  

  $1$ & & $ .57699$ & $ .57698$ &      $ .57639$ &      $ .51213$ \\
  $2$ & & $ .68483$ & $ .68484$ &      $ .68416$ &      $ .57483$ \\
  $3$ & & $ .70345$ & $ .70344$ &      $ .70262$ &      $ .57728$ \\
  $4$ & & $ .70651$ & $ .70650$ &      $ .70519$ &      $ .57735$ \\
  $5$ & & $ .70701$ & $ .70700$ &      $ .70534$ &      $ .57735$ \\
  $6$ & & $ .70709$ & $ .70708$ &      $ .70535$ &      $ .57735$ \\
  $7$ & & $ .70710$ & $ .70709$ &      $ .70535$ &      $ .57735$ \\
\hline  
\end{tabular} \vspace{.5cm}\\  
$T/T_S$ \vspace{.5cm}\\ 
\begin{tabular}{|c|ccccc|}
 \hline
  $n$   & $x_{\rm H}$ & 
$ .0$     &      $ .01$    &      $ .1$     &      $1.0$     \\
\hline  
  $1$ & & $ .39365$ & $ .39601$ &      $ .41750$ &      $ .64915$ \\
  $2$ & & $ .16649$ & $ .16900$ &      $ .19299$ &      $ .58048$ \\
  $3$ & & $ .06773$ & $ .07028$ &      $ .09870$ &      $ .57744$ \\
  $4$ & & $ .02738$ & $ .03000$ &      $ .07256$ &      $ .57735$ \\
  $5$ & & $ .01106$ & $ .01388$ &      $ .07059$ &      $ .57735$ \\
  $6$ & & $ .00446$ & $ .00809$ &      $ .07054$ &      $ .57735$ \\
  $7$ & & $ .00180$ & $ .00711$ &      $ .07053$ &      $ .57735$ \\
\hline  
\end{tabular}
\end{center} 
\vspace{1.cm} 
{\bf Table 1}\\
The mass $\mu(\infty)$, dilaton charge $D$ and 
Hawking temperature $T/T_{\rm S}$
for the SU(2) EYMD regular solutions ($x_{\rm H}= 0$) and black hole 
solutions ($x_{\rm H}=$ $.01$, $.1$ and $1$)
 with up to seven nodes $n$
and dilaton coupling constant 
$\gamma = 1$. 
\end{table}
\newpage

\begin{table}[p!]
\begin{center}
\begin{tabular}{|c|cccc|} \hline
\multicolumn{1}{|c|} { $n$ }&
\multicolumn{4}{ c|}  {$\gamma$ }  \\   
 \hline
          &   $0.1$    &    $0.5$    &    $1.0$   & $2.0$ \\
 \hline 
 $1$      &  $1.53846$ &   $1.42382$ &  $1.35402$ & $1.62119$ \\ 
 $2$      &  $1.08289$ &   $ .80649$ &  $ .53637$ & $ .44233$ \\  
 $3$      &  $ .95933$ &   $ .52922$ &  $ .21751$ & $ .10992$ \\  
 $4$      &  $ .92289$ &   $ .36395$ &  $ .08791$ & $ .02606$ \\  
 $5$      &  $ .90328$ &   $ .25272$ &  $ .03550$ & $ .00612$ \\  
 $6$      &  $ .88668$ &   $ .17577$ &  $ .01434$ & $ .00143$ \\  
 $7$      &  $ .87082$ &   $ .12229$ &  $ .00579$ & $ .00034$ \\  
 \hline  
 $c_{\gamma}$ &  $.018$   &   $ .363$   &  $ .907$   & $1.451$ \\ 
 $d_{\gamma}$ &  $.988$   &   $1.550$   &  $3.308$   & $8.661$ \\          
\hline
\end{tabular}
\end{center} 
\vspace{1.cm} 
{\bf Table 2}\\
The location of the innermost node $z_n^{(1)}(\gamma)$
for the SU(2) EYMD regular
solutions up to seven nodes $n$ for several values of the dilaton
coupling constant $\gamma$. 
The nodes are well approximated by 
$z_n^{(1)}(\gamma) = d_{\gamma} e^{-c_{\gamma} n}$,
with constants $d_{\gamma}$, $c_{\gamma}$ also shown.
\end{table}

\begin{table}[p!]
\begin{center}
\begin{tabular}{|c|ccc|ccc|} \hline
\multicolumn{1}{|c|} {}&
\multicolumn{3}{ c|}  {$0 < x_{\rm H} \leq 1$ } &
\multicolumn{3}{ c|}  {$10 \leq x_{\rm H} < \infty $ } \\
          &  $\mu(\infty)$ &    $D$    &    $T/T_S$  &  
             $\mu(\infty)$ &    $D$    &    $T/T_S$\\
 \hline 
 $\delta$ &  $1.916$      & $1.951$ &  $2.941$ & 
             $1.001$      & $1.002$ &  $2.006$  \\ 
 $b$      &  $1.156$      & $2.960$ &  $3.361$ & 
             $0.782$      & $1.594$ &  $1.607$  \\ 
 $\alpha$ &  $3.623$      & $3.594$ &  $3.568$ & 
             $3.591$      & $3.630$ &  $3.624$  \\  
 \hline
\end{tabular}
\end{center} 
\vspace{1.cm} 
{\bf Table 3}\\
The constants $\delta$, $b$, $\alpha$ of the least square fits for
$\Delta \mu_n = \mu_\infty(\infty) - \mu_n(\infty)
= \frac{b_\mu}{(x_{\rm H})^{\delta_\mu}} e^{-\alpha_\mu n}$
(and analogous for $D$ and $T/T_S$ )
for the SU(2) EYMD sequence for  $\gamma=1$.
The fits have been done separately for the regions 
$0 < x_{\rm H} \leq 1$ and 
$10 \leq x_{\rm H} < \infty $.
\end{table}
\newpage

\begin{table}[p!]
\begin{center}
\begin{tabular}{|c|cccc|} \hline
\multicolumn{1}{|c|} { $(j,j+n)$ }&
\multicolumn{4}{ c|}  {$x_{\rm H}$ }  \\   
 \hline
              &   $ 0.$    &    $0.5$    &    $1.0$   & $2.0$ \\
 \hline    
 $(0,1)$      &  $ .90853$ &   $1.02708$ &  $1.16497$ & $1.49442$ \\ 
 $(0,2)$      &  $1.14153$ &   $1.20944$ &  $1.30297$ & $1.57396$ \\
 $(1,1) *$    &  $1.15397$ &   $1.25989$ &  $1.38135$ & $1.67341$ \\
 $(1,1)$      &  $1.18797$ &   $1.27399$ &  $1.38437$ & $   -   $ \\ 
 $(0,3)$      &  $1.20383$ &   $1.24529$ &  $1.32140$ & $1.58069$ \\
 $(0,4)$      &  $1.21958$ &   $1.24966$ &  $1.32279$ & $1.58111$ \\
 $(0,5)$      &  $1.22347$ &   $1.24998$ &  $1.32287$ & $1.58114$ \\
 $(1,2)$      &  $1.32130$ &   $1.37840$ &  $1.46227$ & $1.71399$ \\ 
 $(1,3)$      &  $1.36195$ &   $1.39923$ &  $1.47209$ & $1.71729$ \\
 $(2,2) *$    &  $1.36967$ &   $1.42005$ &  $1.49382$ & $1.73014$ \\ 
 $(1,4)$      &  $1.37300$ &   $1.40151$ &  $1.47278$ & $1.71750$ \\
 $(2,2)$      &  $1.37417$ &   $   -   $ &  $   -   $ & $   -   $ \\ 
 $(2,3)$      &  $1.39866$ &   $1.43109$ &  $1.49813$ & $1.73148$ \\ 
\hline
\end{tabular}
\end{center} 
\vspace{1.cm} 
{\bf Table 4}\\
The mass of the lowest regular ($x_{\rm H} = 0$) and black hole 
($x_{\rm H} = 0.5$, $1$ and $2$) SO(3) EYMD solutions
with a total number of nodes 
$n_T \leq 5$ and dilaton coupling constant 
$\gamma = 1$.
The asterix indicates the scaled SU(2) solutions.
Note that the genuine SO(3) black hole solutions with node structure
$(n,n)$ do not exist for all horizons. 
\end{table}
\newpage
\begin{table}[p!]
\begin{center}
\begin{tabular}{|c|ccc|} \hline
   $(j,j+n)$      &  $\mu({\infty})$ &    $D$    &    $T/T_S$   \\
 \hline 
 $(0,1)     $      &  $1.16497$      & $ .87027$ &  $ .58942$  \\ 
 $(0,2)     $      &  $1.30297$      & $1.08770$ &  $ .43054$  \\  
 $(0,3)     $      &  $1.32140$      & $1.12979$ &  $ .38322$  \\  
 $(0,4)     $      &  $1.32279$      & $1.13364$ &  $ .37830$  \\  
 $(0,5)     $      &  $1.32287$      & $1.13388$ &  $ .37798$  \\ 
 $(0,\infty)$      &  $1.32288$      & $1.13389$ &  $ .37796$  \\ 
 \hline
 $\alpha_\mu^0$    &  $2.798$        & $ 2.786$  &  $ 2.780$   \\
 $a_\mu^0     $    &  $6.529$        & $17.582$  &  $22.235$   \\   
 \hline
 \hline 
 $(1,2)     $      &  $1.46227$      & $1.26493$ &  $ .39467$  \\ 
 $(1,3)     $      &  $1.47209$      & $1.28839$ &  $ .36740$  \\  
 $(1,4)     $      &  $1.47278$      & $1.29029$ &  $ .36499$  \\ 
 $(1,5)     $      &  $1.47283$      & $1.29041$ &  $ .36484$  \\ 
 $(1,6)     $      &  $1.47283$      & $1.29042$ &  $ .36483$  \\
 $(1,\infty)$      &  $1.47283$      & $1.29042$ &  $ .36483$  \\  
 \hline 
 $\alpha_\mu^1$    &  $2.773$        & $2.748$   &  $2.727$    \\
 $a_\mu^1     $    &  $ .180$        & $ .443$   &  $ .527$    \\   
 \hline 
 \hline 
 $(2,3)     $      &  $1.49813$      & $1.32829$ &  $ .33968$  \\ 
 $(2,4)     $      &  $1.49843$      & $1.32917$ &  $ .33853$  \\ 
 $(2,5)     $      &  $1.49845$      & $1.32922$ &  $ .33846$  \\ 
 $(2,6)     $      &  $1.49845$      & $1.32922$ &  $ .33845$  \\ 
 $(2,7)     $      &  $1.49845$      & $1.32922$ &  $ .33845$  \\ 
 $(2,\infty)$      &  $1.49845$      & $1.32922$ &  $ .33845$  \\ 
 \hline 
 $\alpha_\mu^2$    &  $2.806$        & $2.801 $  &  $2.803$    \\
 $a_\mu^2     $    &  $ .005$        & $ .015 $  &  $ .020$    \\   
 \hline
 \hline   
 $(\infty,\infty)$     &  $1.5$      & $1.33333$ &  $ .33333$  \\   
 \hline   
\end{tabular}
\end{center} 
\vspace{1.cm} 
{\bf Table 5  }\\
The mass $\mu(\infty)$, dilaton charge $D$ and Hawking Temperatur
$T/T_S$ for the SO(3) EYMD black hole solutions of the 
sequences $(j,j+n)$ with $0 \leq j \leq 2$, $1 \leq n \leq 5$ 
for dilaton coupling constant $\gamma =1$ and horizon $x_{\rm H}=1$.
For each sequence the limiting values 
$\mu_\infty^j(\infty)$, $D_\infty^j$ and 
$(T/T_S)_\infty^j$ are given.
The masses $\mu_n(\infty)$ are approximated by
$\mu_\infty(\infty) - \mu_n(\infty) =
a_\mu^j e^{-\alpha_\mu^j n}$
(and analogously $D$ and $T/T_S$),
with the constants $a$, $\alpha$ also shown.
The last line contains the values obtained 
for $\mu_\infty^j(\infty)$,
 $D_\infty^j$ and $(T/T_S)_\infty^j$ in the limit
  $j \longrightarrow \infty$.
\end{table}

\newpage

\begin{table}[p!]
\begin{center}
\begin{tabular}{|c|c|crrr|rr|}
 \hline
&  $n$  & $j$ = & $0$  & $1$ & $2$ & $c_n$        & $d_n$   \\
\hline
\hline           
& $1$   & & $2.47869$ & $1.86167$ & $1.69430$ & $ .094$ & $2.046$ \\ 
& $2$   & & $1.86074$ & $1.64685$ & $1.58248$ & $ .040$ & $1.714$ \\ 
& $3$   & & $1.64914$ & $1.56969$ & $1.53269$ & $ .024$ & $1.608$ \\ 
$\gamma= 0.1$
& $4$   & & $1.57468$ & $1.53460$ & $1.50527$ & $ .019$ & $1.565$ \\ 
& $5$   & & $1.53986$ & $1.50966$ & $1.48215$ & $ .018$ & $1.538$ \\
\cline{2-8} 
& $c_j$ & & $ .034$ & $ .019$  & $ .017$        &              & \\ 
& $d_j$ & & $1.821$ & $1.663$  & $1.611$        &              & \\  
\hline
\hline           
& $1$   & & $2.29374$ & $1.34077$ & $ .89474$ & $ .404$ & $2.009$ \\
& $2$   & & $1.43827$ & $ .94746$ & $ .65374$ & $ .371$ & $1.373$ \\
& $3$   & & $1.01694$ & $ .69874$ & $ .48400$ & $ .367$ & $1.009$ \\
$\gamma= 0.5$
& $4$   & & $ .75328$ & $ .52643$ & $ .36697$ & $ .361$ & $ .755$ \\
& $5$   & & $ .56598$ & $ .39571$ & $ .27491$ & $ .364$ & $ .570$ \\
\cline{2-8} 
& $c_j$ & & $ .293$ & $ .284$ & $ .283$        &              & \\
& $d_j$ & & $2.443$ & $1.640$ & $1.133$        &              & \\
 \hline 
\hline    
& $1$   & & $2.18639$ & $ .85307$ & $ .34698$ & $ .920$ & $2.171$ \\
& $2$   & & $1.02365$ & $ .42121$ & $ .17227$ & $ .891$ & $1.025$ \\
& $3$   & & $ .50393$ & $ .20783$ & $ .08377$ & $ .897$ & $ .506$ \\
$\gamma= 1$
& $4$   & & $ .24955$ & $ .10352$ & $ .04215$ & $ .889$ & $ .250$ \\
& $5$   & & $ .12364$ & $ .05102$ & $ .02055$ & $ .897$ & $ .124$ \\
\cline{2-8}  
& $c_j$ & & $ .716$ & $ .704$ & $ .706$        &              & \\
& $d_j$ & & $4.373$ & $1.722$ & $ .704$        &              &\\
\hline
\hline  
& $1$   & & $2.62983$ & $ .68046$ & $ .16794$ & $1.376$ & $2.651$ \\
& $2$   & & $ .91273$ & $ .23353$ & $ .05661$ & $1.390$ & $ .921$ \\
& $3$   & & $ .31115$ & $ .07745$ & $ .01818$ & $1.420$ & $ .314$ \\
$\gamma= 2$
& $4$   & & $ .10281$ & $ .02540$ & $ .00605$ & $1.416$ & $ .103$ \\
& $5$   & & $ .03357$ & $ .00824$ & $ .00193$ & $1.428$ & $ .034$ \\
\cline{2-8}  
& $c_j$ & & $1.091$ & $1.105$ & $1.117$        &              &\\
& $d_j$ & & $8.001$ & $2.096$ & $ .520$        &              &\\
\hline  
\end{tabular}
\end{center} 
\vspace{1.cm} 
{\bf Table 6}\\
The location of the innermost node $z_n^{(1)}$
for the SO(3) EYMD regular
solutions with node structure $(j,j+n)$ 
with $j \leq 2$ and $n \leq 5$
for the dilaton
coupling constants $\gamma = 0.1$, $0.5$, $1$, $2$.
The nodes are approximated by
 $z_n^{(1)}(j) = d_{j} e^{-c_{j} n}$
 and $z_j^{(1)}(n) = d_{n} e^{-c_{n} j}$,
with the constants $d_j$, $c_j$ and $d_n$, $c_n$ also shown. 
\end{table}
\newpage

\begin{table}[p!]
\begin{center}
$10 \leq x_{\rm H} < \infty $ \vspace{1.cm}\\
\begin{tabular}{|c|ccc|ccc|ccc|} \hline
\multicolumn{1}{|c|} {}&
\multicolumn{3}{ c|}  {$j = 0$ } &
\multicolumn{3}{ c|}  {$j = 1$ } &
\multicolumn{3}{ c|}  {$j = 2$ } \\
          &  $\mu(\infty)$ &    $D$    &    $T/T_S$ & 
             $\mu(\infty)$ &    $D$    &    $T/T_S$ &            
             $\mu(\infty)$ &    $D$    &    $T/T_S$\\
 \hline 
 $\delta$ &  $1.006$      & $1.009$ &  $2.017$ & 
             $1.008$      & $1.016$ &  $2.024$ &
             $1.008$      & $1.018$ &  $2.025$  \\              
 $b$      &  $2.553$      & $5.302$ &  $5.439$ & 
             $0.071$      & $0.148$ &  $0.157$ & 
             $0.002$      & $0.003$ &  $0.004$  \\              
 $\alpha$ &  $2.795$      & $2.804$ &  $2.795$ & 
             $2.807$      & $2.804$ &  $2.808$ & 
             $2.808$      & $2.697$ &  $2.810$  \\              
 \hline
\end{tabular}
\end{center} 
\vspace{1.cm} 
{\bf Table 7}\\
The constants $\delta$, $b$, $\alpha$ 
of the least square fits for 
$\Delta \mu_n = \mu_\infty(\infty) - \mu_n(\infty)
= \frac{b_\mu}{(x_{\rm H})^{\delta_\mu}} e^{-\alpha_\mu n}$
(and analogous for $D$ and $T/T_S$)
for the SO(3) EYMD sequences
$(0,n)$, $(1,1+n)$ and $(2,2+n)$ for $\gamma=1$
and $x_{\rm H} \geq 10$. For  $x_{\rm H} \leq 1$ see
\cite{fitsmallxh}.
\end{table}
\clearpage
\begin{fixy}{0}
%XXXXXXXXXXXXXXXXXXXXXX Figure 1 (a-e) XXXXXXXXXXXXXXXXXXXXXXXXXX

\newpage

\begin{figure}
\centering
\epsfysize=11cm
\mbox{\epsffile{
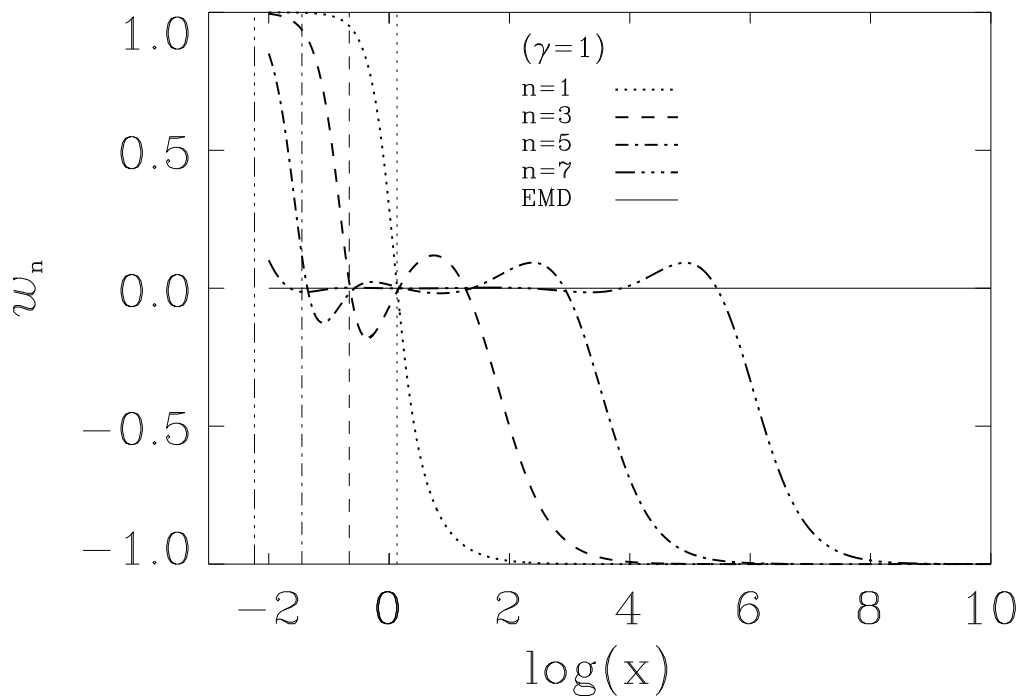
}}
\caption{\label{fig1a}
The SU(2) EYMD gauge field functions $w_n(x)$ with node number
$n=1$ (dotted), 
$n=3$ (dashed), 
$n=5$ (dot-dashed)
and $n=7$ (tripledot-dashed)
 for the black hole solutions with horizon $x_{\rm H}=0.01$
and dilaton coupling constant $\gamma=1$.
The solid line shows the gauge field function of the limiting  
EMD solution with the same horizon and dilaton coupling constant. 
The thin vertical lines
indicate the location of the innermost node of the corresponding 
regular solutions. 
}
\end{figure}
\newpage

\begin{figure}
\centering
\epsfysize=11cm
\mbox{\epsffile{
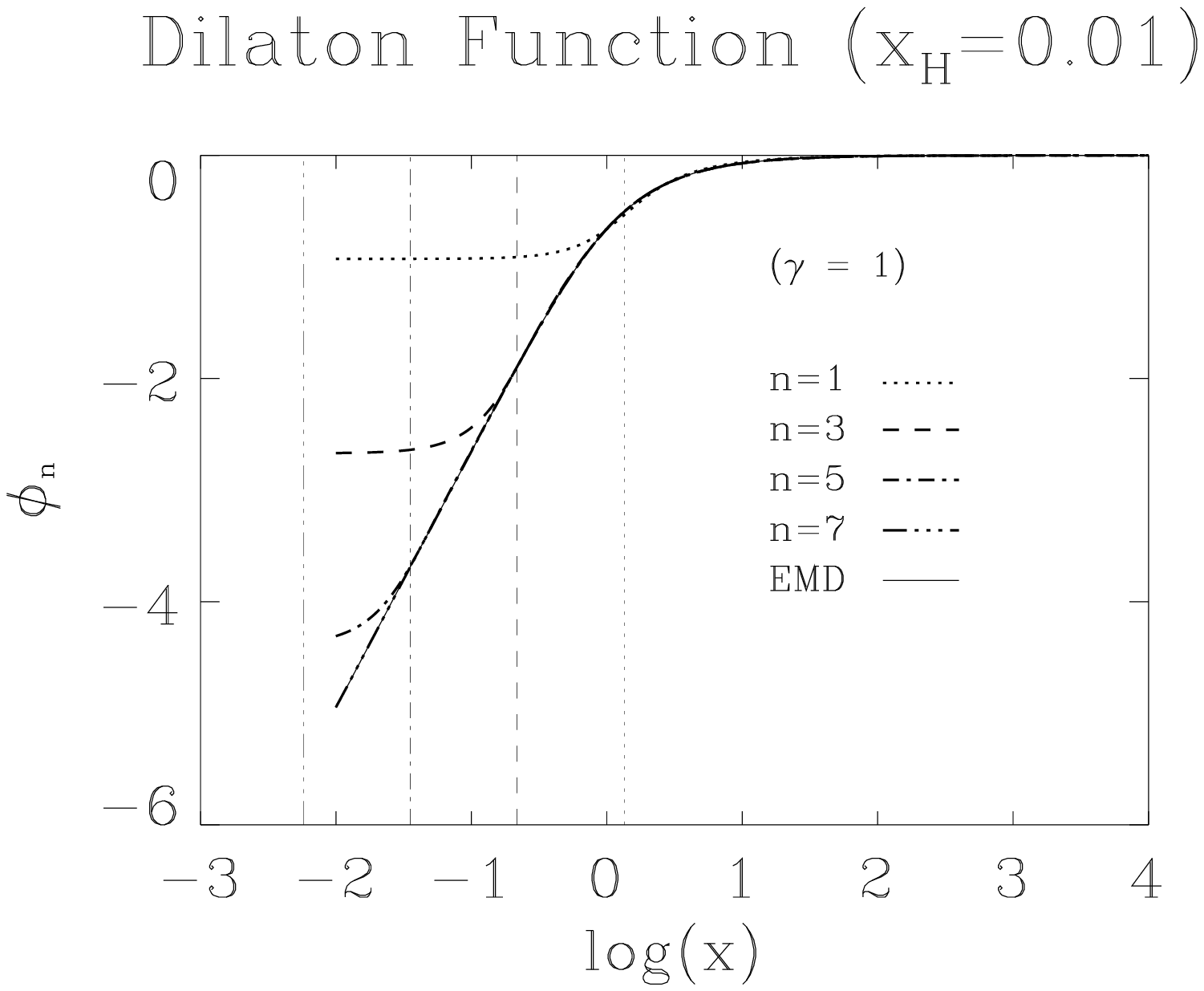
}}
\caption{\label{fig1b}
Same as Figure \ref{fig1a} for the dilaton functions $\phi_n(x)$.
}
\end{figure}
\newpage

\begin{figure}
\centering
\epsfysize=11cm
\mbox{\epsffile{
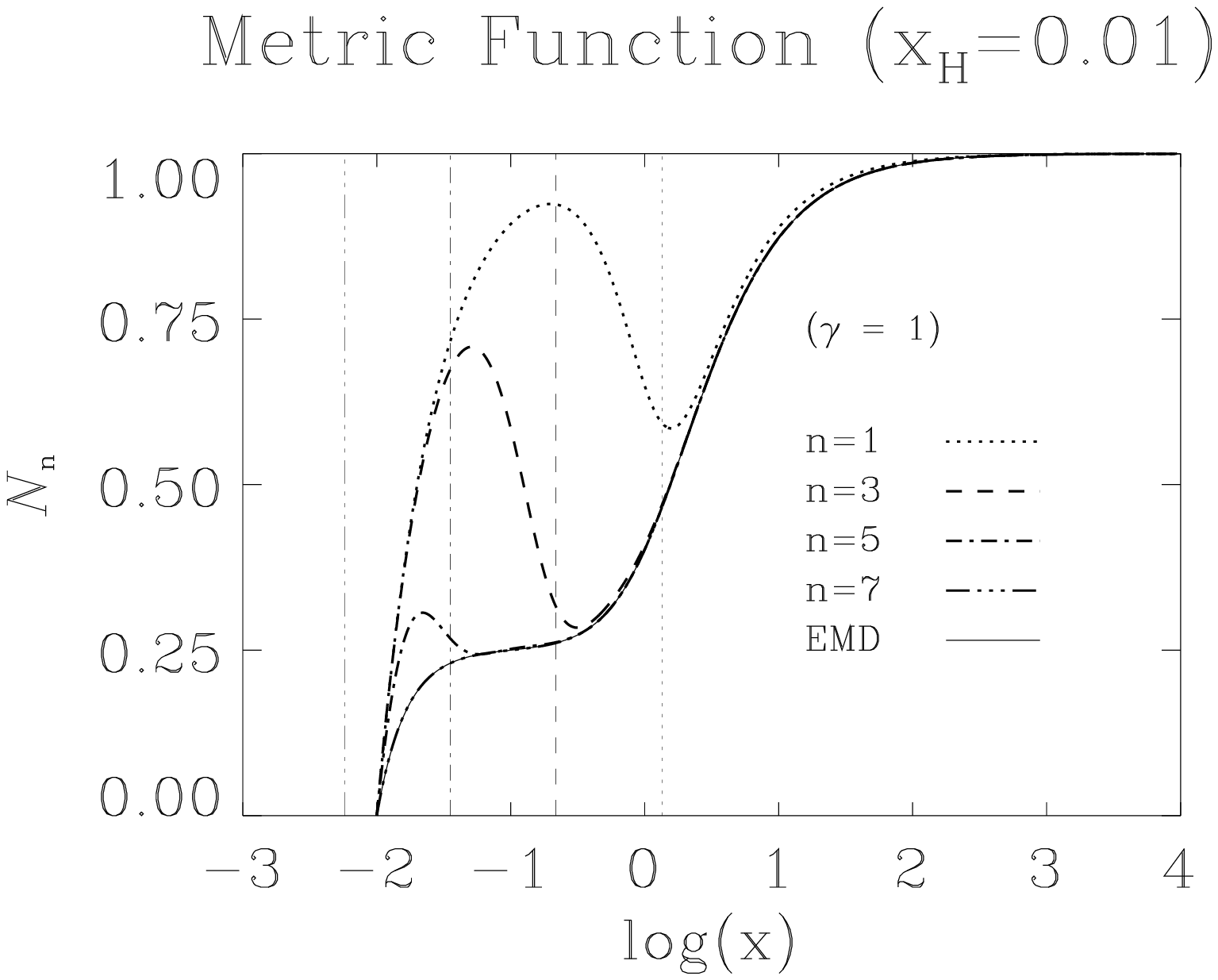
}}
\caption{\label{fig1c}
Same as Figure \ref{fig1a} for the metric functions $N_n(x)$.
}
\end{figure}
\newpage

\begin{figure}
\centering
\epsfysize=11cm
\mbox{\epsffile{
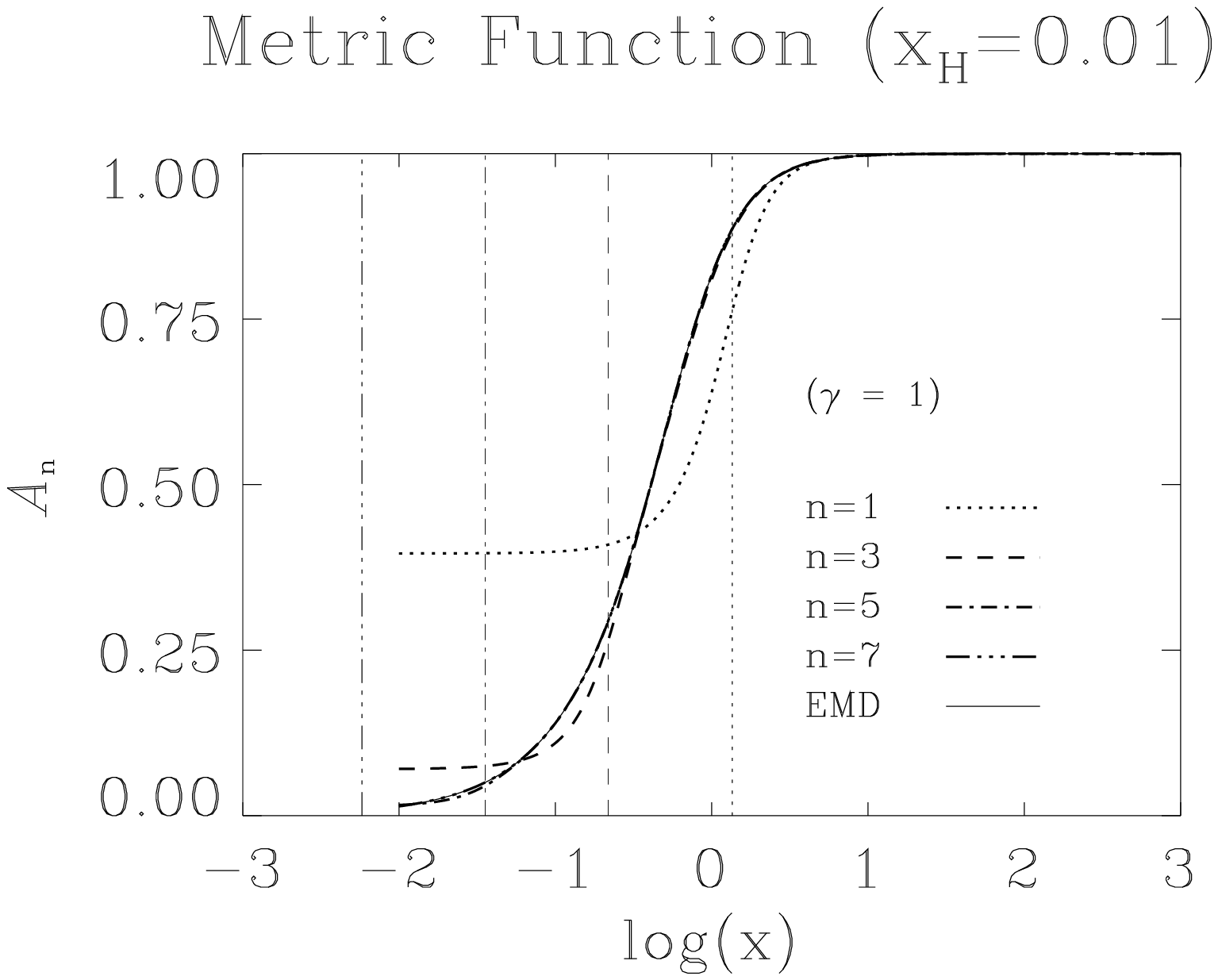
}}
\caption{\label{fig1d}
Same as Figure \ref{fig1a} for the metric functions $A_n(x)$.
}
\end{figure}
\newpage

\begin{figure}
\centering
\epsfysize=11cm
\mbox{\epsffile{
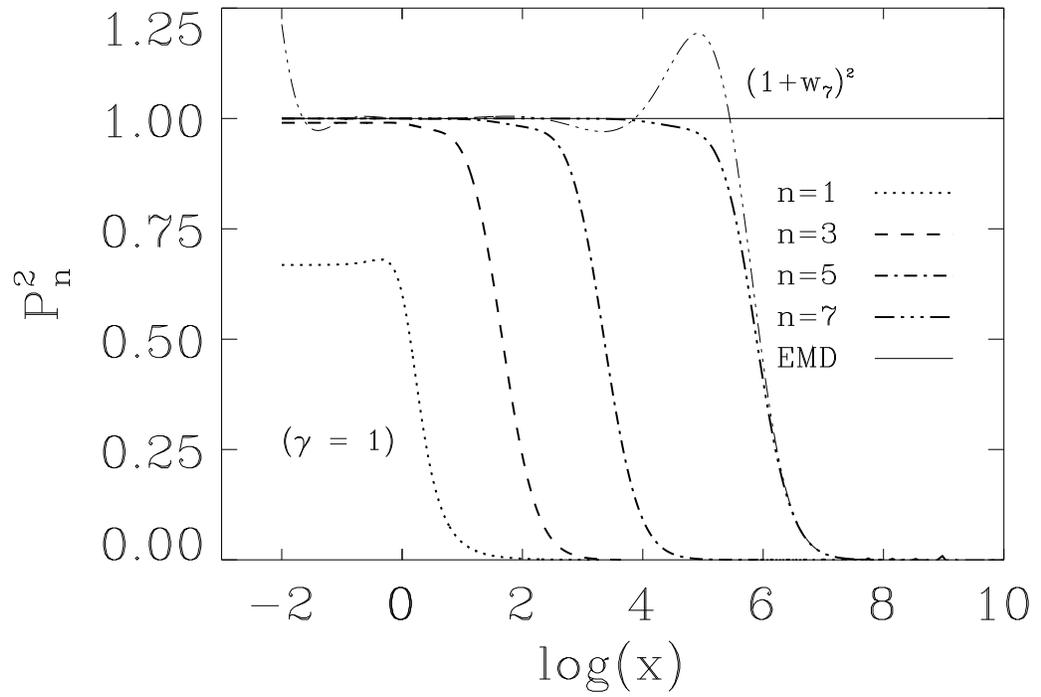
}}
\caption{\label{fig1e}
Same as Figure \ref{fig1a} for the squared charge functions 
$P^2_n(x)$.
Also the function $(1+w_7(x))^2$ 
is shown (thin tripledot-dashed). 
}
\end{figure}
\end{fixy}

%XXXXXXXXXXXXXXXXXXXXXX Figure 2 (a-b) XXXXXXXXXXXXXXXXXXXXXXXXXX
\begin{fixy}{0}
\newpage

\begin{figure}
\centering
\epsfysize=11cm
\mbox{\epsffile{
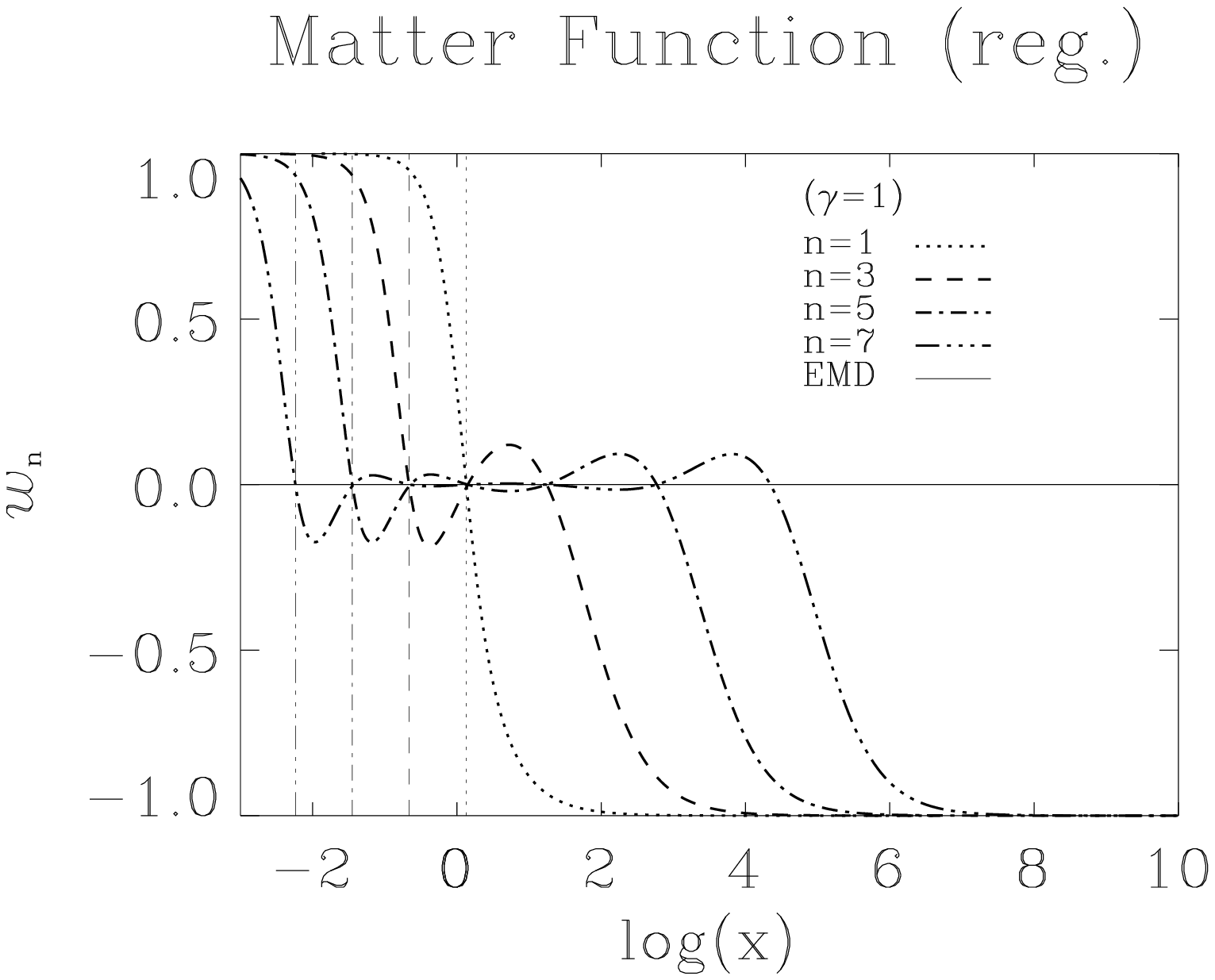
}}
\caption{\label{fig2a}
The SU(2) EYMD gauge field functions $w_n(x)$ with node number
$n=1$ (dotted), 
$n=3$ (dashed), 
$n=5$ (dot-dashed)
and $n=7$ (tripledot-dashed)
for the regular solutions
with dilaton coupling constant $\gamma=1$.
The solid line shows the gauge field function of the 
``extremal" EMD solution with the same dilaton coupling constant. 
The thin vertical lines
indicate the location of the innermost nodes.
}
\end{figure}

\newpage

\begin{figure}
\centering
\epsfysize=11cm
\mbox{\epsffile{
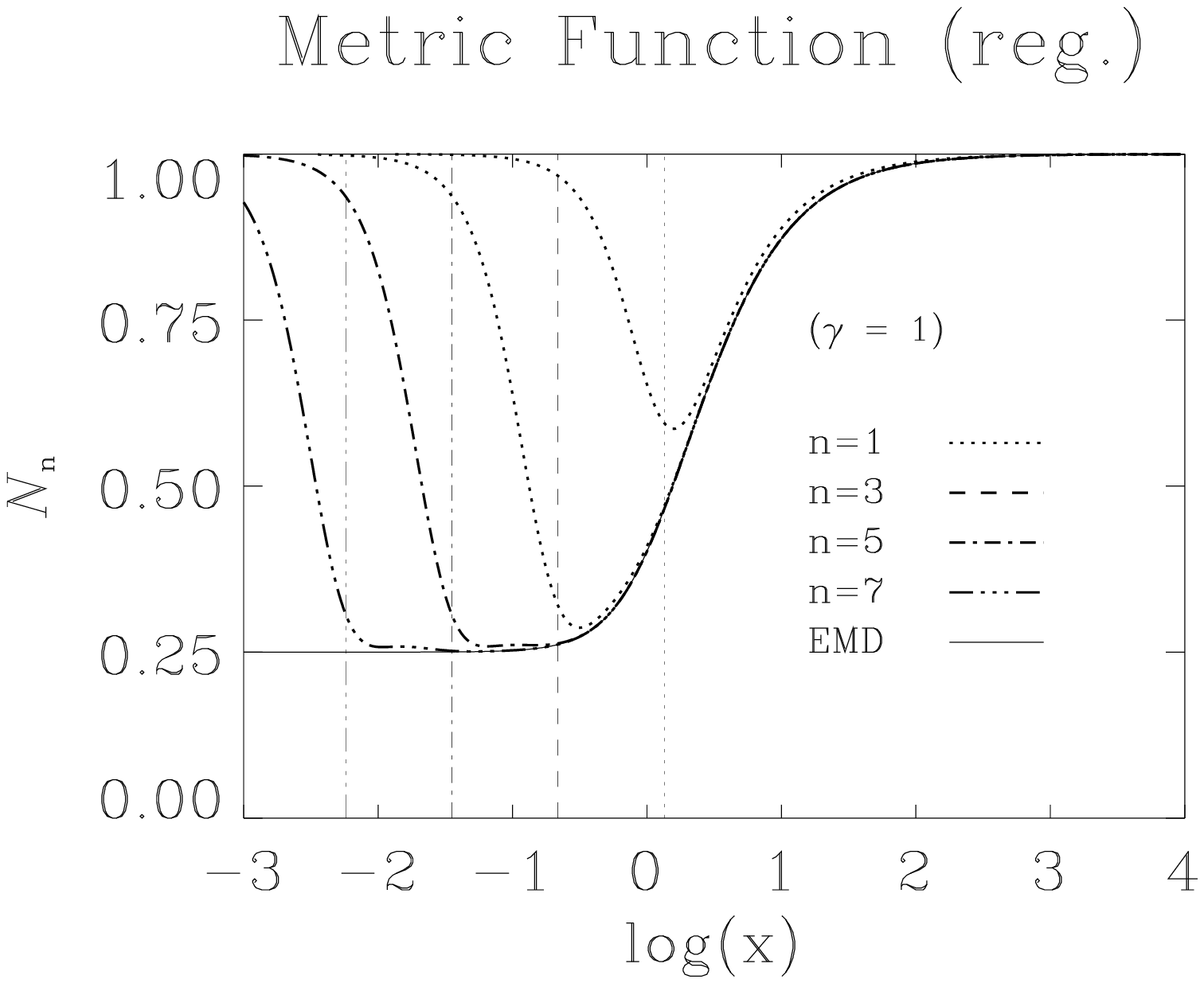
}}
\caption{\label{fig2b}
The same as Figure \ref{fig2a} for the metric functions  $N_n(x)$.
}
\end{figure}
\end{fixy}
\begin{fixy}{-1}
%XXXXXXXXXXXXXXXXXXXXXX Figure 3 XXXXXXXXXXXXXXXXXXXXXXXXXXXXXXXXXX

\newpage

\begin{figure}
\centering
\epsfysize=11cm
\mbox{\epsffile{
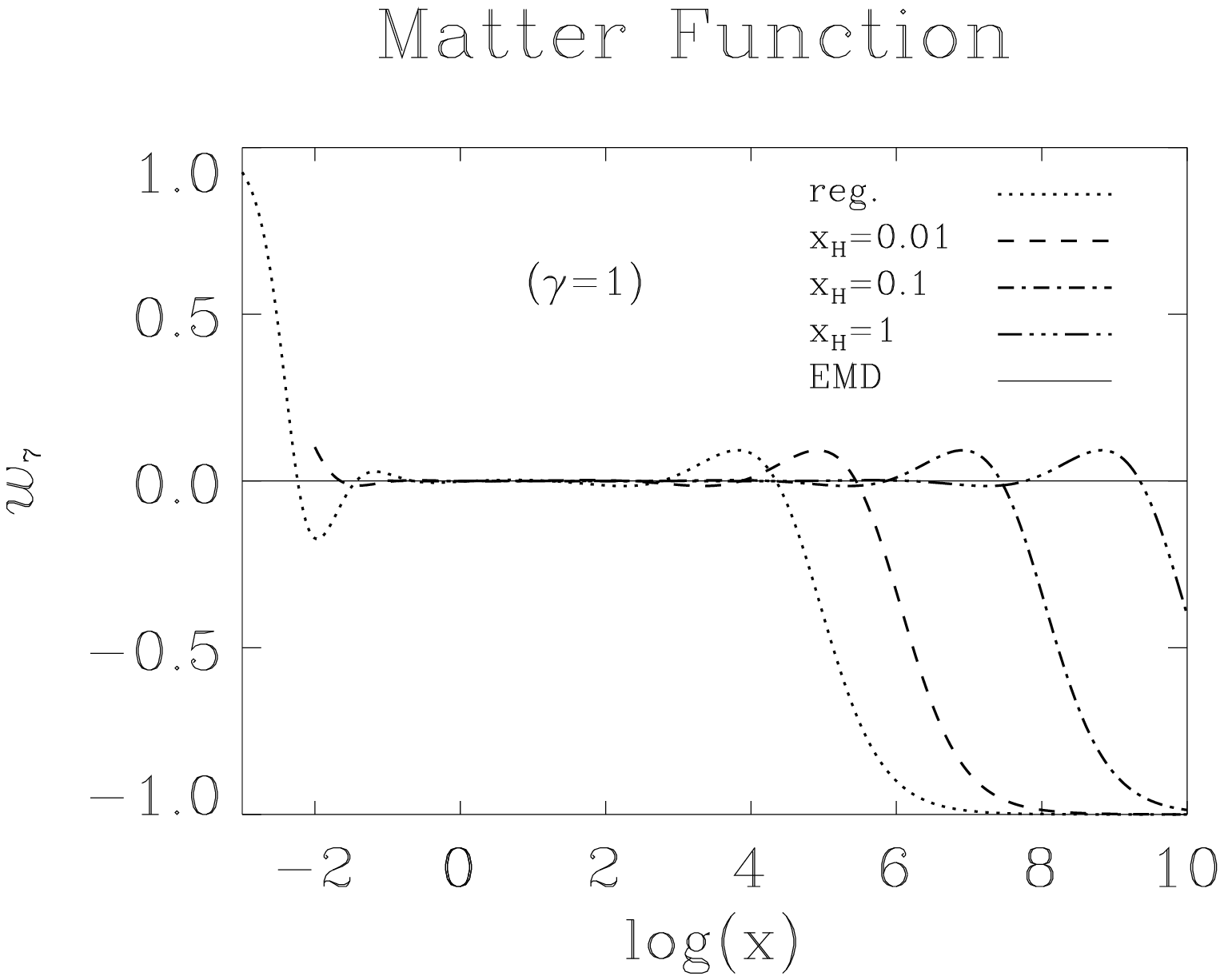
}}
\caption{\label{fig3}
The SU(2) EYMD gauge field functions $w_7(x)$ 
for the 
regular solution (dotted)
and the black hole solutions  with horizon 
$x_{\rm H} = 0.01$ (dashed), 
$x_{\rm H} = 0.1$  (dot-dashed) and
$x_{\rm H} = 1$   (tripledot-dashed) 
for the dilaton coupling constant $\gamma=1$.
The solid line shows the gauge field function of the 
``extremal" EMD solution.  
}
\end{figure}
\end{fixy}
\begin{fixy}{0}
%XXXXXXXXXXXXXXXXXXXXXX Figure 4 (a-b) XXXXXXXXXXXXXXXXXXXXXXXX

\newpage

\begin{figure}
\centering
\epsfysize=11cm
\mbox{\epsffile{
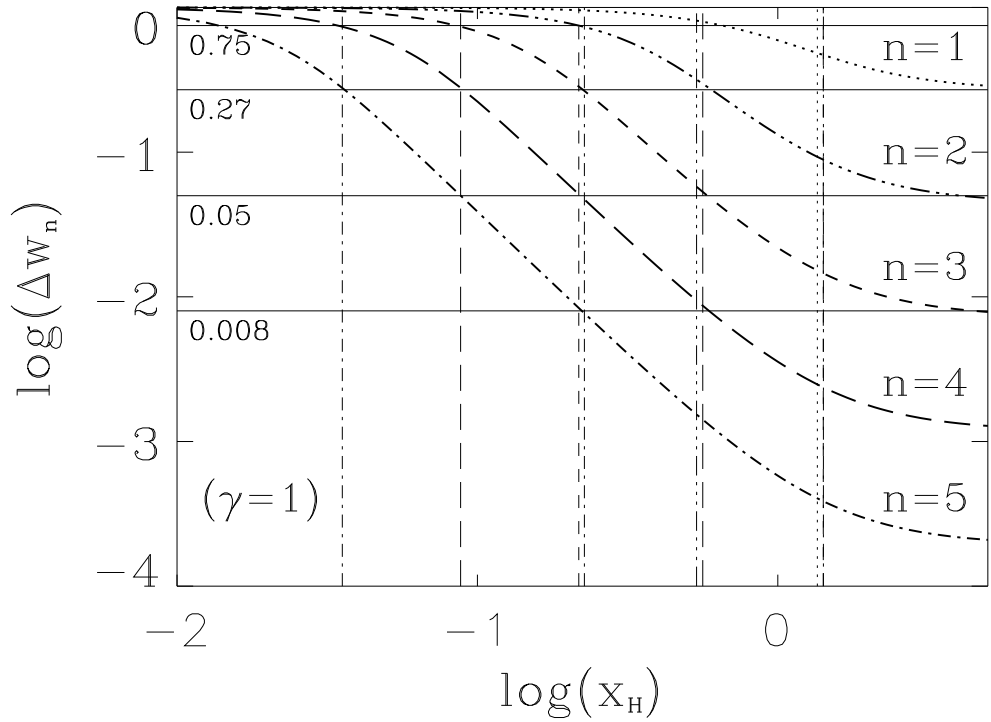
}}
\caption{\label{fig4a}
The deviation at the horizon 
$\Delta w_n(x_{\rm H}) = | w_n(x_{\rm H})- w_\infty (x_{\rm H}) |$
for the SU(2) EYMD gauge field functions 
with up to five nodes  
as a function of the 
horizon $x_{\rm H}$ for the dilaton coupling constant $\gamma=1$.
The vertical lines indicate the location of the innermost and 
second nodes of the corresponding regular solutions.
The horizontal lines show the deviations
of $0.75$, $0.27$, $0.05$ and $0.008$ from the limiting value.
}
\end{figure}

\newpage

\begin{figure}
\centering
\epsfysize=11cm
\mbox{\epsffile{
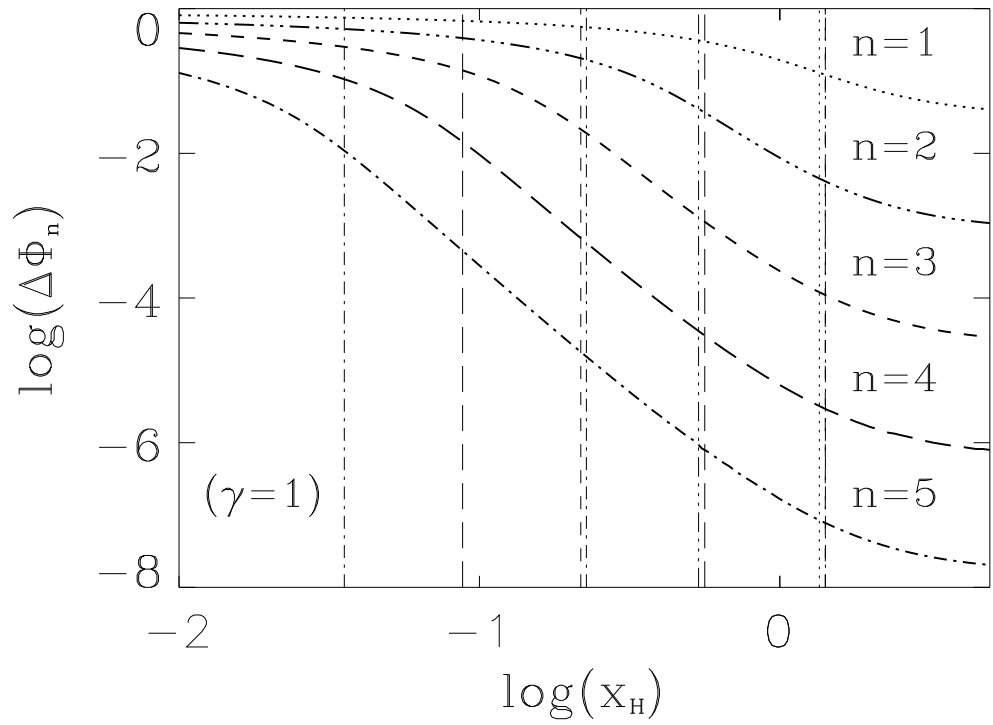
}}
\caption{\label{fig4b}
The relative deviation at the horizon
$\Delta \phi_n(x_{\rm H}) =
 | (\phi_n(x_{\rm H})- \phi_\infty (x_{\rm H}))/\phi_\infty 
 (x_{\rm H}) |$
for the  SU(2) EYMD dilaton functions with up to five nodes
as a function of the 
horizon $x_{\rm H}$ for the dilaton coupling constant $\gamma=1$. 
The vertical lines indicate the location of the innermost and 
second nodes 
of the corresponding regular solutions.
}
\end{figure}
\end{fixy}
\begin{fixy}{-1}
%XXXXXXXXXXXXXXXXXXXXXX Figure 5 XXXXXXXXXXXXXXXXXXXXXXXXXXXXXXXXX

\newpage

\begin{figure}
\centering
\epsfysize=11cm
\mbox{\epsffile{
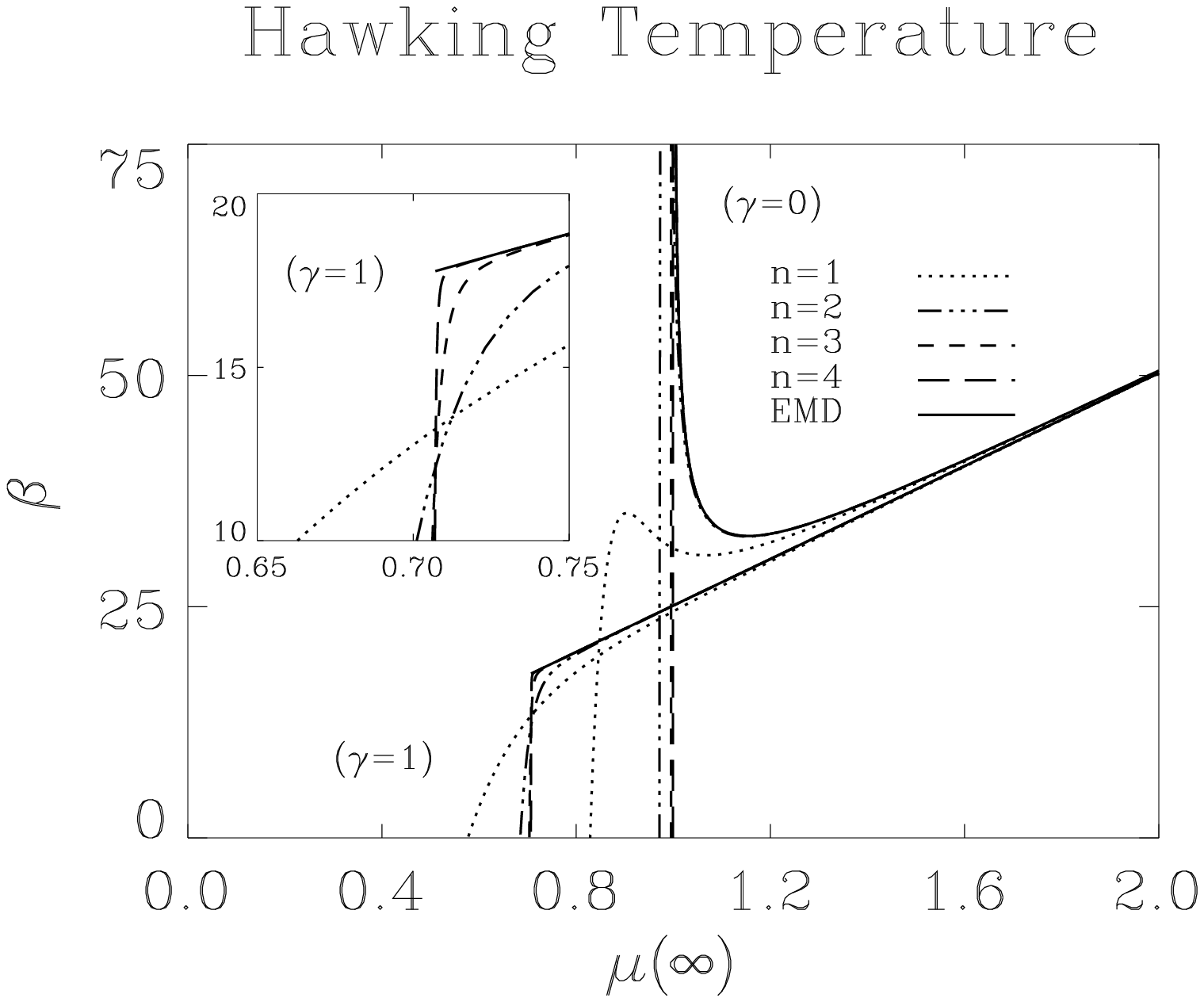
}}
\caption{\label{fig5}
The inverse Hawking temperature $\beta$ as a function of the
 mass $\mu (\infty)$ 
for the SU(2) EYMD solutions 
with node number 
$n=1$ (dotted), 
$n=2$ (tripledot-dashed),
$n=3$ (dashed)
and
$n=4$ (long dashed)
 for dilaton coupling constants $\gamma=1$ and $\gamma=0$.
The solid lines show the inverse Hawking temperature of
the EMD and RN solutions, respectively. 
}
\end{figure}
\end{fixy}
\begin{fixy}{-1}

%XXXXXXXXXXXXXXXXXXXXXX Figure 6 XXXXXXXXXXXXXXXXXXXXXXXXXXXXXXXX

\newpage

\begin{figure}
\centering
\epsfysize=11cm
\mbox{\epsffile{
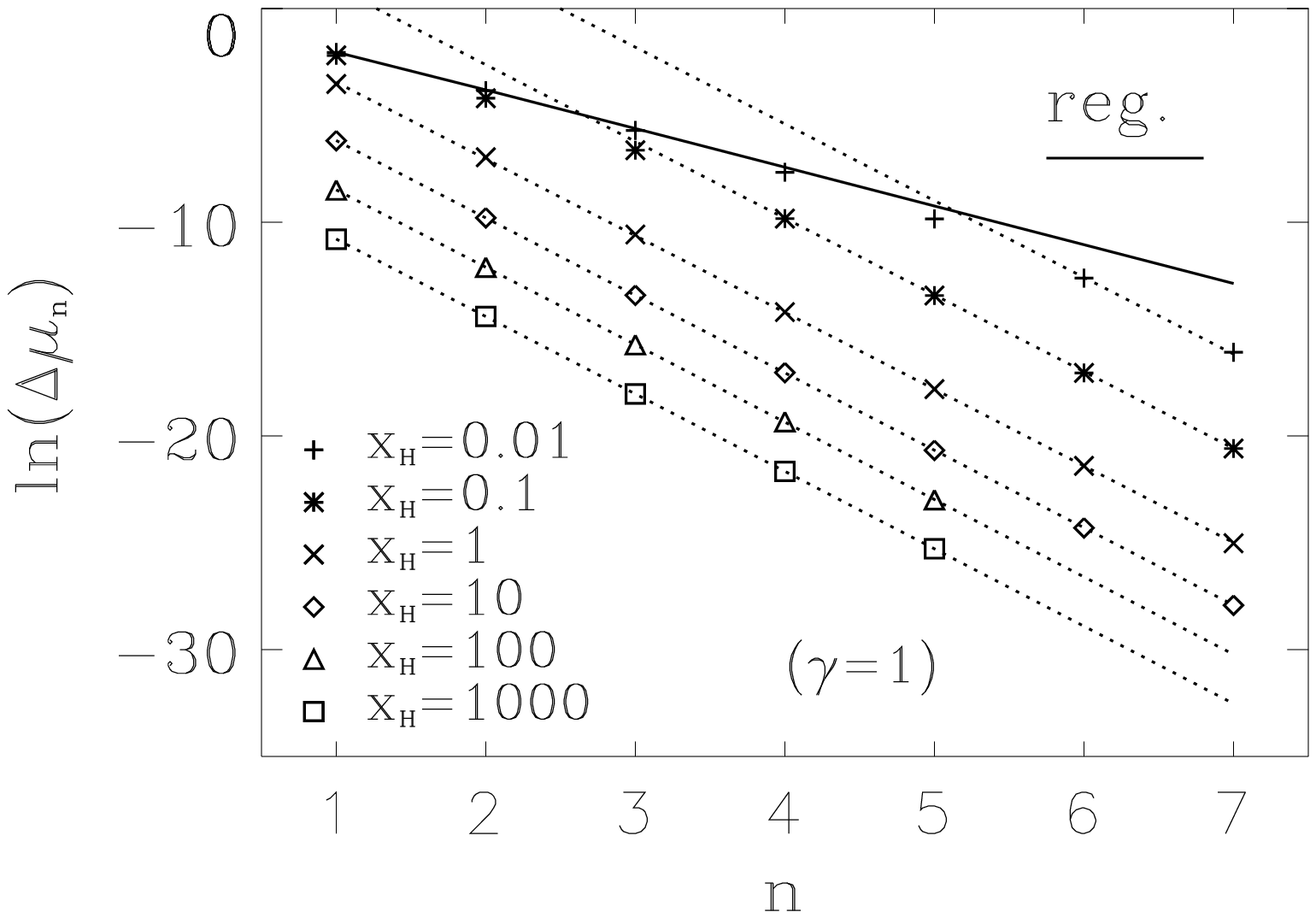
}}
\caption{\label{fig6}
The logarithm of the absolute deviation from the limiting solution 
$\Delta \mu_n = \mu_\infty (\infty) -\mu_n(\infty)$
for the SU(2) EYMD masses as a function of the node number $n$
with dilaton coupling constant $\gamma=1$ for the regular
solution (solid line) and the black hole solutions
($x_{\rm H} = $ $0.01$, $0.1$, $1$, $10$, $100$ and $1000$).
The dotted lines indicate the least square fits with fit constants 
given in Table~3.
}
\end{figure}
\end{fixy}
\begin{fixy}{0}

%XXXXXXXXXXXXXXXXXXXXXX Figure 7 (a-b) XXXXXXXXXXXXXXXXXXXXXXXXXX

\newpage

\begin{figure}
\centering
\epsfysize=11cm
\mbox{\epsffile{
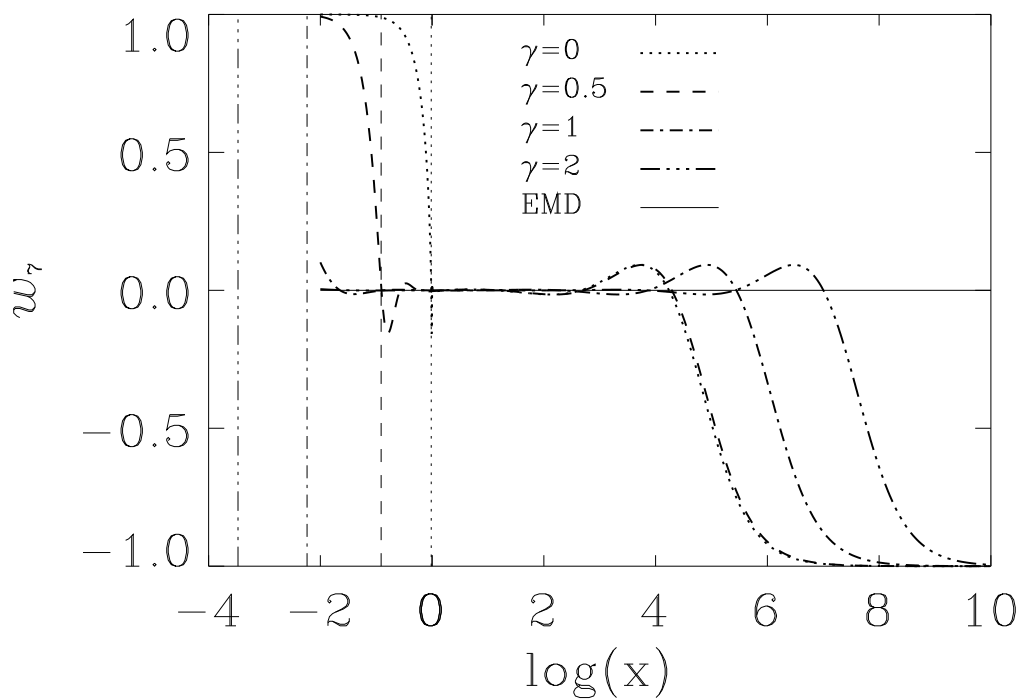
}}
\caption{\label{fig7a}
The SU(2) EYMD gauge field function  $w_7$  
of the black hole solutions with horizon $x_{\rm H} =0.01$ 
and dilaton coupling constants 
$\gamma = 0$   (dotted), 
$\gamma = 0.5$ (dashed), 
$\gamma = 1$   (dot-dashed) 
and
$\gamma = 2$   (tripledot-dashed). 
The solid line shows the gauge field function of the EMD solution.
The thin vertical lines
indicate the location of the innermost node of the corresponding 
regular solutions. 
}
\end{figure}

\newpage

\begin{figure}
\centering
\epsfysize=11cm
\mbox{\epsffile{
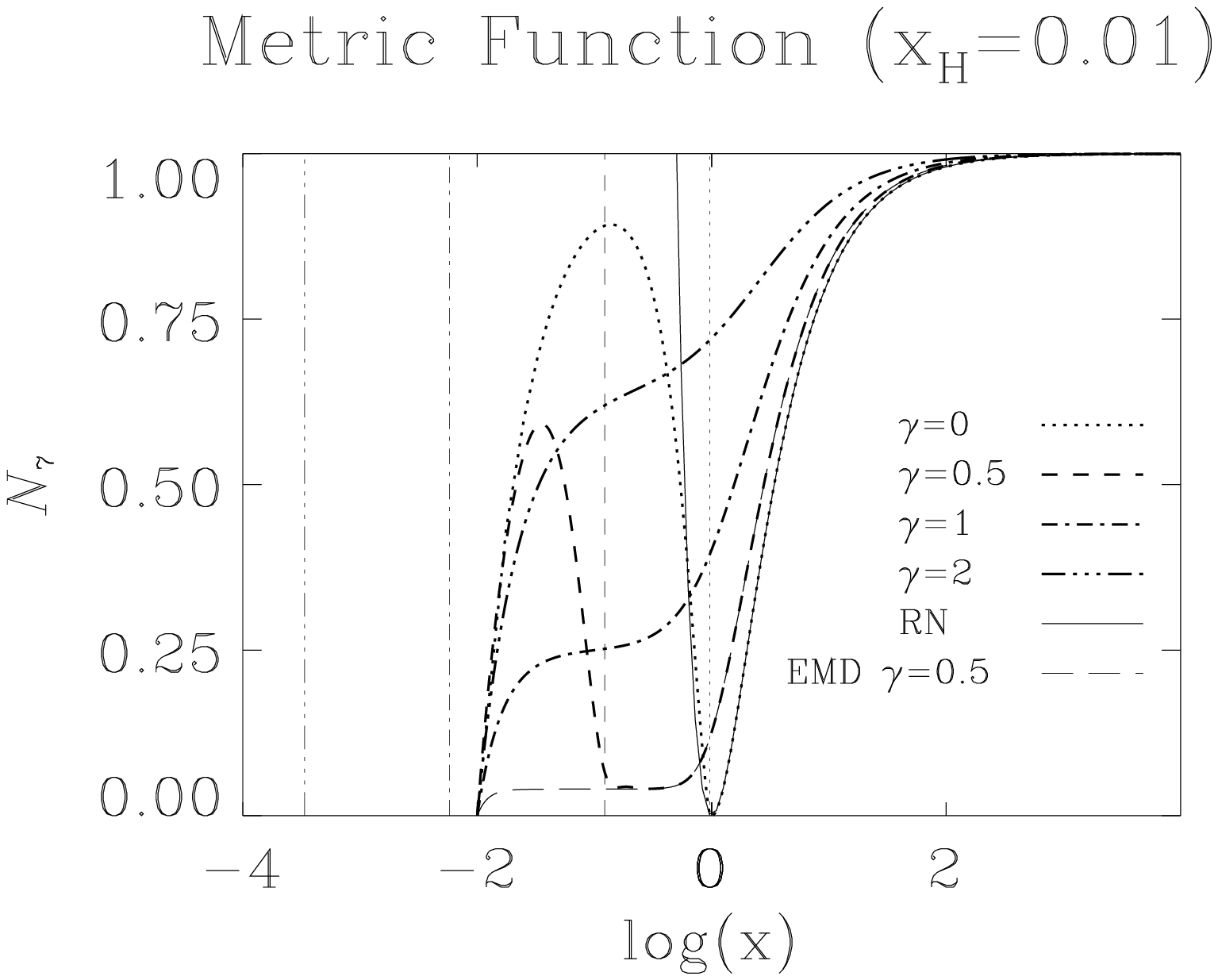
}}
\caption{\label{fig7b}
The SU(2) EYMD metric functions $N_7(x)$ 
of the black hole solutions  
with horizon $x_{\rm H} =0.01$  
and dilaton coupling constants 
$\gamma = 0$   (dotted), 
$\gamma = 0.5$ (dashed), 
$\gamma = 1$   (dot-dashed) 
and
$\gamma = 2$   (tripledot-dashed). 
The solid line shows the metric function of the Reissner-Nordstr\o m 
solution and the long dashed line the metric function of the
EMD solution with $\gamma=0.5$.
For $\gamma = 1$ and $2$ the functions $N_7(x)$ and their limiting 
EMD functions fall on top of each other and are indistinguishable.
The thin vertical lines
indicate the location of the innermost node of the corresponding 
regular solutions. 
}
\end{figure}
\end{fixy}
\begin{fixy}{-1}

%XXXXXXXXXXXXXXXXXXXXXX Figure 8 XXXXXXXXXXXXXXXXXXXXXXXXXXXXXXXX

\newpage

\begin{figure}
\centering
\epsfysize=11cm
\mbox{\epsffile{
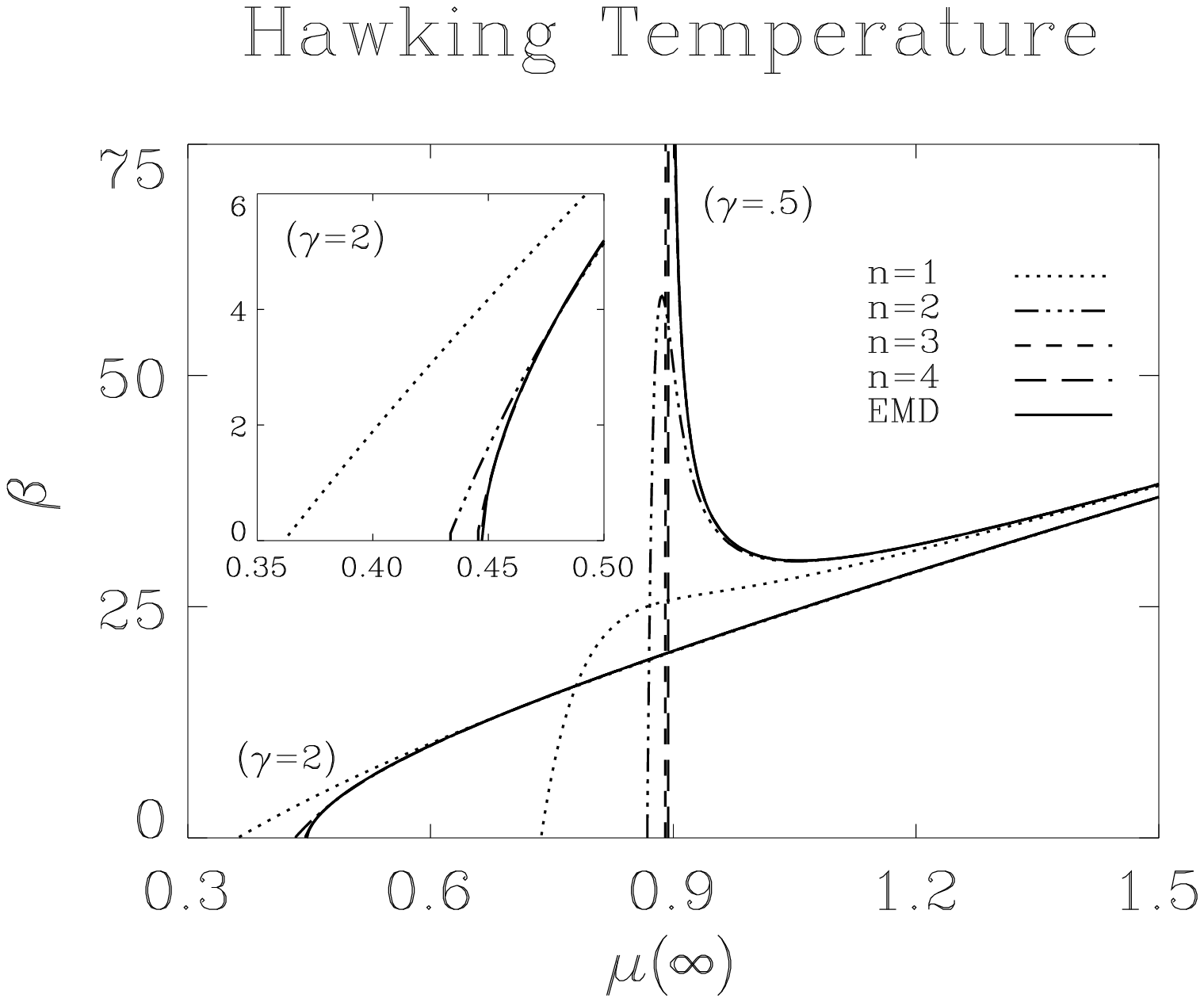
}}
\caption{\label{fig8}
The inverse Hawking temperature $\beta$ as a function of the
 mass $\mu (\infty)$ 
for the SU(2) EYMD solutions 
with node number 
$n=1$ (dotted), 
$n=2$ (tripledot-dashed),
$n=3$ (dashed)
and
$n=4$ (long dashed)
for the dilaton coupling constants $\gamma=0.5$ and $\gamma=2$.
The solid lines show the inverse Hawking temperature of
the corresponding limiting EMD solutions. 
}
\end{figure}
\end{fixy}
\clearpage
\begin{fixy}{-1}
%XXXXXXXXXXXXXXXXXXXXXX Figure 9 XXXXXXXXXXXXXXXXXXXXXXXXXXXXXXXXXX

\newpage

\begin{figure}
\centering
\epsfysize=11cm
\mbox{\epsffile{
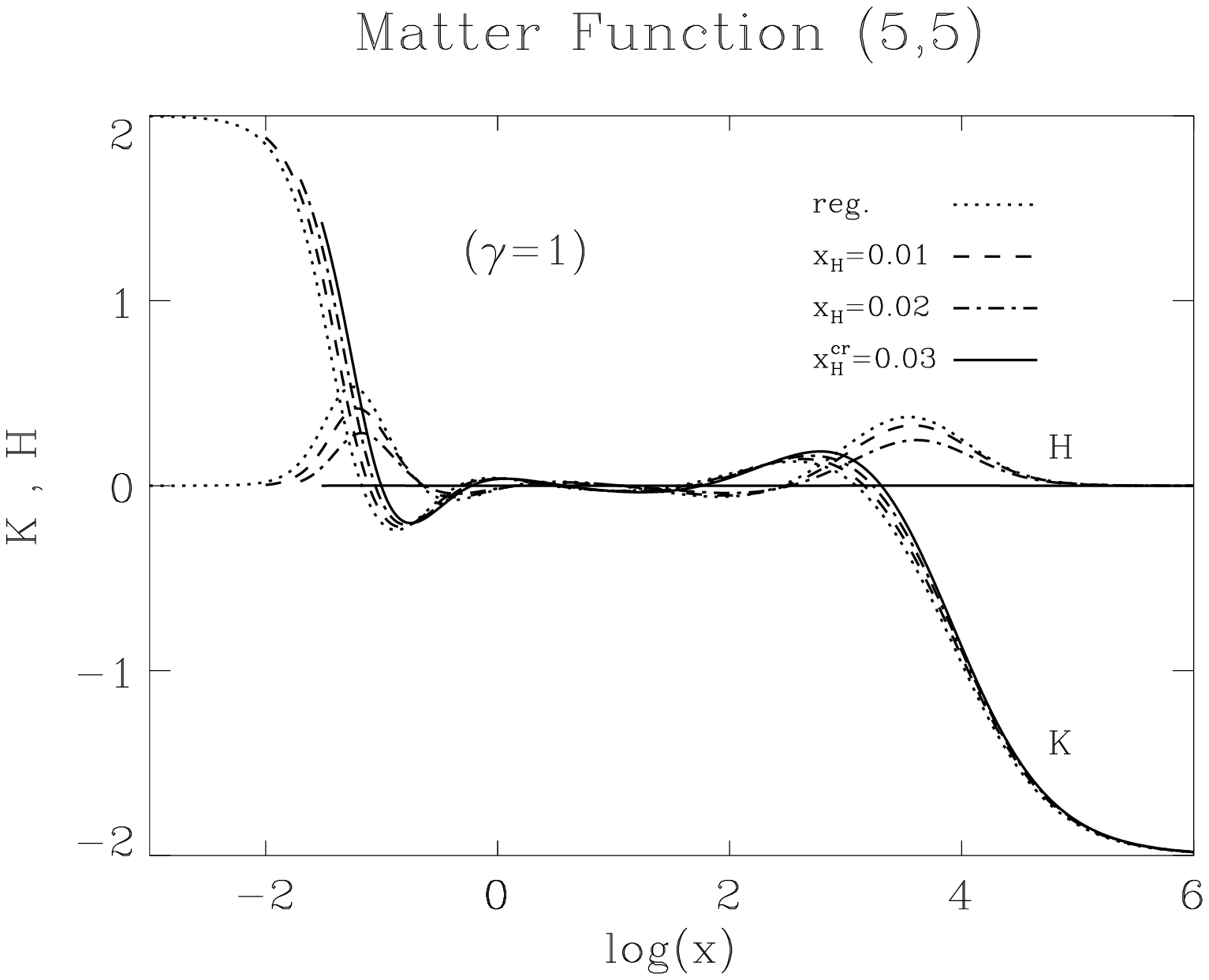
}}
\caption{\label{fig9}
The SO(3) EYMD gauge field functions $K_5(x)$ and $H_5(x)$  
for the
regular solution (dotted)
and black hole solutions with horizon 
$x_{\rm H} = 0.01$ (dashed), 
$x_{\rm H} = 0.02$  (dot-dashed) and
$x_{\rm H}^{\rm cr} = 0.03$ (solid)
with node structure $(5,5)$ and  
dilaton coupling constant $\gamma=1$. 
}
\end{figure}
\end{fixy}
%XXXXXXXXXXXXXXXXXXXXXX Figure 10 XXXXXXXXXXXXXXXXXXXXXXXXXXXXXXXXXX

\newpage
\begin{fixy}{-1}
\begin{figure}
\centering
\epsfysize=11cm
\mbox{\epsffile{
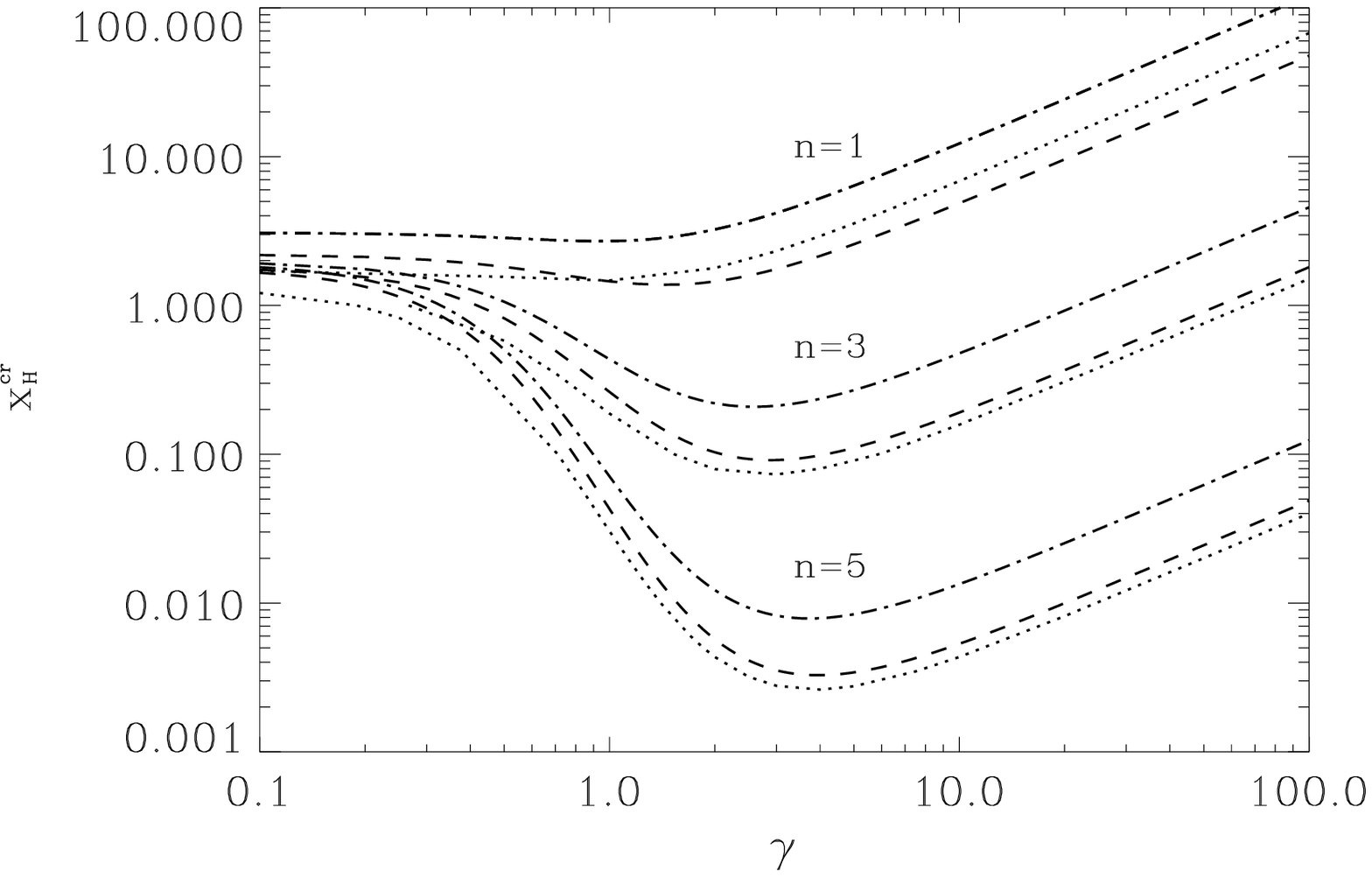
}}
\caption{\label{fig10}
The critical horizon $x_{\rm H}^{\rm cr}(\gamma)$ as a function of 
the dilaton coupling constant $\gamma$
of the SO(3) EYMD solutions with node structure $(n,n)$ 
(dotted) together with the location of the innermost nodes 
of the corresponding regular solutions (dashed).
Also shown are the locations of the innermost nodes 
of the scaled SU(2) solutions (dot-dashed) 
with the same node structure.
}
\end{figure}
\end{fixy}
%XXXXXXXXXXXXXXXXXXXXXX Figure 11a XXXXXXXXXXXXXXXXXXXXXXXXXXXXXXXXXX

\newpage
\begin{fixy}{0}
\begin{figure}
\centering
\epsfysize=11cm
\mbox{\epsffile{
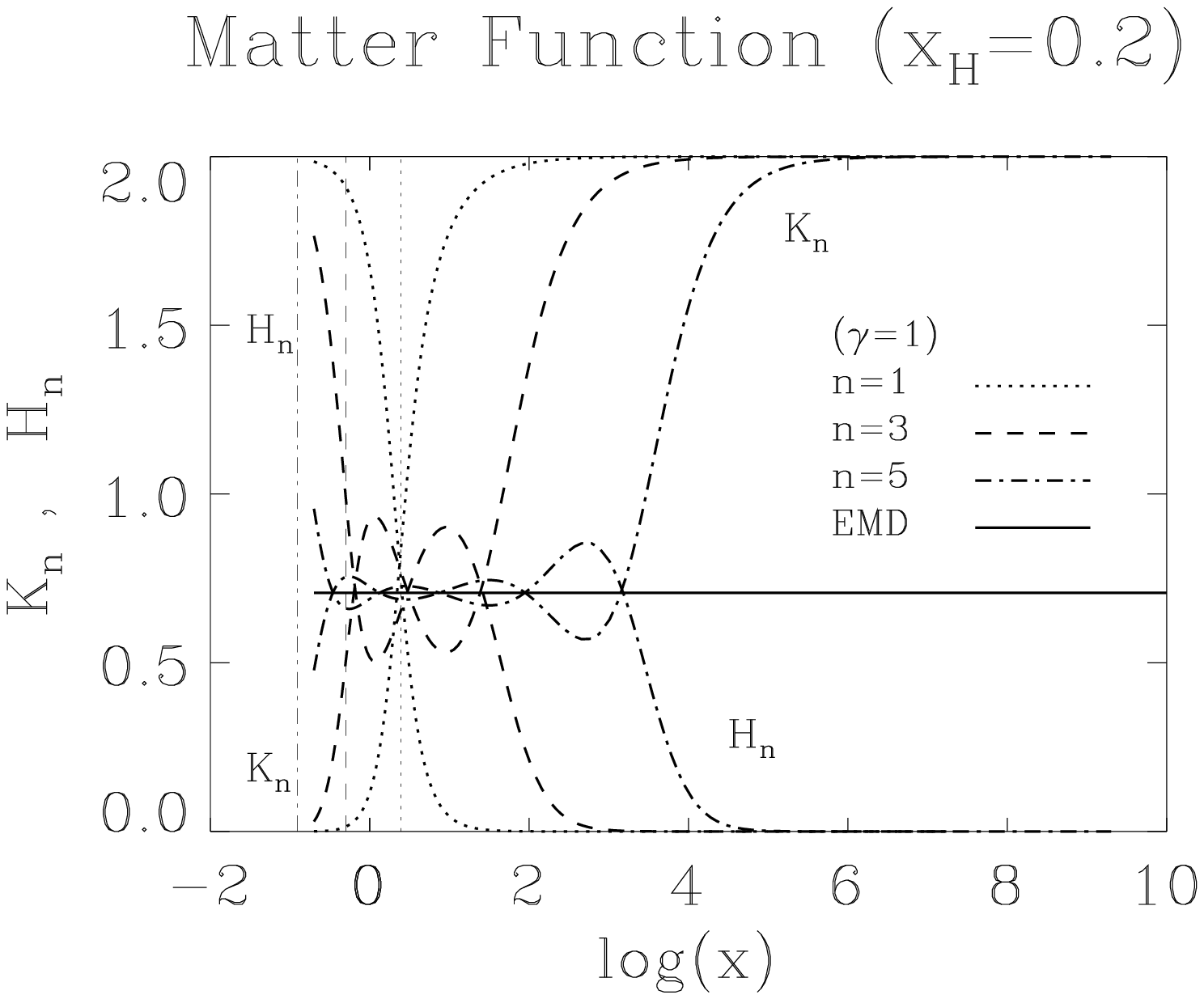
}}
\caption{\label{fig11a}
The SO(3) EYMD gauge field functions $K_n(x)$ and $H_n(x)$  
of the solutions with node structure $(0,n)$ for   
dilaton coupling constant $\gamma=1$ 
and horizon $x_{\rm H} = 0.2$, for
$n=1$ (dotted), 
$n=3$ (dashed) and 
$n=5$ (dot-dashed).
The solid line shows the gauge field function of the
limiting EMD solution
with  magnetic charge 
$P^2 = 3$
 and the same dilaton coupling constant and horizon.
The thin vertical lines
indicate the location of the innermost node of the corresponding 
regular solutions. 
}
\end{figure}

%XXXXXXXXXXXXXXXXXXXXXX Figure 11b XXXXXXXXXXXXXXXXXXXXXXXXXXXXXX

\newpage

\begin{figure}
\centering
\epsfysize=11cm
\mbox{\epsffile{
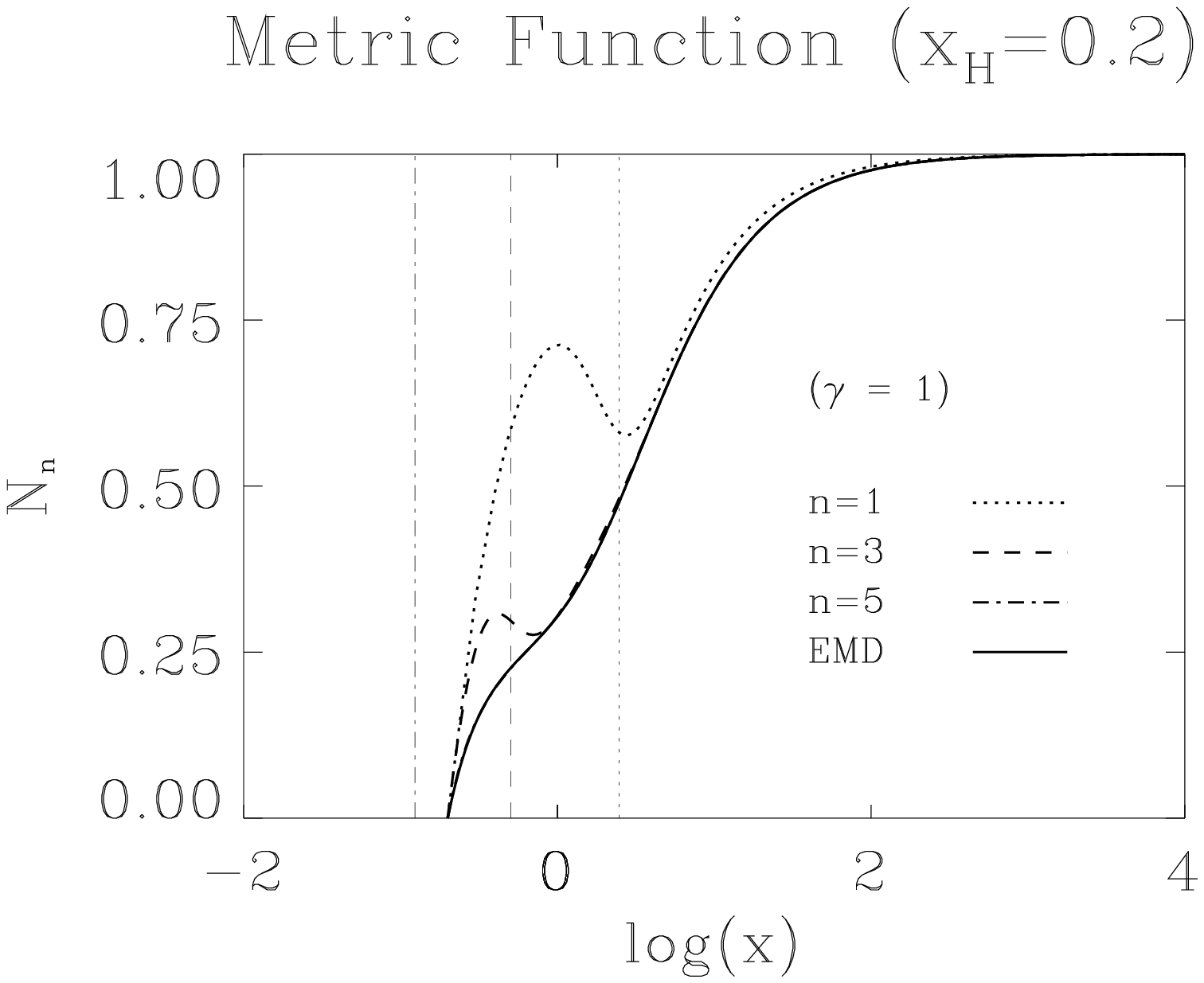
}}
\caption{\label{fig11b}
The same as Figure \ref{fig11a} for the metric functions $N_n(x)$.
}
\end{figure}

%XXXXXXXXXXXXXXXXXXXXXX Figure 11c XXXXXXXXXXXXXXXXXXXXXXXXXXXXXXX

\newpage

\begin{figure}
\centering
\epsfysize=11cm
\mbox{\epsffile{
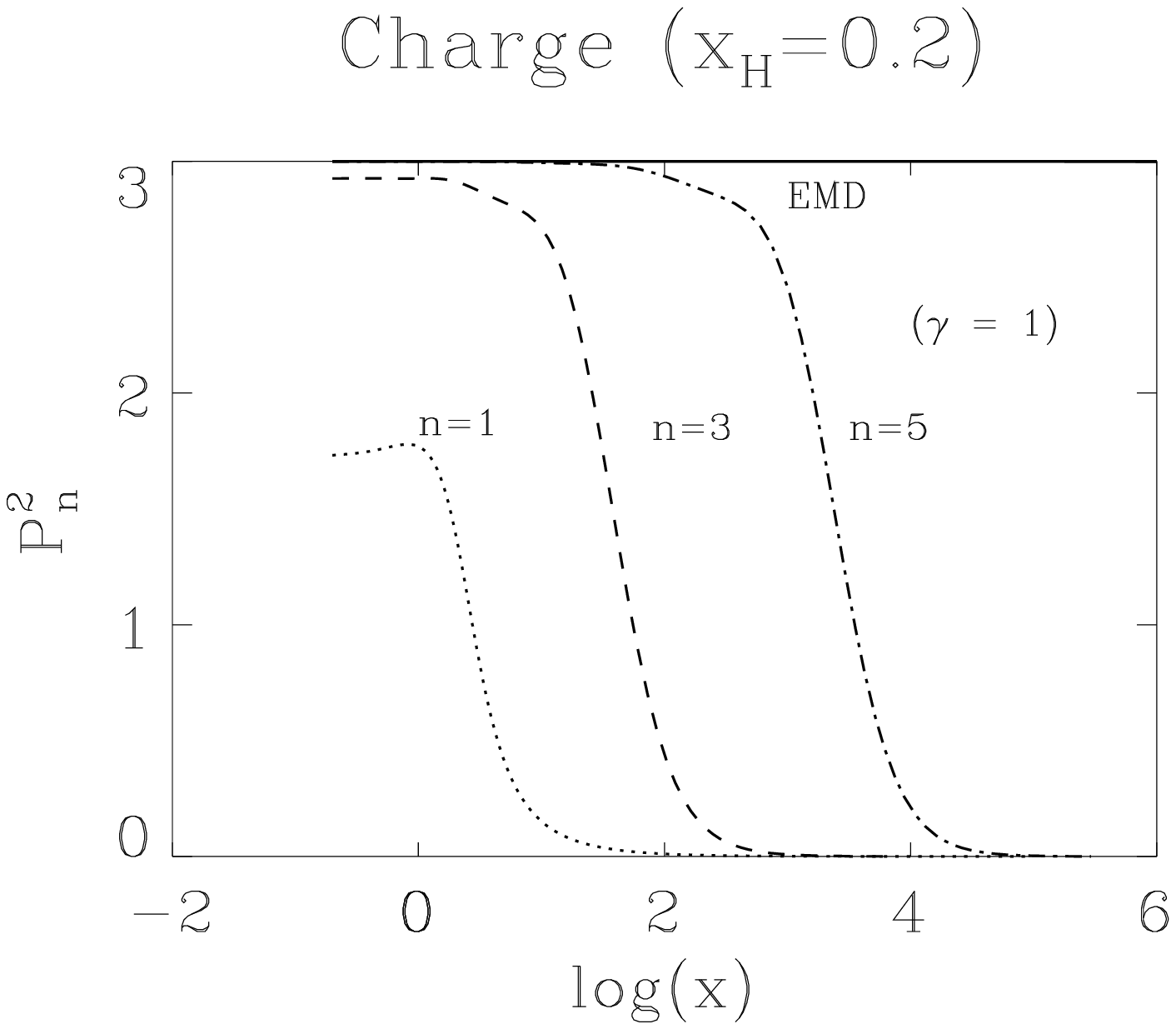
}}
\caption{\label{fig11c}
The same as Figure \ref{fig11a} for the
squared charge functions $P^2_n(x)$.
}
\end{figure}
\end{fixy}
%XXXXXXXXXXXXXXXXXXXXXX Figure 12 XXXXXXXXXXXXXXXXXXXXXXXXXXXXXXXXXX

\newpage
\begin{fixy}{-1}
\begin{figure}
\centering
\epsfysize=11cm
\mbox{\epsffile{
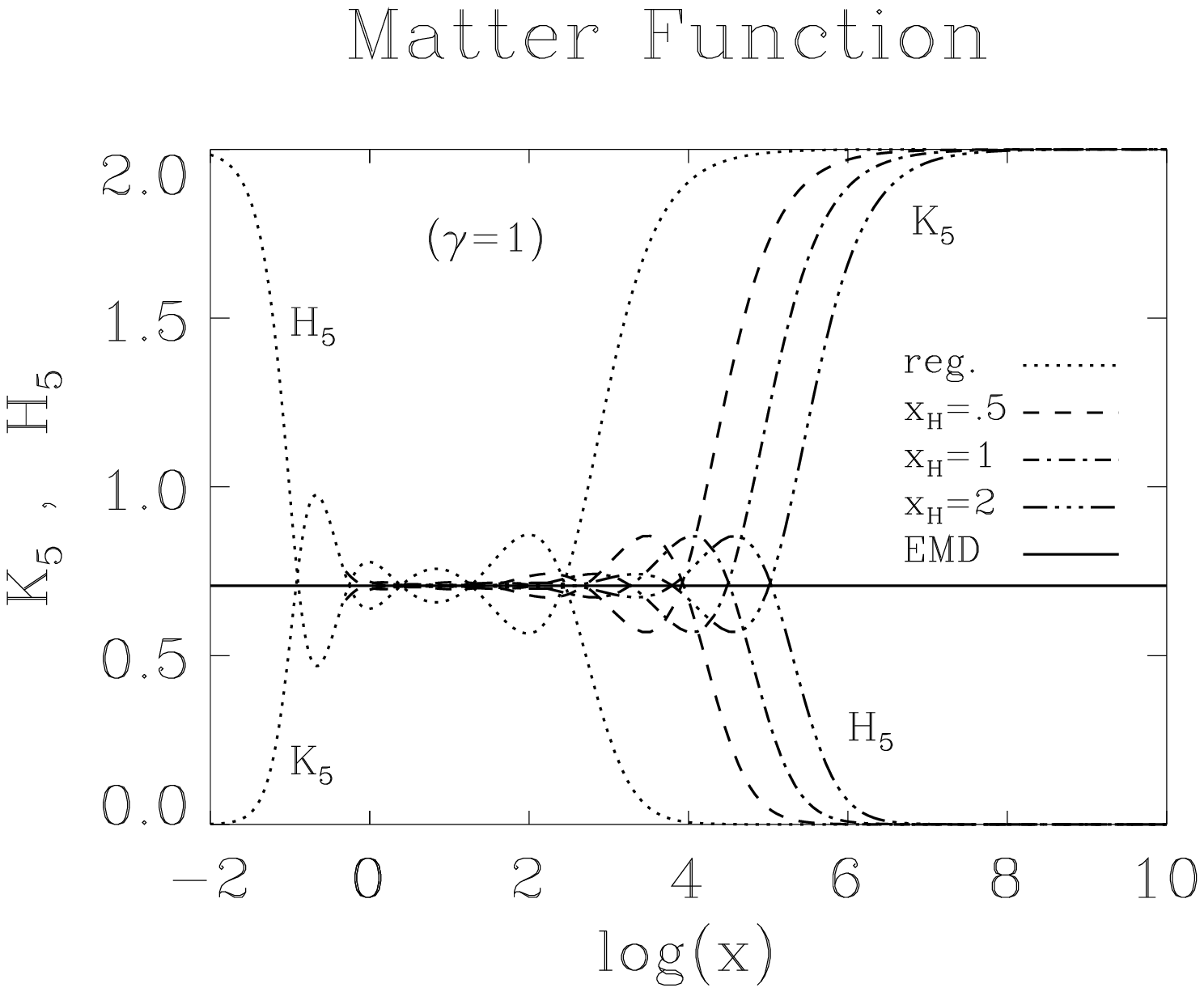
}}
\caption{\label{fig12}
The SO(3) EYMD gauge field functions $K_5(x)$ and $H_5(x)$  
with node structure $(0,5)$ and   
dilaton coupling constant $\gamma=1$ for the regular solution 
(dotted) and black hole solutions
$x_{\rm H} = .5$ (dashed), 
$x_{\rm H} = 1$ (dot-dashed) and 
$x_{\rm H} = 2$ (tripledot-dashed).
The solid line shows the gauge field function of the
``extremal"  EMD solution
with magnetic charge 
$P^2 = 3$
 and the same dilaton coupling constant. 
}
\end{figure}
\end{fixy}

%XXXXXXXXXXXXXXXXXXXXXX Figure 13 XXXXXXXXXXXXXXXXXXXXXXXXXXXXXXXX

\newpage
\begin{fixy}{-1}
\begin{figure}
\centering
\epsfysize=11cm
\mbox{\epsffile{
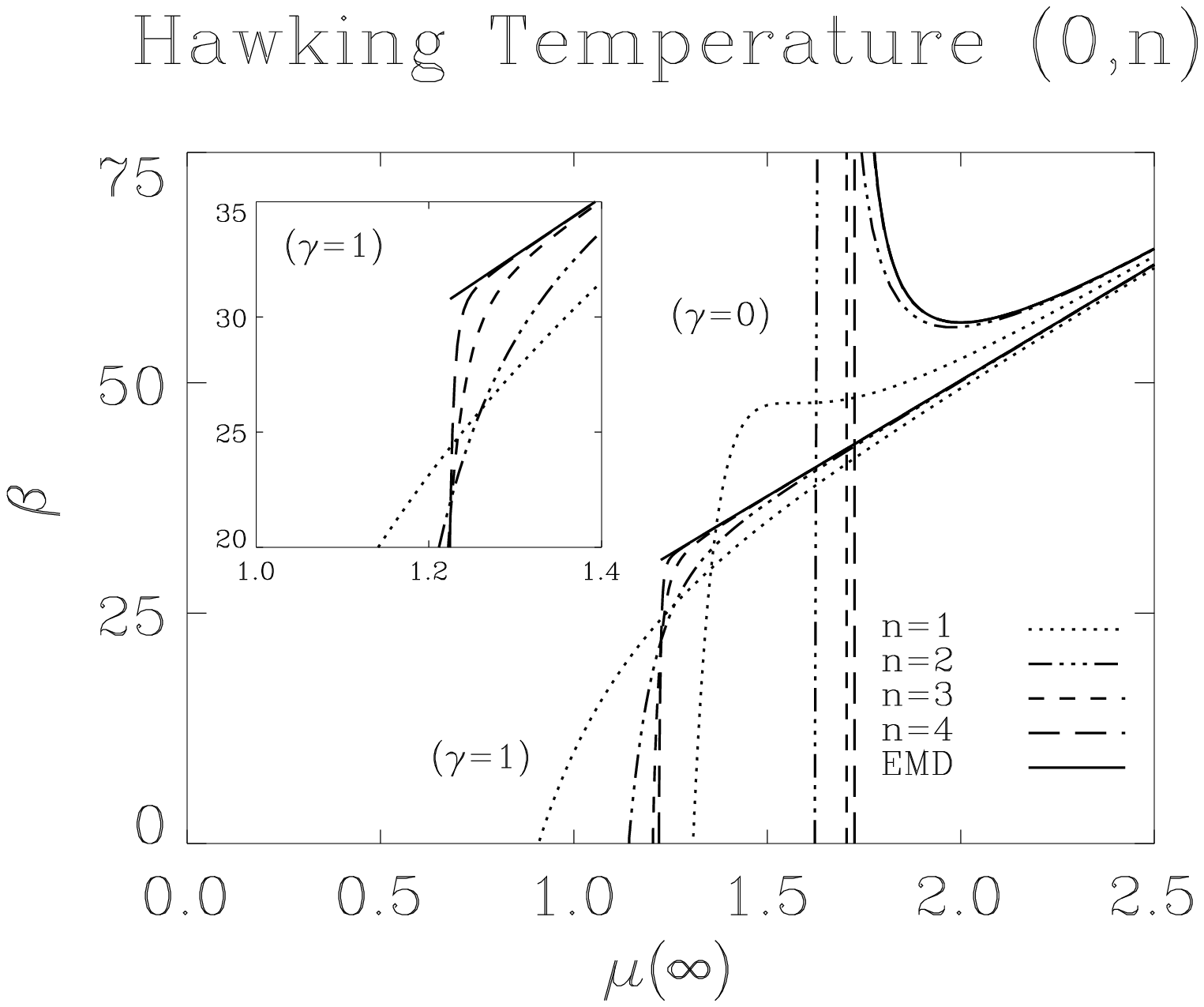
}}
\caption{\label{fig13}
The inverse Hawking temperature $\beta$ as a function of the
 mass $\mu (\infty)$ 
for the SO(3) EYMD solutions 
with node structure $(0,n)$
 for the dilaton coupling constants $\gamma=0$ and $\gamma=1$
and node number 
$n=1$ (dotted), 
$n=2$ (tripledot-dashed),
$n=3$ (dashed)
and
$n=4$ (long dashed).
Also shown is the inverse Hawking temperature of
the limiting EMD and RN solutions (solid). 
}
\end{figure}
\end{fixy}

%XXXXXXXXXXXXXXXXXXXXXX Figure 14a XXXXXXXXXXXXXXXXXXXXXXXXXXXXXX

\newpage
\begin{fixy}{0}
\begin{figure}
\centering
\epsfysize=11cm
\mbox{\epsffile{
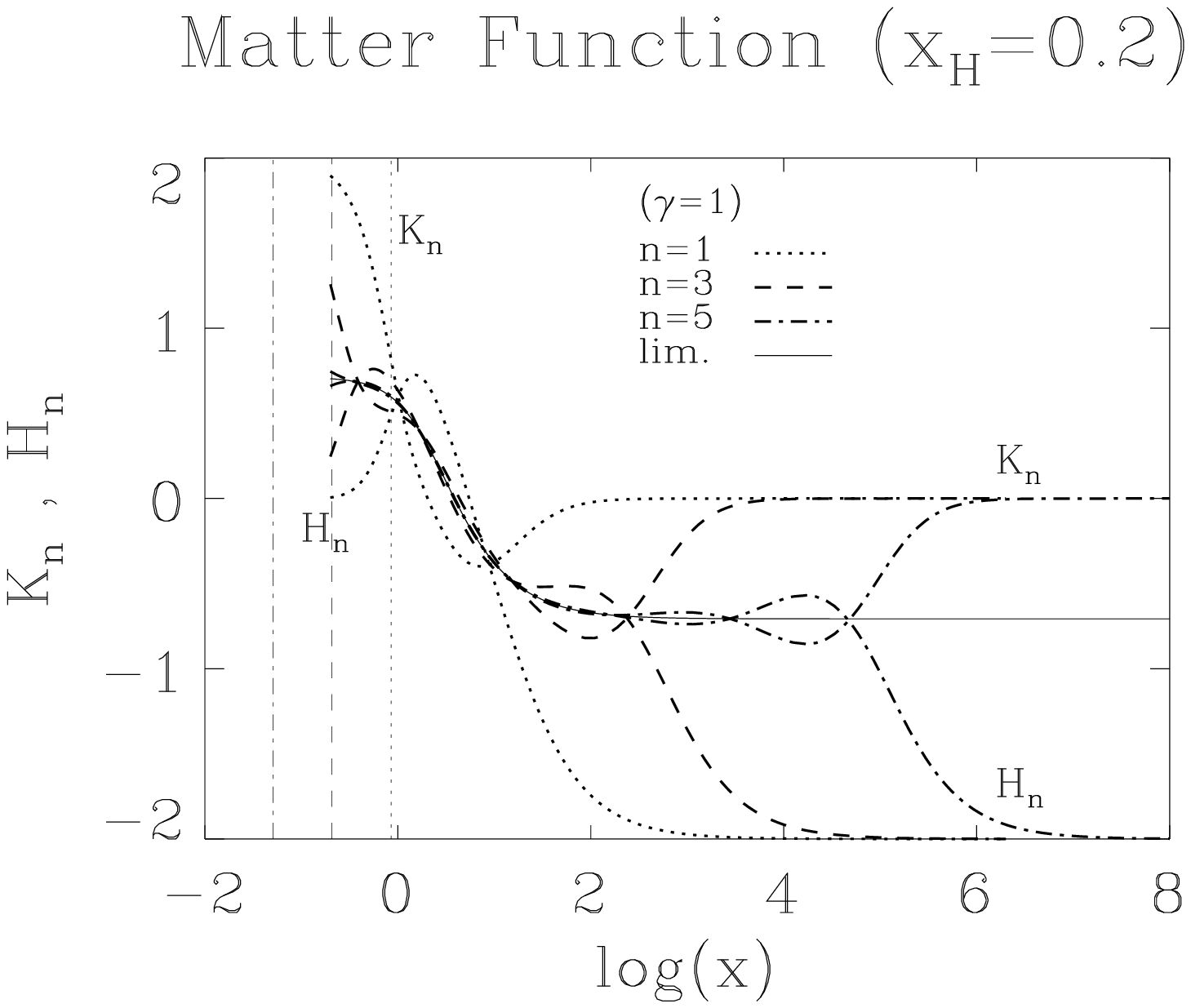
}}
\caption{\label{fig14a}
The SO(3) EYMD gauge field functions $K_n(x)$ and $H_n(x)$  
of the solutions with node structure $(1,1+n)$ for the   
dilaton coupling constant $\gamma=1$ 
and horizon $x_{\rm H} = 0.2$, for
$n=1$ (dotted), 
$n=3$ (dashed) and 
$n=5$ (dot-dashed).
The solid line shows the gauge field function of the limiting
charged  SU(3) EYMD solution with one node, $j=1$,
and magnetic charge $P^2 = 3$.
The thin vertical lines
indicate the location of the innermost node of the corresponding 
regular solutions. 
}
\end{figure}

%XXXXXXXXXXXXXXXXXXXXXX Figure 14b XXXXXXXXXXXXXXXXXXXXXXXXXXXXXX

\newpage

\begin{figure}
\centering
\epsfysize=11cm
\mbox{\epsffile{
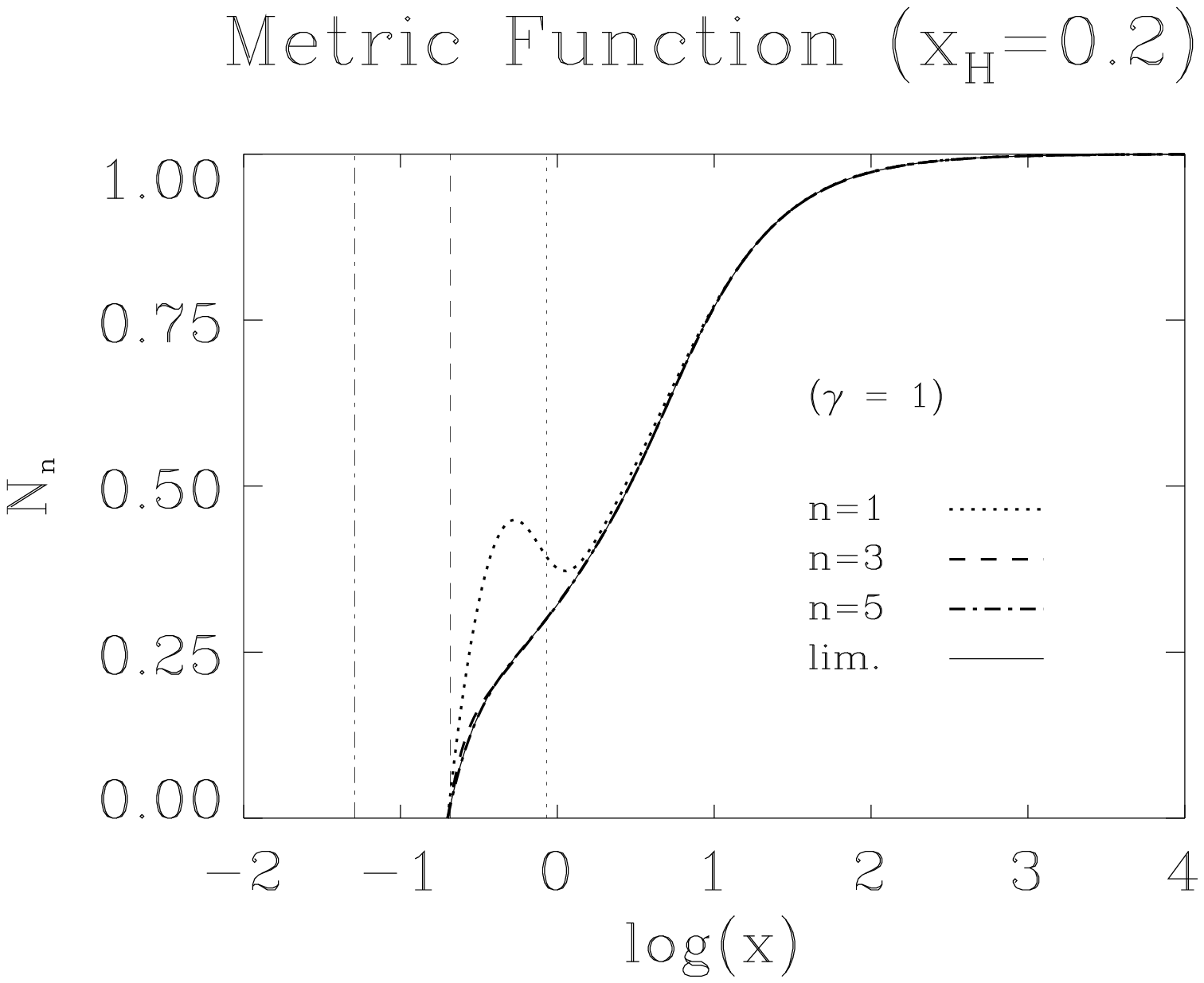
}}
\caption{\label{fig14b}
The same as Figure \ref{fig14a} for the metric functions $N_n(x)$.
}
\end{figure}

%XXXXXXXXXXXXXXXXXXXXXX Figure 14c XXXXXXXXXXXXXXXXXXXXXXXXXXXXXXX

\newpage

\begin{figure}
\centering
\epsfysize=11cm
\mbox{\epsffile{
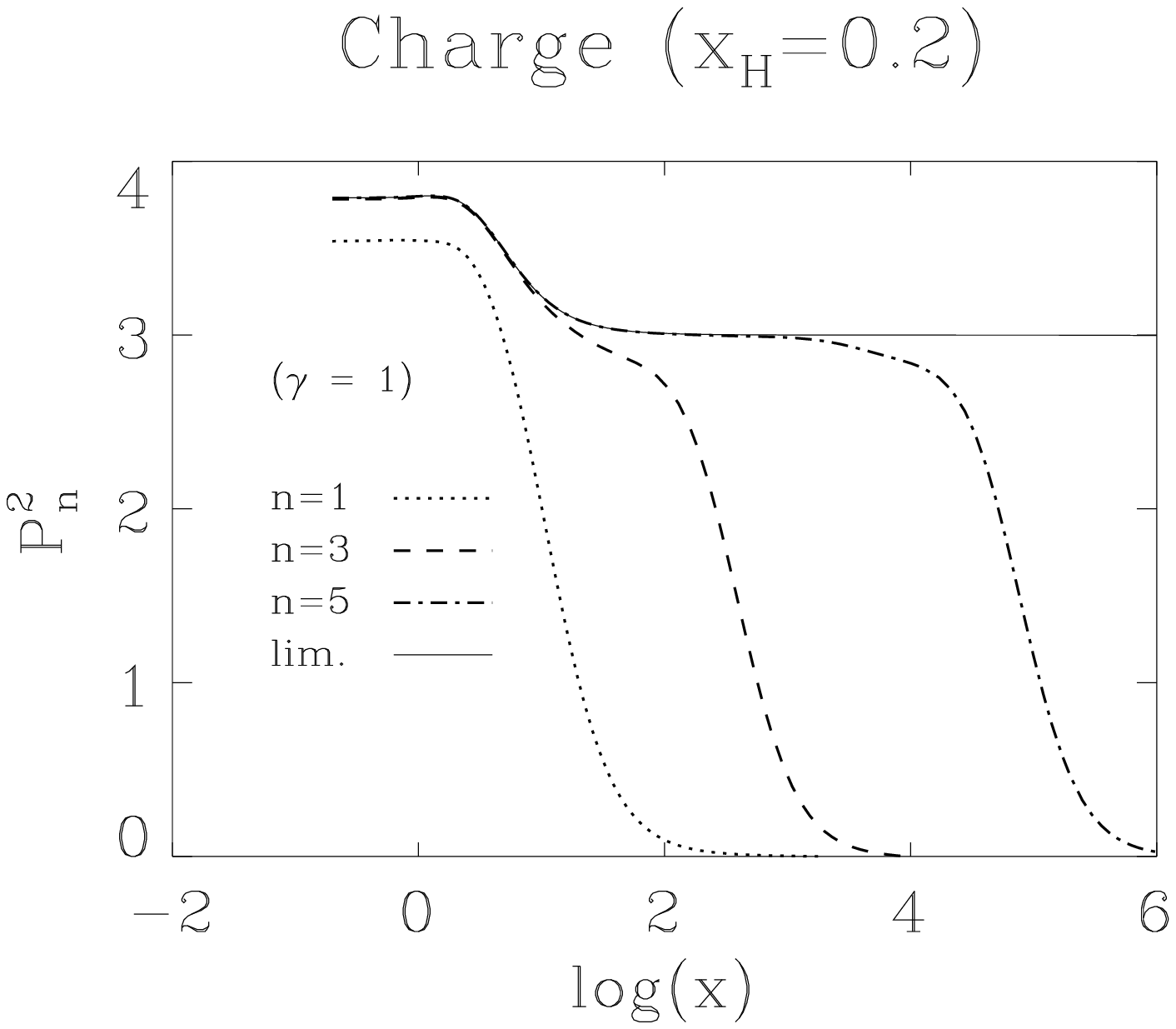
}}
\caption{\label{fig14c}
The same as Figure \ref{fig14a} for the
squared charge functions $P^2_n(x)$.
}
\end{figure}
\end{fixy}

%XXXXXXXXXXXXXXXXXXXXXX Figure 15 XXXXXXXXXXXXXXXXXXXXXXXXXXXXXXXXX

\newpage
\begin{fixy}{-1}
\begin{figure}
\centering
\epsfysize=11cm
\mbox{\epsffile{
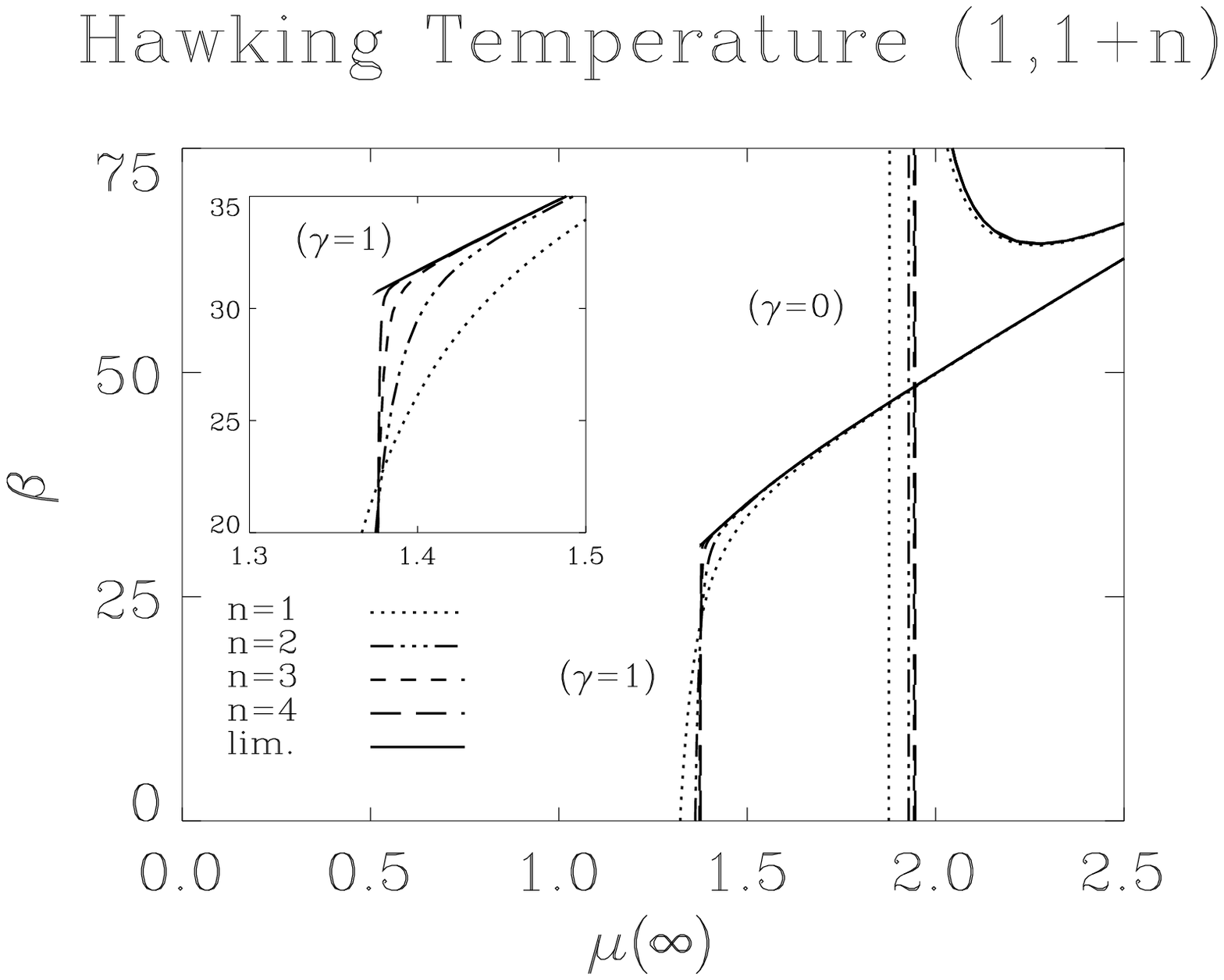
}}
\caption{\label{fig15}
The inverse Hawking temperature $\beta$ as a function of the
 mass $\mu (\infty)$ 
for the SO(3) EYMD solutions 
with node structure $(1,1+n)$
 for the dilaton coupling constants $\gamma=0$ and $\gamma=1$
and node number 
$n=1$ (dotted), 
$n=2$ (tripledot-dashed),
$n=3$ (dashed)
and
$n=4$ (long dashed).
Also shown is the inverse Hawking temperature of
the limiting
charged SU(3) EYMD and EYM solutions with one node, $j = 1$ and
magnetic charge $P^2 = 3$.
}
\end{figure}
\end{fixy}

%XXXXXXXXXXXXXXXXXXXXXX Figure 16a XXXXXXXXXXXXXXXXXXXXXXXXXXXXXXX

\newpage
\begin{fixy}{0}
\begin{figure}
\centering
\epsfysize=11cm
\mbox{\epsffile{
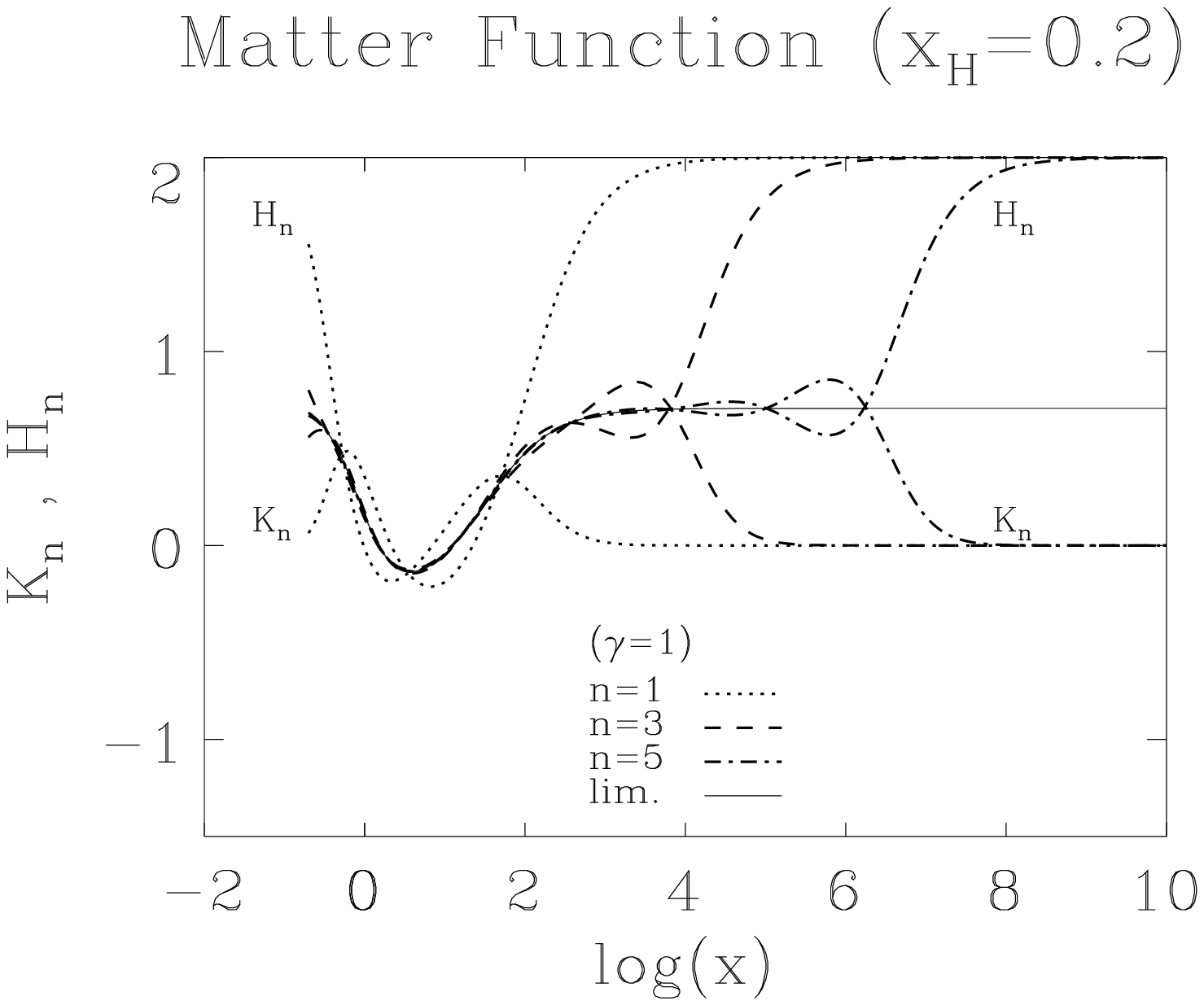
}}
\caption{\label{fig16a}
The SO(3) EYMD gauge field functions $K_n(x)$ and $H_n(x)$  
of the solutions with node structure $(2,2+n)$ for the   
dilaton coupling constant $\gamma=1$ 
and horizon $x_{\rm H} = 0.2$, for
$n=1$ (dotted), 
$n=3$ (dashed) and 
$n=5$ (dot-dashed).
The solid line shows the gauge field function of the
limiting charged  SU(3) EYMD 
solution with two nodes, $j=2$, 
and magnetic charge $P^2 = 3$.
}
\end{figure}

%XXXXXXXXXXXXXXXXXXXXXX Figure 16b XXXXXXXXXXXXXXXXXXXXXXXXXXXXXXX
\newpage

\begin{figure}
\centering
\epsfysize=11cm
\mbox{\epsffile{
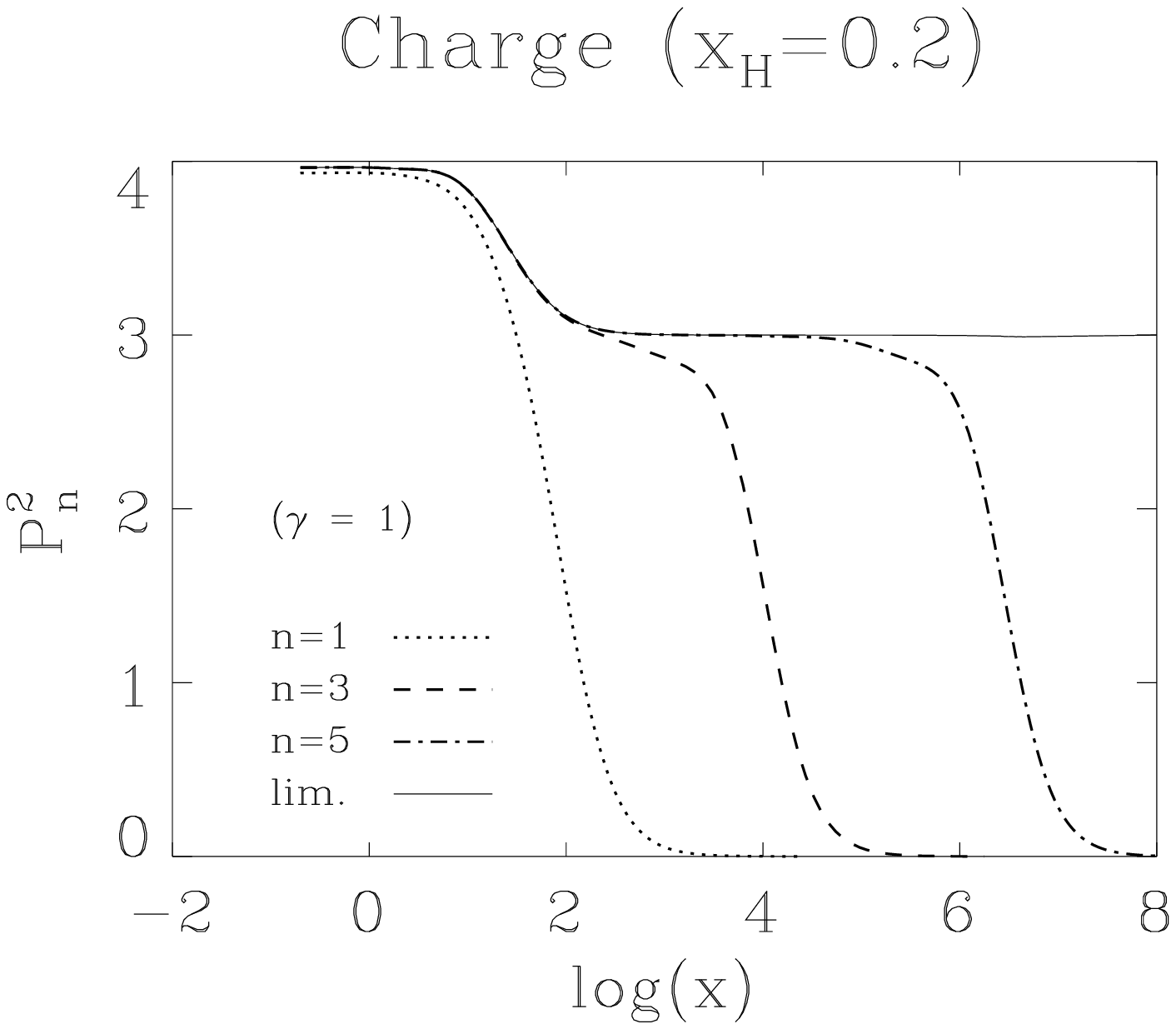
}}
\caption{\label{fig16b}
The same as Figure \ref{fig16a} for the
squared charge functions $P^2_n(x)$.
}
\end{figure}
\end{fixy}

%XXXXXXXXXXXXXXXXXXXXXX Figure 17 XXXXXXXXXXXXXXXXXXXXXXXXXXXXXXXXX

\newpage
\begin{fixy}{-1}
\begin{figure}
\centering
\epsfysize=11cm
\mbox{\epsffile{
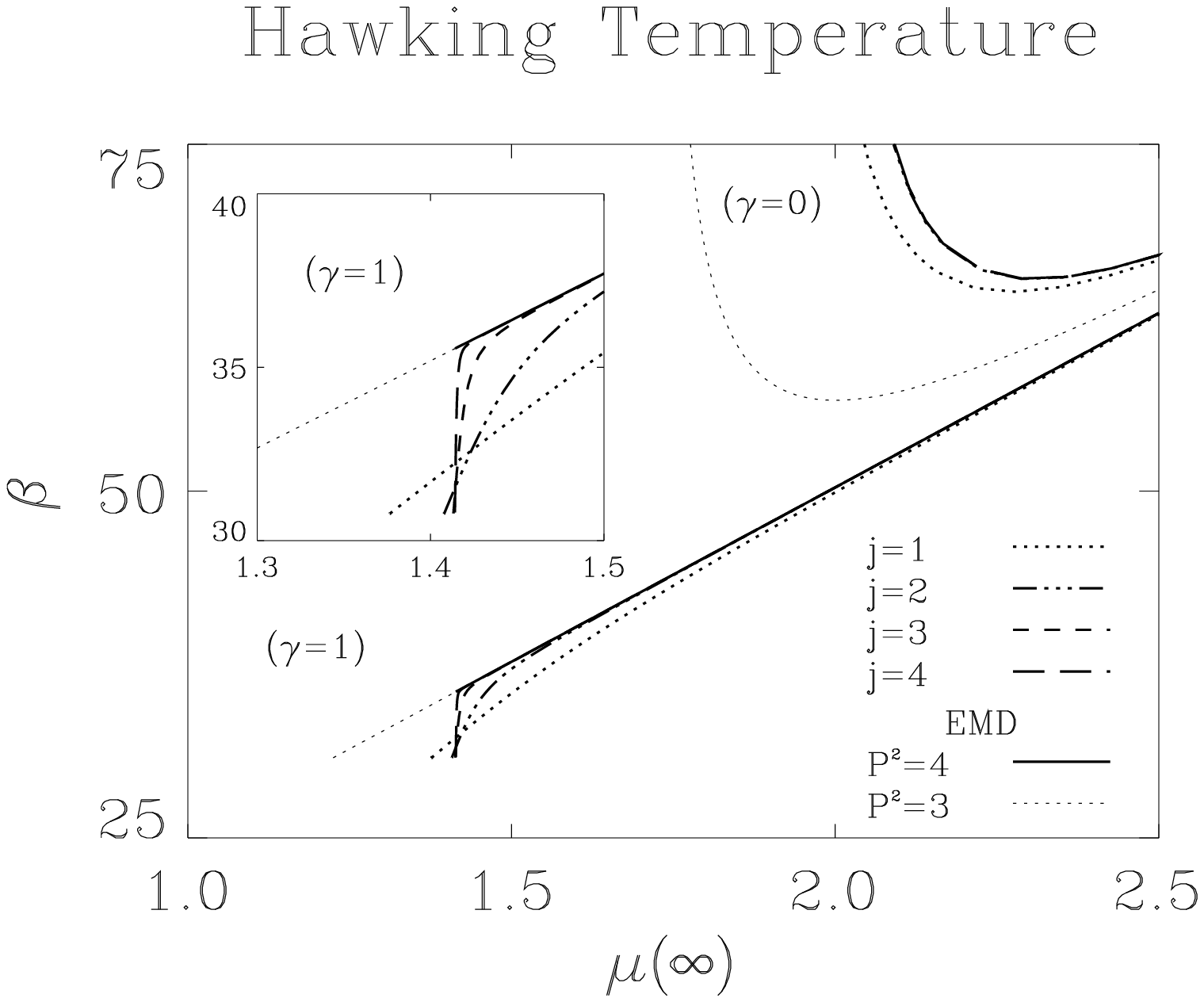
}}
\caption{\label{fig17}
The inverse Hawking temperature $\beta$ as a function of the
 mass $\mu (\infty)$ 
for the  charged SO(3) EYM and EYMD solutions with magnetic charge 
$P^2 = 3$ for dilaton coupling constants $\gamma=0$ and $\gamma=1$
and node number 
$j=1$ (dotted), 
$j=2$ (tripledot-dashed),
$j=3$ (dashed)
and
$j=4$ (long dashed).
Also shown is the inverse Hawking temperature of
the RN and EMD solutions with magnetic charge $P^2 = 4$ (solid) and 
$P^2 = 3$ (thin dotted).
}
\end{figure}
\end{fixy}

\begin{thebibliography}{000}
\bibitem{fn0}
 The equations of motion are invariant under a shift 
 $\phi \rightarrow \phi + \phi_0$ together with a
 rescaling $x \rightarrow x e^{\gamma \phi_0}$.
 Therefore solutions regular at infinity
 can always be chosen to satisfy $\phi(\infty)=0$ \cite{don,lav2}.
\bibitem{fn1}
 An equivalent expression for SU(2) black holes for $\gamma=1$ 
 was obtained in \cite{neill}.
\bibitem{fn2}
 An exception are the ``extremal'' magnetically charged
 EMD and EYMD solutions (see Appendices A and B).
\bibitem{fn7}
 Relation (\ref{p2eq}) for $P^2$ can be generalized for 
 $\gamma \ne 1$.
In order to define the charge function for $\gamma \neq 0$, we
start from 
$-x^2 g_{tt}^{{\rm (EMD) }} =(X -X_-)(X-X_+)$. We introduce the 
variable $\xi(X)=X-\gamma^2 \frac{P^2}{X_+}=X-\gamma D$, and employ
$X_+ + X_- = 2 ( \mu(\infty)+ \gamma D)$ and
$X_- = \frac{1+\gamma^2}{\gamma} D$. 
We then find
$-x^2 g_{tt}^{{\rm (EMD) }} = P^2 +\xi^2 -D^2 - 2 \xi \mu(\infty)$, 
and we define the charge function $P(x)$
for the EYMD system by 
$$
P^2(x) = -x^2 g_{tt} -\xi^2 (x) +2 \xi(x)  \mu(\infty) +D^2 \ ,
$$
with the function $\xi(x)$ implicitly given.
With $y^2=\xi^2 - D^2$ the equation for the charge function can 
be put
into the form 
$
-x^2 g_{tt} = P^2(y) + y^2 - 2  \mu(\infty) \sqrt{y^2 + D^2} \ ,
$
which coincides with (\ref{p2eq}) for $\gamma = 1$ 
(since $y^2 = x^2$).
\bibitem{fn5}
 The coupling of ref.~\cite{maeda}
 is related to ours via $\kappa = \sqrt{2}\gamma$.
\bibitem{fn4}
 The masses $\mu(\infty)$ of the regular solutions
 are well approximated by the formula
 $\mu(\infty,\gamma,n)=[\sqrt{3}-(\pi/2) \exp(-4n/3)]
             /(\sqrt{1+\gamma^2})$ \cite{kks3}.
\bibitem{fitsmallxh}
For $\gamma = 1$ and $x_{\rm H} \leq 1$ many data points of the 
sequences $(0,n)$, $(1,1+n)$ and $(2,2+n)$
for $n \leq 5$ 
do not yet fall on straight parallel lines.
This makes it difficult 
to extract the asymptotic behaviour for large $n$.
Employing
 $$\Delta \mu_{n,j}(\infty) = \mu_{\infty,j}(\infty) - \mu_{n,j}
 (\infty)
 = \frac{b_\mu}{x_{\rm H}^2} e^{-(\alpha_\mu n +\bar{\alpha}_\mu j)} 
 \ , $$ 
 (analogously for D) and 
  $$\Delta T_{n,j} = 
  (T/T_s)_{n,j} - (T/T_s)_{\infty,j}
 = \frac{b_T}{x_{\rm H}^3} 
 e^{-(\alpha_T n +\bar{\alpha}_T j)} \ ,  $$ 
we find for the coefficients 
$b_\mu \approx 6.7$, $b_D \approx 19.3$ and $b_T \approx 25.6$, 
and for the exponents 
$\alpha_\mu \approx \alpha_D \approx \alpha_T \approx 2.8$ and 
$\bar{\alpha}_\mu \approx \bar{\alpha}_D \approx  \bar{\alpha}_T
 \approx 3.7$.
\bibitem{fn8}
 Table~6 shows, that the innermost nodes of the regular solutions
 decrease approximately exponentially
 with $j$ for fixed $n$.
\bibitem{fn6}
 The equations of \cite{gal,volkov3} contain a scale factor $R_e$. 
 They agree with our equations (\ref{eqmu1})-(\ref{eqdil1})
  for $R_e=2$. 
 The black hole solutions of \cite{gal} are presented for $R_e=1$, 
 and therefore tend to a scaled limiting solution with charge $P=1$.
\end{thebibliography}
\end{document}